\documentclass[a4paper, 12pt]{article}
\pdfoutput=1

\usepackage{latexsym,amsmath,amsfonts,amssymb}
\usepackage{tikz}
\usetikzlibrary{
  calc,
  fadings,
  shadings,
  decorations.pathmorphing,
  decorations.markings
}
\usetikzlibrary{arrows,decorations.pathmorphing,cd,decorations.markings,calc}
\usepackage{mathrsfs}
\usepackage[latin1]{inputenc}
\usepackage[american]{babel}
\usepackage{graphicx}
\usepackage{bbm}
\usepackage{cite}
\usepackage[colorlinks=true, citecolor=blue, linkcolor=blue, linktocpage=true]{hyperref}
\usepackage{tcolorbox}
\usepackage{skak}
\usepackage{enumerate}
\tikzset{snake it/.style={decorate, decoration=snake}}

\usepackage[a4paper,
  left=1.75cm,
  right=1.75cm,
  top=1.75cm,
  bottom=2cm
]{geometry}

\newcommand{\ket}[1]{\ensuremath{| #1 \rangle}}
\newcommand{\bra}[1]{\ensuremath{\langle #1 |}}

\newcommand{\ds}{\displaystyle}

\numberwithin{equation}{section}

\newcommand{\be}{\begin{equation}} \newcommand{\ee}{\end{equation}}
\newcommand{\bea}{\begin{equation} \begin{aligned}} \newcommand{\eea}{\end{aligned} \end{equation}}

\newcommand{\cA}{\mathcal{A}}

\newcommand{\cC}{\mathcal{C}}

\newcommand{\cF}{\mathcal{F}}

\newcommand{\cI}{\mathcal{I}}

\newcommand{\cL}{\mathcal{L}}

\newcommand{\cM}{\mathcal{M}}
\newcommand{\calM}{\mathscr{M}}
\newcommand{\cN}{\mathcal{N}}
\newcommand{\cO}{\mathcal{O}}

\newcommand{\cR}{\mathcal{R}}
\newcommand{\cS}{\mathcal{S}}
\newcommand{\cT}{\mathcal{T}}

\newcommand{\cZ}{\mathcal{Z}}
\newcommand{\bA}{\mathbb{A}}
\newcommand{\bB}{\mathbb{B}}

\newcommand{\bH}{\mathbb{H}}

\newcommand{\bN}{\mathbb{N}}

\newcommand{\bR}{\mathbb{R}}

\newcommand{\bZ}{\mathbb{Z}}

\newcommand{\unit}{\mathbbm{1}}

\def\repa{\raise4pt\hbox{$\square$}\mkern-14mu\raise-4pt\hbox{$\square$}}
\def\repab{\overline{\raise4pt\hbox{$\square$}\mkern-14mu\raise-4pt\hbox{$\square$}\mkern-1mu}}

\DeclareMathOperator{\sign}{sign}

\DeclareMathOperator{\diag}{diag}

\DeclareMathOperator{\abA}{\mathbb{A}}

\DeclareMathOperator{\calD}{\mathscr{D}}

\makeatletter
\DeclareFontFamily{OMX}{MnSymbolE}{}
\DeclareSymbolFont{MnLargeSymbols}{OMX}{MnSymbolE}{m}{n}
\SetSymbolFont{MnLargeSymbols}{bold}{OMX}{MnSymbolE}{b}{n}
\DeclareFontShape{OMX}{MnSymbolE}{m}{n}{
    <-6>  MnSymbolE5
   <6-7>  MnSymbolE6
   <7-8>  MnSymbolE7
   <8-9>  MnSymbolE8
   <9-10> MnSymbolE9
  <10-12> MnSymbolE10
  <12->   MnSymbolE12
}{}
\DeclareFontShape{OMX}{MnSymbolE}{b}{n}{
    <-6>  MnSymbolE-Bold5
   <6-7>  MnSymbolE-Bold6
   <7-8>  MnSymbolE-Bold7
   <8-9>  MnSymbolE-Bold8
   <9-10> MnSymbolE-Bold9
  <10-12> MnSymbolE-Bold10
  <12->   MnSymbolE-Bold12
}{}

\let\llangle\@undefined
\let\rrangle\@undefined
\DeclareMathDelimiter{\llangle}{\mathopen}%
                     {MnLargeSymbols}{'164}{MnLargeSymbols}{'164}
\DeclareMathDelimiter{\rrangle}{\mathclose}%
                     {MnLargeSymbols}{'171}{MnLargeSymbols}{'171}
\makeatother

\tikzset{
  ->-/.style={
    decoration={markings, mark=at position 0.55 with {\arrow{Latex}}},
    postaction={decorate}
  },
  dot/.style={
    circle,
    draw=black,
    fill=white,
    inner sep=1.2pt
  }
}

\newcommand{\symparticle}{\text{\tikz[baseline=-0.1ex, x=0.8ex, y=0.8ex, line cap=round, line join=round]{
  \useasboundingbox (-1.7, -1.2) rectangle (1.7, 1.2);
  \draw[line width=0.08em] (-1.5, -0.5) to[out=0, in=180] (0, 1) to[out=0, in=180] (1.5, -0.5);
}}}

\newcommand{\symkink}{\text{\tikz[baseline=-0.1ex, x=0.8ex, y=0.8ex, line cap=round, line join=round]{
  \useasboundingbox (-1.7, -1.2) rectangle (1.7, 1.2);
  \draw[line width=0.08em] (-1.5, -0.9) 
    -- (-0.6, -0.9) 
    .. controls (-0.1, -0.9) and (0.1, 0.9) .. 
    (0.6, 0.9) 
    -- (1.5, 0.9);
}}}

\newcommand{\symbreather}{\text{\tikz[baseline=-0.1ex, x=0.8ex, y=0.8ex, line cap=round, line join=round]{
  \useasboundingbox (-1.7, -1.2) rectangle (1.7, 1.2);
  \draw[line width=0.08em] (-1.5, 0) 
    to[out=0, in=180] (-0.75, 0.8) 
    to[out=0, in=180] (0, -1) 
    to[out=0, in=180] (0.75, 0.8) 
    to[out=0, in=180] (1.5, 0);
}}}

\begin{document}

\thispagestyle{empty}

\vspace*{20mm}  
\begin{center}
	{\LARGE 
    \textbf{A Twist on Scattering from Defect Anomalies}
  }
	\\[13mm]
   {\large Andrea Antinucci$^{\symparticle}$, Christian Copetti$^{\symparticle}$ \\[0.6em]
        Giovanni Galati$^{\symbreather}$ and Giovanni Rizi$^{\symkink}$
        }

	\bigskip
	{\it
		 $\symparticle$ Mathematical Institute, University
of Oxford, Woodstock Road, Oxford, OX2 6GG, UK \\[.6em]
    $\symbreather$  Physique Theorique et Mathematique and International Solvay Institutes
Universite Libre de Bruxelles, C.P. 231, 1050 Brussels, Belgium \\[.6em]
${\symkink}$ Institut des Hautes Etudes Scientifiques, 91440 Bures-sur-Yvette, France
	} 
\end{center}

\bigskip

 \begin{abstract}
 \noindent
In the presence of extended defects, familiar incoming particles can scatter into exotic outgoing states created by twist operators. We show that one possible mechanism driving these ``categorical scattering" processes is the presence of localized 't Hooft anomalies on the defect's worldvolume. Defect anomalies trap non-trivial charges at junctions between the symmetry lines and the interface, opening new transmission channels that would naively appear to violate selection rules. 
After outlining the general mechanism, we investigate several concrete examples with defects, interfaces, and boundaries. For models of massless chiral fermions already studied in the literature, we show that the emergence of twist operators can be understood as a consequence of defect anomalies. We then introduce new massive integrable theories in which a similar phenomenon occurs, and we explicitly solve the associated scattering problem, obtaining new integrable solutions. Finally, we construct lattice spin chains with defects where similar physics is expected to arise.
 \end{abstract}

\newpage

 \tableofcontents

\newpage

\section{Introduction}

Scattering amplitudes are among the most important observables in quantum systems. They encode the probability of detecting a given set of outgoing particles, after preparing a specified incoming state. This information is collected in the S-matrix, whose entries depend on the quantum numbers of the external particles involved in the process, and it is constrained by unitarity and global symmetries of the theory. The study of S-matrices has a long and successful history, providing a powerful framework for predicting scattering processes of fundamental particles across a wide range of energy scales, and discovering new kinds of particles, bound states, and resonances.

Although the standard setup usually describes processes involving only dynamical particles, one may also consider more general situations in which very heavy particles interact with light degrees of freedom. At low energies, such processes can be effectively described by modeling the heavy particles as extended defects, localized at fixed positions in space and extending along the time direction. Studying such processes is particularly useful from an effective field theory perspective, as it may provide indirect access to aspects of high-energy physics. The same framework is also well suited to describe experimental setups in which particles scatter off impurities, defects, or boundaries in a given material, and is therefore relevant in a broad range of physical contexts.\footnote{The study of extended defects has also applications beyond scattering experiments. For instance they give rise to new universality classes, and phase transitions. See \cite{Billo:2016cpy,Herzog:2017xha,Lauria:2020emq,Cuomo:2021rkm, Cuomo:2022xgw,Rodriguez-Gomez:2022gbz,Aharony:2022ntz,Aharony:2023amq,Barkeshli:2025cjs,Popov:2025cha,Komargodski:2025jbu,Diatlyk:2026gab} for a non-exhaustive list of references.} A striking aspect of this problem is that, in the presence of a defect, familiar incoming states associated with particles of the bulk theory may scatter into outgoing states containing excitations that are exotic from the bulk perspective. These states are created by operators attached to topological lines, or surfaces, ending on the defect. We refer to such operators as \emph{twist operators}. Even though this can sound exotic, it was recently pointed out that in some realistic situations, such as the scattering of electrons off heavy magnetic monopoles (the so--called Callan--Rubakov effect \cite{Callan,Rubakov_1988,CallanPhysRevD.25.2141,RUBAKOV1982311}), this is the only consistent channel compatible with symmetries \cite{vanBeest:2023dbu}. Following this observation, several related setups have been studied in which the same phenomenon takes place \cite{vanBeest:2023mbs,Loladze:2024ayk,Bolognesi:2024kkb,Loladze:2025jsq,Ueda:2025ecm,Tachikawa:2026cxd}.
In all of these works, the exotic scattering channel is related to the structure of boundary conditions for (1+1)d fermionic systems \cite{Callan:1994ub,Maldacena:1995pq,Tong:2019bbk,Smith:2019jnh}.

The purpose of this work is to shed light on the mechanism underlying this phenomenon. We mainly focus on $1+1$-dimensional quantum systems in which a single-particle state scatters off a localized interface or boundary. This setup has a long history, due to the rich structure of scattering processes in $1+1$ dimensions; see e.g. \cite{Dorey:1996gd} for a review. One of its main advantages is that, in the presence of integrability, two-dimensional scattering amplitudes can sometimes be computed exactly, while still encoding non-trivial dynamics and interesting physics. Moreover, two-dimensional theories are perfect playgrounds to study processes involving topological lines and their endpoints, due to the richness of generalized symmetry in $2$d \cite{Chang:2018iay}.\footnote{See \cite{Schafer-Nameki:2023jdn,Shao:2023gho,Kaidi:2026urc} for reviews on generalized symmetries.} Finally, the two--dimensional scattering of particles off boundaries has also applications to higher--dimensional scattering processes when restricted to the s-wave channels.

\begin{figure}[t]
\centering
\begin{tikzpicture}[scale=1]

\begin{scope}[xshift=0cm]

   
    \node at (0.3,2.6) {$B$};

    \draw[line width=1.1pt, black] (0,-2.4) -- (0,2.5);

    \draw[dashed, line width=0.9pt] (-2.2,-1.7) -- (0,0);
    \draw[red, line width=0.9pt, decorate,
      decoration={snake, amplitude=1.5pt, segment length=8pt}](0,0) -- (-2.2,1.7);

\fill[gray!15] (0,-2.4) rectangle (1.0,2.5);

    \fill[black] (-2.2,-1.7) circle (2.2pt) node[below right] {};
    \fill[black] (-2.2,1.7) circle (2.2pt) node[above right] {};

   
    \node at (0.3,2.6) {$B$};
    \draw[line width=1.0pt, ->] (-2.2,-1.7) -- (-1.9,-1.47);
    \draw[line width=1.0pt, ->] (-2.2,1.7) -- (-2.55,1.98);

    \draw[line width=1.0pt, ->] (2.7,-1.7) -- (2.7,-0.7) node[below right]{$t$};

    \fill[red] (0,0) circle (2.4pt) node[below right]{\small$q_{\calD}$};

    \node[align=center] at (-3.,-1.25) {in--going\\particle};
     \node[align=center] at (-3.,1.0) {reflected\\exotic particle};

\end{scope}

\begin{scope}[xshift=8cm]

        \node at (0.3,2.6) {$\calD$};

    \draw[line width=1.1pt, black] (0,-2.4) -- (0,2.5);

    \draw[dashed, line width=0.9pt] (-2.2,-1.7) -- (0,0);
   
    \draw[red, line width=0.9pt, decorate,
      decoration={snake, amplitude=1.5pt, segment length=8pt}]
      (0,0) -- (2.2,1.7);

    \fill[black] (-2.2,-1.7) circle (2.2pt) node[below right]{};
   
    \fill[black] (2.2,1.7) circle (2.2pt) node[above left] {};

    \fill[red] (0,0) circle (2.4pt) node[below right]{\small$q_{\calD}$};

    \draw[line width=1.0pt, ->] (-2.2,-1.7) -- (-1.9,-1.47);
    
    \draw[line width=1.0pt, ->] (2.2,1.7) -- (2.55,1.98);

    \node[align=center] at (-3.,-1.25) {in--going\\particle};
     \node[align=center] at (3.35,1.20) {transmitted\\exotic particle};
\end{scope}

\end{tikzpicture}
\caption{Scattering of a single particle state created by a genuine local operator off a boundary (Left) or an interface (Right). In both cases, the out--going state can be created by a non-genuine local operator attached to a topological line. The presence of a defect anomaly implies that some of the global charge is stored in the topological junction and localized on the defect.}
\label{fig: intro}
\end{figure}
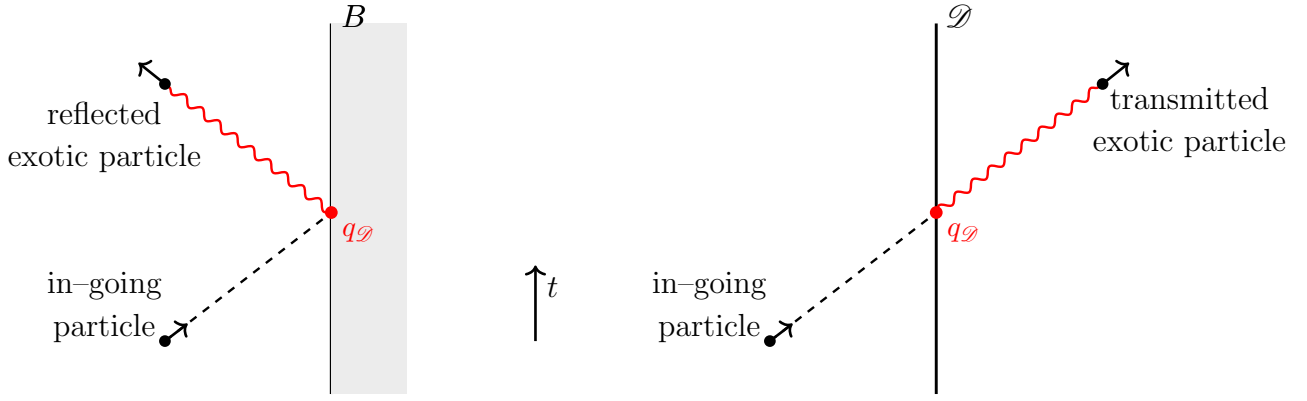

The main insight of this work is that these exotic scattering processes, which we dub \emph{categorical scattering}, are intimately tied to the realization of global symmetries in a theory with extended defects. Indeed, recent developments have shown that symmetries can be realized in the presence of interfaces or boundaries in ways that are richer than one might naively expect \cite{Antinucci:2024izg,Komargodski:2025jbu}. There are two main concepts that are crucial for our analysis, which we review in Section \ref{sec:symdef}.
\begin{itemize}
\item The first is the one of \emph{symmetry reflecting defects}: in the presence of interfaces, global symmetry charges can be conserved independently on the two sides of the interface. This implies that topological lines generating the symmetry can end topologically on the interface, and therefore twist operators can be connected to the defect via a topological line. Even though this property is universal for symmetric boundary conditions, when we discuss interfaces, the distinction between symmetric and symmetry reflecting interfaces becomes crucial. As we will make clear in Section \ref{sec: defect sel rules}, the symmetry reflecting property is what makes the seemingly unconventional scattering outcomes to become come conventional, in the sense that they belong to the same superselection sector as the incoming state.
\item The second concept is the one of \emph{defect anomalies}, namely 't Hooft anomalies localized on the defect. Among their various consequences, we argue that defect anomalies imply that topological junctions between symmetry lines and the defect carry non-trivial charges. This property plays a crucial role: it opens new scattering channels in which twist operators, creating particle states, can emerge after an ordinary particle scatters off the interface or boundaries. For example, for symmetry-reflecting interfaces, Ward identities forbid a charged particle from transmitting charged energy states. By contrast, the transmission of a twist operator through the interface can be compatible with charge conservation, because of the charged topological junction left behind on the interface, see Figure~\ref{fig: intro}.
\end{itemize} 

After a brief review of Tube and Strip algebra given in Section \ref{sec: tubestrip}, in Section \ref{sec: defect sel rules} we analyze the underlying mathematical structure which organizes particle states, and operators creating them into various multiplets with the same symmetry charge. The main concept is the one of \emph{Defect Strip algebra}, which slightly modifies the more usual concept of Strip algebra to take into account both an asymptotic boundary and a physical defect. We show that the mechanism by which defect anomalies open new scattering channels emerges naturally within this formalism, which also accommodates more general non-invertible symmetries described by fusion categories. In particular, the symmetry structure carried by the defect can make particle states that would belong to distinct superselection sectors in the absence of the defect part of the same symmetry multiplet. Consequently, such states are allowed to mix in scattering processes.

After this general discussion, we move onto the analysis of specific models where categorical scattering is expected. In Section \ref{sec: massless scattering} we analyze the two models already described in the literature: the $3450$--model and the fermion--rotor system. Here, we focus primarily on the interplay between boundaries and defects, and the bulk global symmetries of the two theories. We show that these objects carry mixed defect anomalies for such symmetries. These anomalies are responsible for the emergence of twist operators among the outgoing states of scattering amplitudes.

In Section~\ref{sec:soliton}, we then turn to new models in which the same phenomenon takes place. We focus on massive theories, for which the scattering matrix is well defined and free from IR divergences. Our first example is the Ising field theory, without a magnetic field, in the presence of an interface separating the ordered and disordered phases. We show that this interface is symmetry-reflecting with respect to the global $\bZ_2$ symmetry of the Ising model, and that it carries the associated defect anomaly. Using the free-field realization of the theory, we solve the scattering problem exactly and derive explicit reflection and transmission amplitudes as functions of the rapidity of the incoming single-particle state, and the parameters defining the interface. By analyzing the structure of the correlators and form factors in the presence of the interface, we argue that the transmitted radiation is a single-particle state created by the Ising twist operator. Moreover, we show that these amplitudes have poles corresponding to the presence of a zero mode making the defect vacuum state degenerate. This degeneracy is a direct consequence of the defect anomaly.

 Our second example is the tricritical Ising CFT deformed by the integrable perturbation $\sigma '$ (or $\Phi_{2,1}$ in the Kac notation) on a manifold with boundaries. The bulk theory preserves the non-invertible Fibonacci symmetry of the UV CFT, and we fix weakly symmetric boundary conditions so that this symmetry is preserved along the boundary flow as well. By imposing the constraints of the defect strip algebra, we show that incoming breather particles can be transmuted into kinks upon scattering off the boundary. We further show that this process is compatible with integrability, and present a new explicit integrable solution to this scattering problem. 
 
Finally, in Section~\ref{sec: lattice}, we turn to lattice models with impurities. As recently observed in~\cite{Ueda:2025ecm}, the same phenomenon also arises in this setting. We show that defect anomalies provide the underlying mechanism for categorical scattering even on the lattice. We establish this by computing the defect anomaly in the model in~\cite{Ueda:2025ecm}, and by introducing new examples with similar, yet richer, physics: XYZ spin chains enriched by various types of defects, all carrying a defect anomaly. Several Appendices provide details of the general setup and on techniques used in the main text.

To conclude, let us highlight some interesting open directions connected with our results:
\begin{enumerate}
    \item \emph{Higher dimensions:} In this work, we consider only (1+1) dimensional examples. However the motivating problem of fermion-monopole scattering is really an higher dimensional one. Defect anomalies can be described also in higher-dimensional setups \cite{Antinucci:2024izg} so we expect our construction to provide leverage also in these cases.
    In higher dimensions, the role of twist operators is played by the so-called monodromy defects. In this case the description of the twisted Hilbert space and the role played by anomalies needs to be clarified and expanded. Furthermore, monodromy defects also behave very much like a fractional magnetic monopole. Thus, they also provide a natural setup for categorical scattering to take place.
    \item \emph{Non-invertible symmetries:} While here we focus mostly on invertible symmetries (excluding the study of Fib in Section \ref{sec:soliton}), our selection rules can be extended to more general cases involving non-invertible symmetries, too. These should become transparent in a boundary/defect SymTFT description \cite{Bhardwaj:2024igy,Choi:2024tri,Copetti:2024onh}, that is the natural framework for a systematic analysis beyond specific examples. The recent results of \cite{Bartsch:2026wqq} should also be relevant for such an extension.
    \item \emph{Gauge theories in (1+1)d:} Non-Abelian gauge theories  with matter in (1+1)d provides a rich arena for massive RG flows with defects that are symmetry reflecting and have anomalies \cite{Komargodski:2020mxz}. Our general results apply there, but the concrete examples are challenging for the lack of additional structures like integrability.
    \item \emph{Bootstrap:} In \cite{Copetti:2024dcz}, the authors leveraged the representation theory of the Strip algebra to efficiently implement an S-matrix bootstrap in the presence of solitons. It is natural to assume that a similar role can be fulfilled by the defect strip algebra in studying scattering amplitudes in the presence of defects.
    \item \emph{Lattice impurities:} In Section \ref{sec: lattice}, we constructed symmetry reflecting lattice defects in the XYZ model and showed that they carry a defect anomaly. It would be worthwhile to solve the associated scattering problem explicitly, for example along the lines of the analysis presented in \cite{Ueda:2025ecm}, in order to determine whether there are channels that allow twist operators to be transmitted through the defect. In this setup, we have also argued that, at a generic point in the bulk parameter space away from the XX model, the defects we constructed are likely to be factorized, and therefore forbid transmission. Therefore, the only interesting point for the scattering problem is the XX theory. It would be interesting to look for other classes of lattice defects that are not factorized, while still carrying a defect anomaly.
    \item \emph{Topological data from Scattering:} Finally, interfaces carrying defect anomalies provide a natural way to probe distinct (gapless) topological phases \cite{Prembabu:2025qvi}. It would be enticing to understand precisely how such data could be probed by scattering experiments in the presence of the interface.\footnote{See also \cite{brouwer1998scattering,braunlich2010equivalence,Lo:2026sum} for related ideas.}
\end{enumerate}

\paragraph{Authors' Note:} while this work was being finalized, we were made aware of related upcoming work \cite{rishi}, which has some overlap with the discussion in Section \ref{sec: massless scattering}. We thank the authors of \cite{rishi} for coordinating submission.

\section{Symmetries and Defects}\label{sec:symdef}
We consider a bulk theory with a global 0-form symmetry $\cS$. In this section we review how $\cS$ can be realized in presence of a defect $\calD$. If $\calD$ has codimension one, it can act as an interface between two distinct theories. If one of those theories is an SPT phase, $\calD$ effectively serves as a boundary condition. The notions of symmetric and symmetry reflecting defects are completely general, while the definition of defect anomalies is only given if $\cS$ is invertible with some underlying group $G$. On the other hand, in Section \ref{sec: tubestrip}, restricting to the (1+1)d context we will discuss several notions for a completely general fusion category symmetry. 
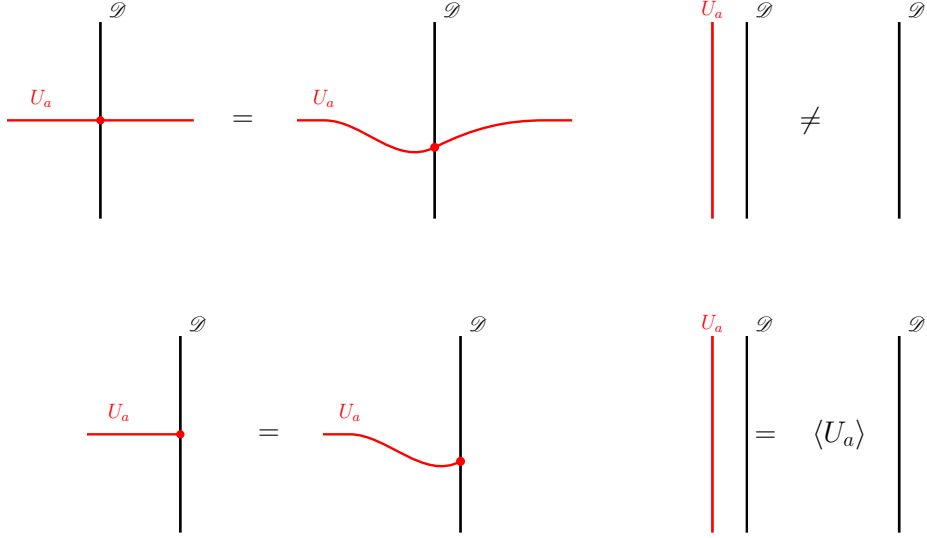
\begin{figure}[t]
\centering

\begin{tabular}{c@{\hspace{1.6cm}}c}

\begin{tikzpicture}[scale=0.65, transform shape]

    \draw[line width=1.0pt, black] (-3,-2.0) -- (-3,2.0);
    \draw[line width=0.9pt, red] (-4.9,0) -- (-1.1,0);
    \fill[red] (-3,0) circle (2.4pt);

    \node at (-2.65,2.25) {$\calD$};
    \node[red] at (-4.2,0.45) {$U_a$};

    \node at (-0.1,0) {\Large $=$};

    \draw[line width=1.0pt, black] (3.8,-2.0) -- (3.8,2.0);

    \draw[line width=0.9pt, red]
        (1.0,0) -- (1.55,0)
        .. controls (2.35,0) and (3.00,-0.95) .. (3.8,-0.55)
        .. controls (4.60,-0.15) and (5.25,0) .. (6.05,0)
        -- (6.6,0);

    \fill[red] (3.8,-0.55) circle (2.6pt);

    \node at (4.15,2.25) {$\calD$};
    \node[red] at (1.55,0.45) {$U_a$};

\end{tikzpicture}

&

\begin{tikzpicture}[scale=0.65, transform shape]

    \draw[line width=1.0pt, red] (-2,-2.0) -- (-2,2.0);
    \draw[line width=0.9pt, black] (-1.3,-2.0) -- (-1.3,2.0);

    \node[red] at (-2,2.25) {$U_a$};
    \node at (-0.95,2.25) {$\calD$};

    \node at (0,0) {\Large $\not=$};

    \draw[line width=1.0pt, black] (1.8,-2.0) -- (1.8,2.0);

    \node at (2.15,2.25) {$\calD$};

\end{tikzpicture}

\\[1.0cm]

\begin{tikzpicture}[scale=0.65, transform shape]

    \draw[line width=1.0pt, black] (-1,-2.0) -- (-1,2.0);
    \draw[line width=0.9pt, red] (-2.9,0) -- (-1,0);
    \fill[red] (-1,0) circle (2.4pt);

    \node at (-0.65,2.25) {$\calD$};
    \node[red] at (-2.25,0.45) {$U_a$};

    \node at (0.8,0) {\Large $=$};

    \draw[line width=1.0pt, black] (4.7,-2.0) -- (4.7,2.0);

    \draw[line width=0.9pt, red]
        (1.9,0) -- (2.45,0)
        .. controls (3.20,0) and (3.95,-0.95) .. (4.7,-0.55);

    \fill[red] (4.7,-0.55) circle (2.6pt);

    \node at (5.05,2.25) {$\calD$};
    \node[red] at (2.45,0.45) {$U_a$};

\end{tikzpicture}

&

\begin{tikzpicture}[scale=0.65, transform shape]

    \draw[line width=1.0pt, red] (-2,-2.0) -- (-2,2.0);
    \draw[line width=0.9pt, black] (-1.3,-2.0) -- (-1.3,2.0);

    \node[red] at (-2,2.25) {$U_a$};
    \node at (-0.95,2.25) {$\calD$};

    \node at (0,0) {\Large $= \quad \langle U_a\rangle$} ;

    \draw[line width=1.0pt, black] (1.8,-2.0) -- (1.8,2.0);

    \node at (2.15,2.25) {$\calD$};

\end{tikzpicture}

\end{tabular}

\caption{Upper figure: A symmetric defect $\calD$. A topological symmetry line $U_a$ can intersect $\calD$ topologically, but it cannot be absorbed by it.  
Lower figure: A symmetry-reflecting defect $\calD$. A topological symmetry line $U_a$ can end topologically on $\calD$, and can also be absorbed by it. }
\label{fig:symdef}

\end{figure}
\subsection{Symmetric and Symmetry Reflecting Defects}
Let us denote by $\gamma_p$ the $p-$dimensional world-volume of the defect $\calD$. The symmetry $\cS$ is implemented by topological operators $U_a(\Sigma)$ supported on a co-dimension one submanifold $\Sigma$ of space-time. In a canonically quantized picture $\Sigma$ is space, and $U_a(\Sigma)$ is an operator that commutes with the Hamiltonian $H$. A defect $\calD$ that extends along time, say $\gamma_p=\gamma_{p-1}\times \bR$, represents an impurity localized in space at $\gamma_{p-1}\subset \Sigma$, giving rise to a local modification of the Hamiltonian supported on $\gamma_{p-1}$, called defect Hamiltonian $H_{\calD}$. Its presence can break the symmetry $\cS$
\begin{equation}
    [U_a(\Sigma),H_{\calD}] \neq 0 \ .
\end{equation}
This non-commutativity only arises at the \emph{junction} $\gamma_{p-1}=\Sigma \cap \gamma_p$ where the defect and the symmetry operator intersect. 

\paragraph{Symmetric defects.} It can sometimes happen that, using the degrees of freedom localized on the defect, one can construct an improved symmetry operator by dressing $U_a(\Sigma)$ with a defect operator $u_a(\gamma_{p-1})$ supported on the junction $\widetilde{U}_a(\Sigma)=U_a(\Sigma) u_a(\gamma_{p-1})$ such that
\begin{equation}
    [\widetilde{U}_a(\Sigma),H_{\calD}]=0\,.
\end{equation}
The existence of $\widetilde{U}_a(\Sigma)$ defines the notion of $\cS-$symmetric defect $\calD$. In a covariant QFT, without any specific reference to a canonically quantized picture, we say that a defect $\calD$ supported on $\gamma_p$ is symmetric if there exist operators $  \widetilde{U}_a(\Sigma)$ implementing $\cS$ that remain topological even if $\gamma_p \cap \Sigma \neq \emptyset$, see Figure \ref{fig:symdef}. Conversely, if this condition is not satisfied the defect is said to be symmetry breaking.

If the symmetry $\cS$ is continuous with an underlying Lie group $G$ and a current $J^\mu$, the discussion above can be rephrased in terms of the tilt operator \cite{Padayasi:2021sik,Drukker:2022pxk, Herzog:2023dop} arising in the (broken) Ward-identities in presence of the defect. This viewpoint will not be used in the present paper, and we refer the interested reader to \cite{Antinucci:2024izg} for the link between the two presentations.

\paragraph{Symmetry reflecting defects.} For co-dimension one ($p=d-1$) symmetric defects we can define a stronger notion. The defect $\calD$ extended along time intersects space $\Sigma$ along $\gamma_{d-2}$, dividing it into two regions $\Sigma=\Sigma_L \cup \Sigma_R$. If it is possible to write the symmetry operator as $U_a(\Sigma)=U_a(\Sigma_L) \otimes U_a(\Sigma_R)$ such that the two localized operators independently commute with the defect Hamiltonian
\begin{equation}[U_a(\Sigma_{L}),H_{\calD}]=[U_a(\Sigma_{R}),H_{\calD}]=0 \ ,
\end{equation}
then $\calD$ is called \emph{symmetry reflecting} \cite{Antinucci:2024izg}. This formulation makes it clear that if $\calD$ is symmetry reflecting, the defect Hamiltonian has two separate symmetries.

In the euclidean QFT picture, a symmetry reflecting defect is characterized by the property that $U_a$ can topologically terminate on it, see Figure \ref{fig:symdef}.
More precisely, we need to define the operator $U_a(\Sigma)$ when $\Sigma$ has a boundary $\partial \Sigma \subset \gamma _{d-1}$. In fact this is always the case: the boundary requires the extra data of an operator living there. Importantly, here the boundary is embedded in the defect world-volume, hence this boundary operator can be constructed using degrees of freedom localized on the defect\footnote{Notice that, unless the symmetry $\cS$ acts trivially in the bulk, $U_a$ ending on the defect has a chance to be topological only if the boundary operator involves defect degrees of freedom. Otherwise the open topological defect would exist already far from the defect, that is incompatible with a faithfully acting symmetry.}. This is completely analogous to the dressed defect discussed above in the definition of a symmetric defect.

  The fact that a symmetry reflecting defect effectively has two separate symmetries, has crucial consequences:
\begin{itemize}
    \item In presence of an $\cS-$symmetry reflecting defect, the selection rules (if $\cS$ is not spontaneously broken) must be satisfied independently on the two sides of $\calD$. In particular if $\cS$ is an ordinary invertible symmetry, charge conservation must be satisfied on the left and on the right independently: the charge cannot be transmitted across the defect. At the level of correlation functions, this means that if $\cO_1^L(x_1),...,\cO_{n_L}^L(x_{n_L})$ and $\cO_1^R(x_1),...,\cO_{n_R}^R(x_{n_R})$ are supported respectively on the left and right of $\calD$, then
    \begin{equation}\label{eq:selection rule basic}
        \big\langle \cO_1^L(x_1)\cdots\cO_{n_L}^L(x_{n_L}) \cO_1^R(x_1)\cdots \cO_{n_R}^R(x_{n_R})  \big\rangle \neq 0 \ , \ \ \ \ \text{only if} \ \ \ \begin{array}{l}
             \cR_1^L\otimes \cdots \otimes \cR_{n_L}^L \supset 1 \\
               \cR_1^R\otimes \cdots \otimes \cR_{n_R}^R \supset 1 
        \end{array}
    \end{equation}
    A different possibility for obtaining a non-vanishing correlator is to also insert defect operators. In fact a defect operator $\cO_{\calD}(x)$ may carry non-trivial charge under the bulk symmetries, and its insertion can then compensate a non-vanishing bulk charge.
    \item The most general background field configuration for $\cS$, in presence of $\calD$, involves two independent background fields $A_L,A_R$. Each of them is defined on one side of the defect, including the defect world-volume. In the topological operators picture, a flat background field corresponds to a mesh of topological defects, and here we are merely saying that the most general mesh includes $\cS-$symmetry operators terminating on $\calD$ from the left/right.
\end{itemize}

Notice that the co-dimension one allows, in general, to have two different theories on the two sides, $\cT_L$ and $\cT_R$. In this case $\calD$ is referred to as an interface, and if $\cT_L$ and $\cT_R$ both have symmetry $\cS$ the definitions above can be immediately applied in this case. A special case is when one of the two sides, say $\cT_R$, is the trivial theory (that can be assumed to be $\cS$ symmetric), hence $\calD$ is a boundary condition for $\cT_L$. In this case if $\calD$ is $\cS$-symmetric, it is automatically symmetry reflecting.

\paragraph{The folding trick.} 
More generally the two sides can have different symmetries $\cS_L$ and $\cS_R$, and this case is more conveniently treated by using the folding trick. An interface $\calD$ between $\cT_L$ and $\cT_R$ can be mapped, by folding, to a boundary condition $B_{\calD}$ for $\cT_L \otimes \overline{\cT}_R$. The folded theory has symmetry $\cS_L \times \cS_R$, and the boundary condition will preserve a subsymmetry $\cS_P$, that has an embedding $\iota : \cS_P\hookrightarrow   \cS_L \times \cS_R$. Upon unfolding, this has the following interpretation. If $\iota(x)=(x_L,x_R) \in \cS_L \times \cS_R$, then the interface $\calD$ admits a topological junction where $x_L$ from the left and $x_R$ from the right meet on the defect. The two special cases considered above are:

\begin{itemize} 
    \item If $x_R=1$ (resp. $x_L=1$) the interface is symmetry reflecting under $x_L$ (resp. $x_R$). In particular, when $\cT_L=\cT_R$, an $\cS$-symmetry reflecting defect has $\cS_P=\cS_L\times \cS_R=\cS\times \cS$.
    \item Suppose that the preserved symmetry is $\cS_P \cong \cS_L$ and there is an embedding $\phi : \cS_L \hookrightarrow \cS_R$ such that $\iota(x)=(x,\phi(x))$. Then the interface $\calD$ is $\cS_L$-symmetric, given that all the symmetry operators from the left can cross $\calD$ topologically, by joining $\phi(x)$ from the right. In the special case $\cT_L=\cT_R$, $\phi$ is an automorphism of $\cS$\footnote{Notice that this is not required to be the identity: the defect is allowed to permute the symmetry operators. The simplest example is a gauge theory with 1-form symmetry $\cS=\Gamma ^{(1)}$ as well as charge conjugation symmetry $\bZ_2$. The topological operator implementing charge conjugation is regarded as a defect $\calD$ preserving $\Gamma ^{(1)}$, and a topological operator $g\in \Gamma ^{(1)}$ intersecting $\calD$ from the left is required to join $g^{-1} \in \Gamma ^{(1)}$ from the right.}.
\end{itemize}

\subsection{Defect Anomalies}
Consider a defect $\calD$ that is symmetric under an ordinary invertible symmetry $\cS$. We consider the partition function $Z_{\calD}[A]$ in the presence of the defect, coupled with a background field $A$ for $\cS$. Gauge invariance may fail by two independent contributions, a bulk one and a defect one:
\begin{equation}
    Z_{\calD}[A^\lambda]=\exp{\left(i\int _{X_d}\alpha(A,\lambda)+i\int _{\gamma_p}\alpha_{\calD}(A,\lambda)\right)}Z_{\calD}[A] \ .
\end{equation}
$\alpha(A,\lambda)$ is the ordinary 't Hooft anomaly, that can be canceled by inflow with a $d+1$ classical bulk action $\omega(A)$ such that $\delta_\lambda \omega =d\alpha(A,\lambda)$. On the other hand $\alpha_{\calD}(A,\lambda)$ is called \emph{defect anomaly} \cite{Antinucci:2024izg} (see also \cite{Brennan:2025acl, Copetti:2025sym, Komargodski:2025jbu, Prembabu:2025qvi} for recent applications and \cite{Aharony:2023amq} for an earlier work). As the bulk anomaly, also the defect anomaly can be canceled by inflow, but the inflow action $\omega_{\calD}(A)$ is now $p+1$ dimensional. In other words, in the presence of a non-trivial background field $A$ the defect $\calD$ is no longer a full-fledged $p$-dimensional object, but it exists as the boundary of a $p+1$ dimensional surface $\gamma_{p+1}$. This surface is almost trivial, in the sense does not have interactions with other operators of the theory, but its presence modifies all correlation functions by a phase $\exp{\left(i\int _{\gamma_{p+1}} \omega_{\calD}(A)\right)}$. 

\begin{figure}[t]
\centering

\begin{tabular}{c@{\hspace{1.6cm}}c}

\begin{tikzpicture}[scale=0.65, transform shape]

    \draw[line width=1.0pt, black] (-3,-2.0) -- (-3,2.0);
    \draw[line width=0.9pt, red] (-4.9,1) -- (-1.1,1);
    \draw[line width=0.9pt, green!60!black] (-4.9,-1) -- (-1.1,-1);

    \node at (-2.65,2.25) {$\calD$};
    \node[red] at (-4.2,1.45) {$U_g$};
    \node[green!60!black] at (-4.2,-1.45) {$U_h$};

    \node at (-0.1,0) {\Large $= \omega(g,h)$};

    \draw[line width=1.0pt, black] (3.55,-2.0) -- (3.55,2.0);
    \draw[line width=0.9pt, blue!80!black] (1.5,0) -- (5.6,0);

    \node at (3.95,2.25) {$\calD$};
    \node[blue!80!black] at (2.,0.45) {$U_{gh}$};

\end{tikzpicture}

&

\begin{tikzpicture}[scale=0.65, transform shape]

    \draw[line width=1.0pt, black] (-3,-2.0) -- (-3,2.0);
    \draw[line width=0.9pt, red] (-4.9,1) -- (-1.1,1);
    \draw[line width=0.9pt, green!60!black] (-4.9,-1) -- (-1.1,-1);

    \node at (-2.65,2.25) {$\calD$};
    \node[red] at (-4.2,1.45) {$U_g$};
    \node[green!60!black] at (-4.2,-1.45) {$U_h$};

    \node at (0.5,0) {\Large $= \chi(g,h)$};
    
    \begin{scope}[xshift=7cm]
        \draw[line width=1.0pt, black] (-3,-2.0) -- (-3,2.0);
        \draw[line width=0.9pt, green!60!black] (-4.9,1) -- (-1.1,1);
        \draw[line width=0.9pt, red] (-4.9,-1) -- (-1.1,-1);

        \node at (-2.65,2.25) {$\calD$};
        \node[green!60!black] at (-4.2,1.45) {$U_{ghg^{-1}}$};
        \node[red] at (-4.2,-1.45) {$U_g$};
    \end{scope} 

\end{tikzpicture}

\\[1.2cm]

\multicolumn{2}{c}{
\begin{tikzpicture}[scale=0.65, transform shape]

    \draw[line width=1.0pt, black] (-3,-2.0) -- (-3,2.0);

    \draw[line width=0.9pt, red] (-4.9,1) -- (-3,1);
    \draw[line width=0.9pt, green!60!black] (-3,-1) -- (-1.1,-1);

    \node at (-2.65,2.25) {$\calD$};
    \node[red] at (-4.2,1.45) {$U_g$};
    \node[green!60!black] at (-1.8,-1.45) {$U_h$};

    \node at (0.5,0) {\Large $= \chi(g,h)$};
    
    \begin{scope}[xshift=7cm]
        \draw[line width=1.0pt, black] (-3,-2.0) -- (-3,2.0);

        \draw[line width=0.9pt, green!60!black] (-3,1) -- (-1.1,1);
        \draw[line width=0.9pt, red] (-4.9,-1) -- (-3,-1);

        \node at (-2.65,2.25) {$\calD$};
        \node[green!60!black] at (-1.8,1.45) {$U_h$};
        \node[red] at (-4.0,-1.45) {$U_g$};
    \end{scope} 

\end{tikzpicture}
}

\end{tabular}

\caption{Manifestation of the defect anomaly for line defects in $1+1$ dimensions. Above: A fusion of two symmetry operators intersecting a symmetric defect can produce an anomalous phase $\omega(g,h)\in H^2(G,U(1))$. This implies that their commutator produces a bi--character
$\chi(g,h):=\frac{\omega(g,h)}{\omega(ghg^{-1},g)}$. Below: how the defect anomaly is realized in a symmetry reflecting defect. In this case, the bi--character $\chi(g,h)$ can be interpreted as the charge of the topological end--points.}
\label{fig:defect anomaly}
\end{figure}
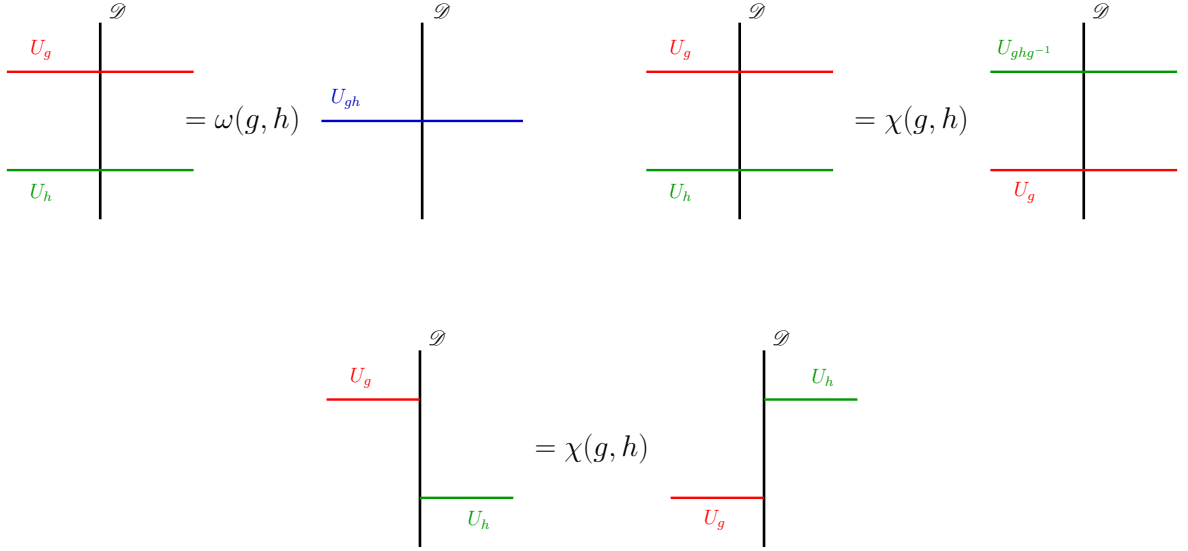

An alternative, yet equivalent viewpoint is to embed $\gamma_{p+1}$ as a submanifold of the $d+1$ space $X_{d+1}$ supporting the bulk inflow action $\omega(A)$, thus giving rise to a stratified inflow:
\begin{equation}
    S_{\text{inflow}}=i\int _{X_{d+1}}\omega(A) +i\int_{\gamma_{p+1}}\omega_{\calD}(A) \ .
\end{equation}
This is useful because it implies that defect anomalies, as their bulk counterparts, are RG invariant and robust against symmetry preserving deformations. Moreover defect anomalies can be classified by the $p+1$ dimensional topological actions that can be written using $A$, hence following the familiar classification of 't Hooft anomalies in $p$ dimensions. For instance bosonic defect anomalies of a $p-$dimensional symmetric defect under a 0-form symmetry $G$ are classified by $H^{p+1}(BG,U(1))$. Finally, in close analogy with ordinary bulk anomalies, defect anomalies can be detected,
via Poincar\'e duality, by considering suitable configurations of topological defects. The case of line defects in two spacetime dimensions is depicted in Figure \ref{fig:defect anomaly}.

\paragraph{Defect anomalies for symmetry reflecting defects.} The classification becomes different if the defect is symmetry reflecting. In this case there are more possible anomalies than for just a symmetric defect. In fact, given that more background field configurations are allowed, it is reasonable to expect that the failure of gauge invariance of $Z_{\calD}[A]$ can happen in more ways. In practice, we should remember that a background gauge field is a pair $A_L,A_R$, hence we look for topological actions $\omega(A_L,A_R)$. This shows that defect anomalies for a symmetry reflecting defect under $\cS$ are the same as 't Hooft anomalies for $\cS\times \cS$ in $d-1$ dimensions. For example bosonic anomalies for a 0-form symmetry $G$ are given by $H^d(B (G\times G),U(1))$. An example is provided by $\bZ_2$ symmetry reflecting lines in (1+1)d. While there is no 't Hooft anomaly for a single $\bZ_2$ on a line (because $H^2(B\bZ_2,U(1))=0$), a defect anomaly is in fact possible and its topological action is simply 
\begin{equation} \label{eq:defect anomaly Ising}
    \frac{2\pi i}{2}\int A_L \cup A_R \ .
\end{equation}
It corresponds to the fact that $H^2(B (\bZ_2 \times \bZ_2),U(1))=\bZ_2$. This defect anomaly is realized for instance by the duality defect in the critical Ising model \cite{Antinucci:2024izg}.

In the general case of an interface separating two theories with symmetries $\cS_L$ and $\cS_R$, the anomalies can be understood more clearly using the folding trick. They become boundary anomalies for the preserved symmetry $\cS_P\subset \cS_L \times \cS_R$. Hence for 0-form symmetries, in the bosonic case, they are classified by $H^d(B\cS_P,U(1))$. There is a particular form of defect anomalies that is of remarkable interest for us. Suppose $\cS_L$ is a finite Abelian 0-form symmetry $\bA$, while $\cS_R$ is also finite and Abelian, with underlying group $\bB$, but of degree $d-2$. Denote by $A_L \in H^1(X,\bA)$ and $B_R\in H^{d-1}(X,\bB)$ the two background fields. Suppose the defect is fully symmetry reflecting under $\cS_L$ and $\cS_R$ (namely $\cS_P=\cS_L\times \cS_R$), and the defect anomaly takes the form
\begin{equation}
    2\pi i k \int _{Y_d}A_L\cup B_R \ .
\end{equation}
Here $\partial Y_d=\gamma_{d-1}$, with the defect located on $\gamma_{d-1}$, and we assumed that there is a pairing $\chi : \bA \times \bB \rightarrow \bR/\bZ$ that we used to define the $\cup$ product of the two background fields. Notice that \eqref{eq:defect anomaly Ising} is a particular case of this, for $d=2$ and $\bA=\bB=\bZ_2$. In any dimension $d$, the degree of the symmetry $\bB$ is such that its topological operators are lines. The important point here, is that this anomaly is equivalent to the statement that the (topological) endpoint of the $\bB$ symmetry generators on $\calD$ carry a non-trivial charge under $\bA$. The charge of $\phi_b(x)$, the topological defect operator at the end of $b\in \bB$, is given by 
\begin{equation}
  e^{2\pi i k \chi(\, . \, ,b)} \in \text{Hom}(\bA,U(1))=\bA^\vee \ .
\end{equation}
Recalling the discussion around \eqref{eq:selection rule basic}, we conclude immediately that in a correlation function involving a $\bB$-line attached to the defect $\calD$, the presence of $\phi_b(x)$ modifies the selection rules. This will be discussed in much more detail in Section \ref{sec: defect sel rules}.

\section{Tube and Strip Algebras}\label{sec: tubestrip}

We are now interested in understanding how symmetries constrain various physical observables, such as correlation functions and scattering amplitudes, in the presence of a defect. In particular, for the rest of the work we will focus on two--dimensional QFTs, where the mathematical framework is more under control.

Before delving into the analysis in presence of defects, let us start by reviewing the more standard mathematical framework to describe the Ward identities imposed by the symmetry $\cS$ on various bulk observables: the Tube algebra \cite{Chang:2018iay}, which describes symmetry action on local (twist) fields and
the Strip algebra \cite{Kitaev:2011dxc,Cordova:2024iti,Copetti:2024dcz}, which instead describes the quantum numbers of particles and kinks. We refer to \cite{Aasen:2020jwb,Copetti:2024dcz} for our conventions about fusion and module categories.

\subsection{The Tube Algebra} 
\begin{figure}
    \centering
    \begin{tikzpicture}[scale=0.75, baseline={(0,0)}, 
    thick,
    ->-/.style={
        decoration={markings, mark=at position 0.55 with {\arrow{Latex}}},
        postaction={decorate}
    },
    dot/.style={
        circle, 
        draw=black, 
        fill=white, 
        inner sep=1.2pt
    }
]
 \draw (0,-2) ellipse (1 and 0.5);    
  \draw (0,2) ellipse (1 and 0.5);    
 \draw (-1,-2) -- (-1,2); \draw (1,-2) -- (1,2);
 \draw[->-] (0,-2.5) -- (0,1.5);
 \node[right] at (0,0) {$a$};
    \end{tikzpicture} \qquad \qquad \qquad \qquad \qquad 
     \begin{tikzpicture}[scale=0.75, baseline={(0,0)}, 
    thick,
    ->-/.style={
        decoration={markings, mark=at position 0.55 with {\arrow{Latex}}},
        postaction={decorate}
    },
    dot/.style={
        circle, 
        draw=black, 
        fill=white, 
        inner sep=1.2pt
    }
]
 \draw (0,-2) ellipse (1 and 0.5);    
  \draw (0,2) ellipse (1 and 0.5);    
 \draw (-1,-2) -- (-1,2); \draw (1,-2) -- (1,2);

  \draw[->-] (0,-0.5) -- (0,1.5);
  \node[right] at (0,0) {$a$};
  \draw[fill=black] (0,-0.5) circle (0.05) node[below] {$\phi_a$};

\end{tikzpicture}
    \caption{(a) a state on the cylinder in the twisted Hilbert space $\bH_a$. (b) a twist field $\phi_a$, which maps the untwisted to the twisted Hilbert space (time flows upwards).}
    \label{fig:hilbertspaces}
\end{figure}
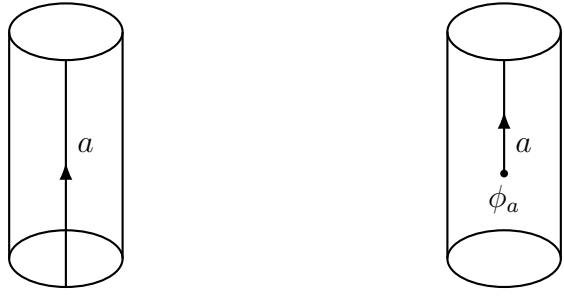

In the context of generalized symmetries, we are naturally led to consider the twisted Hilbert spaces $\bH_a$, in which a topological symmetry defect $U_a$ is placed along the time direction. 
Alternatively (or, by the state operator map in a critical system) we can consider twist \emph{fields}, namely operators $\phi_a$ providing a map $\bH \to \bH_a$, which can also be viewed as a non-topological termination of the symmetry defect $U_a$, see Figure \ref{fig:hilbertspaces}. 
Symmetries act on twist fields via the so-called lasso action $\left[U_{a,x}\right]_b^c : \bH_b \to \bH_c$ \cite{Chang:2018iay}:
\be
\left[ U_{a,x} \right]_b^c \phi_b = \begin{tikzpicture}[scale=0.75, baseline={(0,0)}, 
    thick,
    ->-/.style={
        decoration={markings, mark=at position 0.55 with {\arrow{Latex}}},
        postaction={decorate}
    },
    dot/.style={
        circle, 
        draw=black, 
        fill=white, 
        inner sep=1.2pt
    }
]
    \draw[->-] (0,0) -- (0,1); \draw (0,1)-- (0,2);
    \draw[->-] (0,1) arc (90:-270:1 and 1);
    \node[below] at (0,0) {$\phi_b$};
    \node[left] at (-1,0) {$a$};
    \node[left] at (0,0.5) {$b$};
    \node[above] at (0,2) {$c$};
    \draw[fill=black] (0,1) circle (0.05);
    \node at (0.25,1.25) {$x$};
    \draw[fill=black] (0,0) circle (0.05);
\end{tikzpicture}
= 
\begin{tikzpicture}[scale=0.75, baseline={(0,0)}, 
    thick,
    ->-/.style={
        decoration={markings, mark=at position 0.55 with {\arrow{Latex}}},
        postaction={decorate}
    },
    dot/.style={
        circle, 
        draw=black, 
        fill=white, 
        inner sep=1.2pt
    }
]
    \draw[->-] (0,0) -- (0,1); \draw (0,1)-- (0,2);
    \draw[->-] (0,1) arc (90:-90: 1 and 1);
    \draw (0,-1) arc (-90:-270: 1.25 and 1.25);
    \node[below] at (0,0) {$\phi_b$};
    \node[left] at (-1.25,0.25) {$a$};
    \node[left] at (0,0.5) {$b$};
    \node[above] at (0,2) {$c$};
    \draw[fill=black] (0,1) circle (0.05);
        \draw[fill=black] (0,1.5) circle (0.05);
\node at (0.25,1.25) {$x$};
    \draw[fill=black] (0,0) circle (0.05);
\end{tikzpicture}
=\sum_{\phi_c} [\tau_{x,b}]_{\phi_a}^{\phi_c} \quad  
\begin{tikzpicture}[scale=0.75, baseline={(0,0)}, 
    thick,
    ->-/.style={
        decoration={markings, mark=at position 0.55 with {\arrow{Latex}}},
        postaction={decorate}
    },
    dot/.style={
        circle, 
        draw=black, 
        fill=white, 
        inner sep=1.2pt
    }
]
    \draw[->-] (0,0) -- (0,2);
    \node[below] at (0,0) {$\phi_c$};
    \node[above] at (0,2) {$c$};
    \draw[fill=black] (0,0) circle (0.05);
\end{tikzpicture}
\ee
where we have chosen a basis of four-valent junctions $x$ as depicted above.
 These operators form an associative algebra, dubbed the \emph{Tube algebra} Tube$(\cS)$.\footnote{Explicitly we have:
\be
\left[T_{abf}^{cde}\right]_{x y z} = \sqrt{d_{abfxyzc}} \begin{bmatrix}
    b & y & c \\
    a & x & z 
\end{bmatrix}  \begin{bmatrix}
    b & y & d \\
    z & a & f 
\end{bmatrix}  \begin{bmatrix}
    f & z & d \\
    y & b & a 
\end{bmatrix} \, ,
\ee
where $\begin{bmatrix}
    a & b & e \\
    c & d & f
\end{bmatrix} = d_{ef}^{-1/2} {F^{abc}_d}_{ef}$ are the F-symbols of the underlying bulk symmetry and $d_{abc\dots} = d_a d_b d_c\dots$ is the product of quantum dimensions.}
\be
\left[ U_{b,y} \right]_d^e \, \left[U_{a,x}\right]_c^d = \sum_{\substack{f \in a \otimes b \\ z}} \left[T_{abf}^{cde}\right]_{x y z}  \, \left[U_{f,z}\right]_c^e \, .
\ee
Irreps of this algebra are classified by simple objects $\cL_\mu$ in the so-called Drinfeld center $\cZ(\cS)$ of the symmetry $\cS$ \cite{Aasen:2020jwb,Lin:2022dhv} or, equivalently, by topological lines in the SymTFT for $\cS$ \cite{Bhardwaj:2023ayw,Bartsch:2023pzl}.

The simplest example is the case in which $\cS=G$ is an anomaly-free group. In this case the Tube algebra is easy to derive since all the fusion channels are zero- or one--dimensional and they follow the group multiplication laws. Therefore, we find that $c = a b a^{-1}$. Tube algebra representations are labelled by a pair $([g], \pi)$, where $[g]$ is a conjugacy class in $G$ and $\pi \in \text{Rep}(C_g(G))$ is an irrep of the centralizer of $g$ in $G$. The interpretation is clear: a symmetry line $U_h$ maps a $U_g$ twist field into a $U_{h g h^{-1}}$ twist field. Thus, a multiplet spans a full conjugacy class. Furthermore, if $h$ is in the centralizer of $g$, it acts on the twist operator as a $C_g(G)$ representation $\pi$.

\subsection{The Strip Algebra}
Scattering processes between massive particles often involve a SSB scenario, in which the massive particles (or kinks) may interpolate between different vacua. The quantum fields thus have to satisfy different boundary conditions at $x \to \pm \infty$ which specify the left and right vacua on the spatial slice:
\be
\phi(-\infty) = \phi_m \, \quad \phi(+\infty) = \phi_n \, .
\ee
A convenient way to analyze this scenario is to regulate the infrared by introducing a large cutoff $L$, so that the theory is defined on a finite interval. Wavefunctions satisfying such boundary conditions describe states $|\psi_{mn}\rangle$ in different Hilbert spaces $\bH_{mn}$, called the Strip Hilbert spaces:
\be
\bH_{mn}\;:  
\qquad
\begin{tikzpicture}[
    baseline={(0,0)},
    thick,
    dot/.style={
        circle,
        draw=black,
        fill=white,
        inner sep=1.2pt
    }
]
    \fill[blue!10] (1.5,-1.2) rectangle (1,1.2);
    \fill[blue!10] (-1.5,-1.2) rectangle (-1,1.2);

    \draw[->-] (-1,-1.2) -- (-1,1.2);
    \draw[->-] (1,1.2) -- (1,-1.2);

    \node[below] at (-1,-1.2) {$m$};
    \node[below] at (1,-1.2) {$n$};
\end{tikzpicture}
\ee
Above, we have chosen an orientation of the boundaries. This convention will be kept throughout the paper, although left implicit.
In this work we will deal both with asymptotic boundaries, which encode the choice of asymptotic vacua of the theory, and real (finite-distance) boundaries, which will be key in the description of defects. To avoid confusion, we will use blue and gray fillings for the former and the latter, respectively. Nevertheless, the symmetry data is the same in the two cases.

Bulk symmetry acts on the boundary conditions by topological terminations, which endows them with the structure of a module category $\cM$ over $\cS$ (see e.g. \cite{etingof2016tensor} for a review). Physically, the simple elements $m \in \cM$ are in one-to-one correspondence with the vacua of a $\cS$-symmetric TQFT. These are identified with the gapped vacua of the low energy phase to which the massive bulk theory flows.
Mathematically, a consistent symmetry action is encoded by a multiplication map $\mu$ defined as: 
\be\label{eq:multiplication map}
\begin{tikzpicture}[
    thick, baseline = {(0,0)},
    ->-/.style={
        decoration={markings, mark=at position 0.55 with {\arrow{Latex}}},
        postaction={decorate}
    },
    dot/.style={
        circle, 
        draw=black, 
        fill=white, 
        inner sep=1.2pt
    }, baseline={(0,0)}
]

\filldraw[blue!10] (0,-1) -- (0.5,-1) -- (0.5,1) -- (0,1) -- cycle;
\draw (0,-1) -- (0,1);
\draw[->-] (-1,-0.5) node[left] {$a$} -- (0,-0.5);
\draw[->-] (-1,0.5) node[left] {$b$} -- (0,0.5);
\node[below] at (0,-1) {$m$}; \node[above] at (0,1) {$n$}; \node[right] at (0,0) {$r$};
\end{tikzpicture} \quad = \quad \sum_{c \in a \otimes b}  \left[\mu_{ab}^c\right]_{mnr} 
\begin{tikzpicture}[
    thick, baseline = {(0,0)},
    ->-/.style={
        decoration={markings, mark=at position 0.55 with {\arrow{Latex}}},
        postaction={decorate}
    },
    dot/.style={
        circle, 
        draw=black, 
        fill=white, 
        inner sep=1.2pt
    }, baseline={(0,0)}
]
    \filldraw[blue!10] (0,-1) -- (0.5,-1) -- (0.5,1) -- (0,1) -- cycle;
\draw (0,-1) -- (0,1);
\draw[->-] (-1,-0.5) node[left] {$a$} -- (-0.5,0);
\draw[->-] (-1,0.5) node[left] {$b$} -- (-0.5,0);
\draw[->-] (-0.5,0) -- (0,0);
\node[above] at (-0.25,0) {$c$};
\node[below] at (0,-1) {$m$}; \node[above] at (0,1) {$n$}; 
\end{tikzpicture}
\ee
This is subject to a boundary version of the pentagon equation \cite{etingof2016tensor,Fuchs:2002cm,Choi:2023xjw}, which we do not explicitly write down here. 

Let us proceed with some clarifying comments: generally, not all the topological lines will admit a topological junction which leaves $m$ fixed. Fusing a topological line $U_a$ with a boundary $m$ leads to a NIM representation:
\be \label{eq: nimrep}
U_a \times m = \sum_{n \in \cM} \widetilde{N}_{am}^n \, n \, ,
\ee
where $\widetilde{N}_{am}^n = \text{dim}(\text{Hom}(U_a \times m,n))$. The aforementioned topological junction only exists when $\widetilde{N}_{am}^m \neq 0$, in which case the boundary condition is said to be \emph{weakly} symmetric under $U_a$ \cite{Choi:2023xjw}. A weakly symmetric boundary admits a symmetry action on its Hilbert space. As one can see, most weakly symmetric boundary conditions are not invariant under fusion. Only lines of integer quantum dimension can have this property, called \emph{strong} symmetry. In this work, weak symmetry will be the important concept.

This structure naturally extends to the strip. Considering two different boundary conditions, a symmetry operator $U_a$ defines a map
\be
\left[{U_a}\right]{}_{mn}^{rs} : \bH_{mn} \longrightarrow \bH_{rs} \, , 
\qquad
\begin{tikzpicture}[baseline={(0,0)}
    thick,
    ->-/.style={
        decoration={markings, mark=at position 0.55 with {\arrow{Latex}}},
        postaction={decorate}
    },
    dot/.style={
        circle, 
        draw=black, 
        fill=white, 
        inner sep=1.2pt
    }, baseline={(0,0)}
]
    \fill[blue!10] (1.5, -1.2) rectangle (1, 1.2);
    \fill[blue!10] (-1.5, -1.2) rectangle (-1, 1.2);
    \draw[->-] (-1,0) -- (1,0);
    \draw (-1,-1.2) -- (-1,1.2);
        \draw (1,-1.2) -- (1,1.2);

    \node[below] at (-1,-1.2) {$m$};
     \node[below] at (1,-1.2) {$n$};
      \node[above] at (-1,1.2) {$r$};
       \node[above] at (1,1.2) {$s$};
       \node[above] at (0,0) {$a$};
\end{tikzpicture}
\ee
These maps form an associative algebra under vertical composition, dubbed the \emph{Strip algebra} Strip$_\cM(\cS)$:\footnote{The tensors  $\left[ \widetilde{F}_{ab}^c\right] {}_{mn, rs}^{pq}$ are given in terms of the multiplication map $\mu$ on $\cM$ as:
\be
\left[ \widetilde{F}_{ab}^c\right] {}_{mn, rs}^{pq} = \sqrt{\frac{d_a d_b}{d_c}} \left[ \mu_{ab}^c \right]_{mpr} \left[ \mu_{ab}^c \right]_{nqs} \, .
\ee}
\be
\left[U_a\right] {}_{mn}^{rs} \, \left[ U_b \right] {}_{rs}^{pq} = \sum_{c \in a \otimes b}  \left[ \widetilde{F}_{ab}^c\right] {}_{mn, rs}^{pq} \, \left[U_c\right]_{mn}^{pq} \, .
\ee
A particle (or kink) multiplet carries an irreducible representation of this algebra. These are identified with another set of topological defect lines $K^v\, \in \, \cS^*_{\cM}$:
\be
\text{Rep}(\text{Strip}_\cM(\cS)) = \cS^*_\cM \, ,
\ee
which are uniquely fixed in terms of the bulk symmetry $\cS$ and the choice of boundary multiplet $\cM$. In the language of BCFT, $K^v$ provides the representations of boundary-changing operators under the bulk symmetry, as explained in \cite{Fuchs:2002cm}. We will follow \cite{Copetti:2024dcz} and denote a state carrying a given (fixed) representation of the strip algebra by joining the two boundary conditions and inserting a $K_v$ line, which also allows a simple representation of the symmetry action on the irrep:
\be \label{eq: stripirrepsym}
|K^v_{mn} \rangle \quad  = \quad \begin{tikzpicture}[baseline={(0,0)}, scale=0.75, baseline={(0,-0.5)}
]
 \fill[blue!10] (1.5, -1.2) rectangle (1, 1.2);
    \fill[blue!10] (-1.5, -1.2) rectangle (-1, 1.2);
    \filldraw[blue!10] (-1,-1.2) -- (-1.5,-1.2) arc (-180:0 :1.5 and 1) -- (1,-1.2) arc (0:-180:1 and 0.5) -- cycle;
    \draw[->-] (0,-2.2) node[below] {$v$} -- (0,-1.7);
    \draw (-1,1.2) -- (-1,-1.2) arc (-180:0: 1 and 0.5) -- (1, 1.2);
    \node[above] at (-1,1.2) {$m$};
       \node[above] at (1,1.2) {$n$};    
\end{tikzpicture}
\qquad \qquad 
\left[U_a\right]_{m n}^{ r s} |K^v_{mn} \rangle =
\begin{tikzpicture}[scale=0.75, baseline={(0,-0.5)}]
 \draw[->-] (-0.5,-1.75) arc (180:0: 0.5 and 1);
 \node[above] at (0,-0.75) {$a$};
     \fill[blue!10] (1.5, -1.2) rectangle (1, 1.2);
    \fill[blue!10] (-1.5, -1.2) rectangle (-1, 1.2);
    \filldraw[blue!10] (-1,-1.2) -- (-1.5,-1.2) arc (-180:0 :1.5 and 1) -- (1,-1.2) arc (0:-180:1 and 0.5) -- cycle;
    \draw[->-] (0,-2.2) node[below] {$v$} -- (0,-1.7);
    \draw (-1,1.2) -- (-1,-1.2) arc (-180:0: 1 and 0.5) -- (1, 1.2);
    \node[above] at (-1,1.2) {$m$};
       \node[above] at (1,1.2) {$n$};
       \end{tikzpicture}
       \, .
\ee
The knowledge of the multiplet structure is crucial to study the scattering amplitudes between kinks. Suppose we study a $2 \to 2$ amplitude fixing the initial and final states to be two-kink states $|K_v \ K_w \rangle, \, |K_{v'} \ K_{w'} \rangle$, respectively. At the level of the Strip algebra, the incoming representation factors through a tensor product:
\be
K_v \otimes K_w = \sum_{u} \widetilde{N}_{v w}^{u} K_u \, ,
    \ee
likewise for the outgoing representation. The $S$ matrix then admits a factorization:
\be
S_{v w }^{v' w'}(\beta) = \sum_{u \in v \otimes w \cap v' \otimes w'} A_{u}(\beta)  \, P_{v w}^{v' w'} \, ,
\ee
where $A_u$ are (rapidity-dependent) partial amplitudes, corresponding to the exchange of a charge $K_u$ kink. Such a decomposition automatically satisfies the selection rules imposed by the Strip algebras on kinks: the total charge under the symmetry action must be conserved.

We now give two explicit examples of Strip algebra representations. The first is the maximal SSB of the bulk symmetry $\cS$. In this case $\cM = \text{Reg}(\cS)$ is the so-called regular module, and its elements are simply $\cS$ lines, likewise $\cS^*_{\text{Reg}(\cS)}= \cS$, and the kink charges are classified by the same objects. In this case, we will often simply denote the elements of $\cM$ and the Strip algebra charges $\cS^*_\cM$ by the topological lines of the theory.
For the second example consider $\cS=G$ a group with no 't Hooft anomaly. This can be realized in a symmetry-preserving phase with a single, trivially gapped vacuum. Indeed, the corresponding module category is $\text{Vec}$ and $G^*_{\text{Vec}} = \text{Rep}(G)$: particles carry a representation of the bulk symmetry group.

\section{Defect Selection Rules and Categorical Scattering}\label{sec: defect sel rules} 
We now move onto the main topic of this paper: the constraints imposed by symmetry on the scattering of particles on extended defects and interfaces $\calD$, and the way in which these lead to ``Categorical Scattering". We will focus on the (1+1)d case, and consider a general interface $\calD$ between two different (massive or massless) theories $\cT_L$ and $\cT_R$.\footnote{As always, the definition of an S-matrix for massless theories is problematic, unless they are free. In this case one can think about scattering of wavepackets instead, or, equivalently, correlation functions in the presence of a defect.} 

As shown in several recent works \cite{vanBeest:2023dbu,vanBeest:2023mbs,Loladze:2024ayk,Loladze:2025jsq,Ueda:2025ecm,Tachikawa:2026cxd}, these processes can be quite subtle: the bulk quantum numbers of the incoming particle can sometimes only be matched by ``disorder" type operators in the outgoing state. This matching is highly nontrivial, and the physical mechanism which permits such transmission channel is still mysterious. The purpose of this Section is to link this to the realization of the bulk symmetry on $\calD$:  under certain assumptions we carefully spell out below, the reflection/transmission into a ``nonlocal" particle is inherently tied to the presence of a nontrivial defect anomaly.

In the presence of a defect the correct symmetry algebra to describe this process is a variation of the Strip algebra, which we call the \emph{Defect Strip algebra} dStrip$_{\cM,\cM_{\calD}}(\cS)$.
We will show that the aforementioned processes are tied to the branching of Tube algebra representations into Defect Strip algebra ones, which can be highly nontrivial: the latter provide the correct concept of the symmetry charge preserved in the presence of the defect.
In particular, when two different Tube algebra representations can branch into the same Strip algebra representation, widely different types of particles can appear as the outcome of a scattering process involving a defect. 
In the final part of this Section, we specialize to the case of invertible symmetries. In this case we explain how the appearance of these exotic out--states is only present if the defect realizes a nontrivial defect anomaly on its worldvolume.
Finally, we consider a case where the defect anomaly actually forbids the presence of standard particles in the out-state, by considering symmetry-reflecting defects. 

\subsection{Selection Rules in the Presence of a Defect}\label{sec:half-strip}
We analyze selection rules imposed by symmetry in the presence of a 1d interface $\calD$ between theories $\cT_L$ and $\cT_R$. We will often make use of the folding trick, and treat this as a boundary condition $B_{\calD}$ for the folded theory $\cT_L \otimes \overline\cT_R$. The total bulk symmetry is at least $\cS \equiv \cS_L \otimes \cS_R$.\footnote{As shown in \cite{Antinucci:2025uvj}, following earlier insights \cite{Gaiotto:2012np}, the symmetry of the folded theory can be larger. We will not consider these subtleties in this work.}

In our setup, the spatial slice topology (in the unfolded description) is chosen to be $\bR$. This is the natural choice for the study of scattering but is in contrast with most CFT treatments which use a compact $S^1$ topology. This has important implications for the symmetry structure: on the circle the left and right symmetry defects cannot act independently.
However, this comes at a cost: we must also specify boundary conditions for the fields at $\pm \infty$. In a gapped theory, these prescribe different asymptotic vacua $m_L$ and $m_R$ and allow to discuss kink states. In a gapless theory, instead, the natural boundary condition is one which preserves the symmetry. This is a non-physical boundary condition often referred to as the ``cloaking" boundary condition \cite{Brehm:2024zun}\footnote{We thank B. Rayhaun and Y. Wang for informing us of the existence of this construction.}. We will denote this boundary condition by $\cM$.

We are interested in the action of symmetry on the defect Hilbert space $\bH_{\calD}^{m_L,m_R}$, which is described by extending the $\calD$ worldline in the time direction:
\be
|\psi\rangle \, \in \, \bH_{\calD}^{m_L, m_R} : \qquad \begin{tikzpicture}[baseline={(0,0)}]
    \draw[very thick] (0,-1)  -- (0,1) node[above] {$\calD$};
    \draw[dashed] (-1.5,-1) -- (1.5,-1) ;
    \node[below] at (0,-1) {$|\psi\rangle$};
    \node at (-0.75,0) {$\cT_L$};
     \node at (0.75,0) {$\cT_R$};
     \fill[blue!10] (-1.5,-1)-- (-1.5,1) -- (-2,1) -- (-2,-1) --cycle;
     \draw (-1.5,-1) -- (-1.5,1) node[above] {$m_L$};
         \fill[blue!10] (1.5,-1)-- (1.5,1) -- (2,1) -- (2,-1) --cycle;
     \draw (1.5,-1) -- (1.5,1) node[above] {$m_R$};
     \end{tikzpicture}
     \quad 
     \xrightarrow{\text{Folding}}
\begin{tikzpicture}[baseline={(0,0)}]
    \filldraw[gray!10] (0,-1) -- (0.5,-1) -- (0.5,1) -- (0,1) --  cycle;
    \draw[dashed] (0,-1) -- (-2,-1);
    \draw (0,-1) -- (0,1);
    \node[above] at (0,1) {$B_{\calD}$};
    \node at (-1,0) {$\cT_L \otimes \overline\cT_R$};
    \begin{scope}[shift={(-0.5,0)}]
     \fill[blue!10] (-1.5,-1)-- (-1.5,1) -- (-2,1) -- (-2,-1) --cycle;
     \draw (-1.5,-1) -- (-1.5,1) node[above] {$(m_L, \overline{m}_R) $};
         \end{scope}
         \node[below] at (-1,-1) {$|\psi\rangle$};

\end{tikzpicture}
\ee
In the folded picture, the boundary condition $B_{\calD}$ inherits the structure of a module category $\cM_{\calD}$ over $\cS$. A simple defect $\calD$ corresponds to a simple object in the module category. Given a topological junction connecting a $\cS$ topological defect $U_a$ to $B_{\calD}$ (or, alternatively, a topological junction between defects $U_L \in \cS_L$ and $U_R \in \cS_R$ and $\calD)$ \footnote{Here we assumed $\cS=\cS_L\times \cS_R$, so the topological junction factorizes into the product of two topological endpoints. However, it is straightforward to generalize this argument to the generic case $\cS \subset \cS_L \times \cS_R$.}, we can define a consistent symmetry action $U_a^{\calD}$ on the defect Hilbert space $\bH_{\calD}$ :
\be
U_a^{\calD} :  \bH_{\calD}^{m_L,m_R} \to \bH_{\calD}^{n_L,n_R} \, ,
\ee
as follows:\footnote{In general setups, there might be multiple choices of topological junctions for a given defect. We do not consider such cases in this work.}
\be
U_a^{\calD} |\psi\rangle =  \quad  \begin{tikzpicture}[baseline={(0,0)}]
    \draw[very thick] (0,-1)  -- (0,1) node[above] {$\calD$};
    \draw[dashed] (-1.5,-1) -- (1.5,-1) ;
    \node[below] at (0,-1) {$|\psi\rangle$};
   \draw[blue] (-1.5,0) -- (-0.75,0) node[above,black] {$U_L$} -- (0,0);
    \draw[cyan] (0,0) -- (0.75,0) node[above,black] {$U_R$} -- (1.5,0);
     \fill[blue!10] (-1.5,-1)-- (-1.5,1) -- (-2,1) -- (-2,-1) --cycle;
     \draw (-1.5,-1) node[below] {$m_L$} -- (-1.5,1) node[above] {$n_L$};
         \fill[blue!10] (1.5,-1)-- (1.5,1) -- (2,1) -- (2,-1) --cycle;
     \draw (1.5,-1) node[below] {$m_R$} -- (1.5,1) node[above] {$n_R$};
     \end{tikzpicture}
     \quad 
     \xrightarrow{\text{Folding}}
\begin{tikzpicture}[baseline={(0,0)}]
    \filldraw[gray!10] (0,-1) -- (0.5,-1) -- (0.5,1) -- (0,1) --  cycle;
    \draw[dashed] (0,-1) -- (-2,-1);
    \draw (0,-1) -- (0,1);
    \node[above] at (0,1) {$B_{\calD}$};
    \begin{scope}[shift={(-0.5,0)}]
     \fill[blue!10] (-1.5,-1)-- (-1.5,1) -- (-2,1) -- (-2,-1) --cycle;
     \draw (-1.5,-1) node[below] {$(m_L, \overline{m}_R) $} -- (-1.5,1) node[above] {$(n_L, \overline{n}_R) $};
         \end{scope}
         \node[below] at (-0.5,-1) {$|\psi\rangle$};
    \draw[blue] (0,0) -- (-1,0) node[black, above] {$U_a^{\calD}$} -- (-2,0);
\end{tikzpicture}
\ee
We will refer to the algebra satisfied by $U_a^{\calD}$ as the \emph{Defect strip algebra}. It takes the form:
\be
\left[U_a^{\calD}\right]_{(m_L,m_R)}^{(r_L,r_R)} \times \left[U_b^{\calD}\right]_{(r_L,r_R)}^{(n_L,n_R)} = \sum_c  {D_{ab}^c}{}_{(m_L,m_R) , (n_L,n_R)}^{(r_L,r_R)} \, \left[U_c^{\calD}\right]_{(m_L,m_R)}^{(n_L,n_R)} \, ,
\ee
where ${D_{ab}^c}{}_{(m_L,m_R) , (n_L,n_R)}^{(r_L,r_R)} $ are given in terms of the multiplication maps of the two module categories.\footnote{To be precise:
\be
{D_{ab}^c}{}_{(m_L,m_R) , (n_L,n_R)}^{(r_L,r_R)} = \sqrt{\frac{d_a d_b}{d_c}} \left[ \mu_{L, a b}^c \right]_{m_L,r_L,n_L} \, \left[ \mu_{R, a b}^c \right]_{m_R,r_R,n_R} 
\ee
}
This assumes that $B_{\calD}$ is weakly symmetric under $U_a$. By relaxing this assumption we would instead get a map between simple defects in the same multiplet. In this paper, however, we explicitly focus on weakly symmetric defects.
States $|\psi\rangle \in \bH_{\calD}^{m_L,m_R}$ transform in irreps $R$ of the Defect Strip algebra. As the symmetry action commutes with the defect Hamiltonian we find a selection rule:
\be
\langle \psi_R | \psi_{R'} \rangle = 0 \, , \quad \text{unless} \ R = R' \, .
\ee
Interpreting $|\psi_R\rangle$ and $|\psi_{R'}\rangle$ as the out- (in-) going states in a scattering process involving the defect, this selection rule precisely constrains the quantum numbers of the past and future radiation. In practice, we often build our initial state by acting with local (or suitably smeared) fields $\phi$ on the defect vacuum $|\Omega\rangle$.
The state
\be
\phi(x,0) |\Omega\rangle =\begin{tikzpicture}[baseline={(0,0)}]
    \filldraw[gray!10] (0,-1) -- (0.5,-1) -- (0.5,1) -- (0,1) --  cycle;
    \draw[dashed] (0,-1) -- (-2,-1);
    \draw (0,-1) -- (0,1);
    \node[above] at (0,1) {$B_{\calD}$};
    \draw[fill=black] (-1,0) node[above] {$\phi(x,0)$} circle (0.05);
    \begin{scope}[shift={(-0.5,0)}]
     \fill[blue!10] (-1.5,-1)-- (-1.5,1) -- (-2,1) -- (-2,-1) --cycle;
     \draw (-1.5,-1) -- (-1.5,1) node[above] {$(m_L, \overline{m}_R) $};
         \end{scope}
         \node[below] at (-1,-1) {$|\Omega\rangle$};
\end{tikzpicture}
\quad \in \, \bH_{\calD}^{m_L,m_R} \, 
\ee
will generally have a nontrivial overlap with the single particle states of the theory (in the presence of the defect),
\be
\langle \beta_i | \phi(x,0) |\Omega\rangle  = F_i(x,\beta) \, ,
\ee
where $F_i$ are related to the form factors. This type of construction can also be applied to conformal (gapless) theories, where a precise notion of the $S$-matrix is missing.
Crucially, the weak symmetry of $\calD$ allows us to consider a different type of states, created by acting with twist fields $\phi_a$ and terminating their topological line on $\calD$: 
\be
\phi_a(x,0) |\Omega\rangle = \begin{tikzpicture}[baseline={(0,0)}]
    \filldraw[gray!10] (0,-1) -- (0.5,-1) -- (0.5,1) -- (0,1) --  cycle;
    \draw[dashed] (0,-1) -- (-2,-1);
    \draw (0,-1) -- (0,1);
    \node[above] at (0,1) {$B_{\calD}$};
       \draw[blue] (0,0) -- (-1,0);
    \node[above] at (-0.5,0) {$U_a$};
    \draw[fill=black] (-1,0) circle (0.05) node[below] {$\phi_a(x,0)$};
    \begin{scope}[shift={(-0.5,0)}]
     \fill[blue!10] (-1.5,-1)-- (-1.5,1) -- (-2,1) -- (-2,-1) --cycle;
     \draw (-1.5,-1) -- (-1.5,1) node[above] {$(m_L, \overline{m}_R) $};
         \end{scope}
         \node[below] at (-1,-1) {$|\Omega\rangle$};
\end{tikzpicture}
\quad \in \, \bH_{\calD}^{m_L,m_R} \, .
\ee
This new state \emph{still} belongs to $\bH_{\calD}^{m_L,m_R}$, in sharp contrast with the picture in the absence of a defect in which twist fields provide maps between different superselected (twisted) Hilbert spaces $\bH$ and $\bH_{a}$.
Thus, in the presence of a defect $\calD$, disorder fields behave very much as local particles, creating an excitation in the same Hilbert space which has localized energy and momentum.\footnote{This has been recently noted also in \cite{Tachikawa:2026cxd}.}
We now consider the overlaps of this state with the one-particles states of the theory:
\be
\langle \beta_i | \phi_a(x,0) |\Omega\rangle = f_{i,a}(\beta,x) \, .
\ee
The overlap can be nonzero only if both states transform in the same representation under the Defect Strip algebra.
However, while $|\Omega\rangle$ is already in one such irrep, the bulk symmetry acts on $\phi_a$ via the Tube algebra. Thus, to extract the nontrivial overlaps, we must understand the correct branching rules between these representations.

There is an intermediate (and particularly useful) step, in which we re-express a Tube algebra representation in terms of Strip algebra ones. By matching these we can provide new selection rules directly at the level of the operators $\phi_a$, which are informative even when we do not have direct access to the set of asymptotic states.

\subsection{From Tube to Strip to Defect Strip} 
We now discuss the branching rules, and the constraints imposed by the realization of symmetry on the defect onto the possible in- and out-states.
We will focus on giving sufficient conditions for certain defect scattering channels to be open. These follow from the fact that,
for a given choice of boundary module category $\cM$ for $B_{\calD}$ and tube algebra representation $\{ \phi^\mu_{a,x} \}$ there is a map:
\bea
f_\cM : &\text{Rep}(\text{Tube}(\cS))  &&\longrightarrow  \ \ \ \text{Rep}(\text{Strip})(\cS)_\cM \, , \\
& \ \ \ \ \ f_\cM[\cL_\mu] &&\longrightarrow \ \ \ \sum_v n_\mu^v \, K^v \, ,
\eea
providing a way to branch Tube algebra representations into strip algebra ones. This map was already shown in \cite{Cordova:2024iti} and implicitly used in \cite{Copetti:2024rqj}. In this work, we leverage it to show how it provides nontrivial constraints on scattering of particles off extended defects.
This procedure should be thought of as a symmetry-based form factor analysis: we expand the charges carried by bulk fields into those carried by (single-)particle states. This could be important in bootstrap type studies of defect scattering, following the treatment in \cite{Copetti:2024dcz}.

To understand the map explicitly, consider the twist operator $\phi_a$ at a certain finite distance from the boundary $\cM$, and the action of the Defect Strip algebra on the state $\phi_a |\Omega\rangle$. In this equation we will not be considering a defect preserving junction on $B_{\calD}$ as it makes the algebra cleaner:
\be
\begin{tikzpicture}[
    scale=0.75,
    baseline={(0,1.5)},
    thick,
    ->-/.style={
        decoration={markings, mark=at position 0.55 with {\arrow{Latex}}},
        postaction={decorate}
    }
]
    \def\H{4.6}
    \def\wall{0.75}
    \def\R{4}

    \fill[blue!10] (0,0) rectangle (\wall,\H);
    \fill[gray!10] (\R,0) rectangle (\R+\wall,\H);

    \draw (\wall,0) -- (\wall,\H);
    \draw (\R,0) -- (\R,\H);

    \draw[dashed] (\wall,0) -- (\R,0);

    \draw[->-] (2,2) -- (\R,2);
    \draw[->-] (\wall,3) -- (\R,3);

    \draw[fill=black] (2,2) circle (0.05) node[above] {$\phi_a$};

    \node[above] at (3,2) {$a$};
    \node[above] at (2.5,3) {$b$};
    \node[right] at (\R,2.5) {$r$};

    \node[below] at (0,0) {$(m_L, \overline{m}_R)$};
    \node[above] at (0,\H) {$(n_L,\overline{n}_R)$};

    \node[below] at (\R+\wall/2,0) {$m$};
    \node[above] at (\R+\wall/2,\H) {$n$};
\end{tikzpicture}
= 
\begin{tikzpicture}[
    scale=0.75,
    baseline={(0,1.5)},
    thick,
    ->-/.style={
        decoration={markings, mark=at position 0.55 with {\arrow{Latex}}},
        postaction={decorate}
    }
]
    \def\H{4.6}
    \def\wall{0.75}
    \def\R{4}

    \fill[blue!10] (0,0) rectangle (\wall,\H);
    \fill[gray!10] (\R,0) rectangle (\R+\wall,\H);

    \draw (\wall,0) -- (\wall,\H);
    \draw (\R,0) -- (\R,\H);

    \draw[dashed] (\wall,0) -- (\R,0);

    \draw[->-] (2,2) -- (\R,2);

\draw[->-, thick]
    (\wall,1)
    .. controls (1.20,0.90) and (2.05,0.95) .. (2.75,1.20)
    .. controls (3.15,1.50) and (0.45,2.05) .. (0.95,2.40)
    .. controls (1.65,2.75) and (2.45,2.90) .. (\R,3);

    \draw[fill=black] (2,2) circle (0.05) node[above] {$\phi_a$};

    \node[above] at (3,2) {$a$};
    \node[above] at (2.45,3) {$b$};
    \node[right] at (\R,2.5) {$r$};

    \node[below] at (0,0) {$(m_L, \overline{m}_R)$};
    \node[above] at (0,\H) {$(n_L,\overline{n}_R)$};

    \node[below] at (\R+\wall/2,0) {$m$};
    \node[above] at (\R+\wall/2,\H) {$n$};
\end{tikzpicture}
= 
\sum_s 
\begin{tikzpicture}[
    scale=0.75,
    baseline={(0,1.5)},
    thick,
    ->-/.style={
        decoration={markings, mark=at position 0.55 with {\arrow{Latex}}},
        postaction={decorate}
    }
]
    \def\H{4.6}
    \def\wall{0.75}
    \def\R{4}

    \def\yb{0.65}
    \def\ylow{1.35}
    \def\ya{2.00}
    \def\yhigh{3.25}

    \fill[blue!10] (0,0) rectangle (\wall,\H);
    \fill[gray!10] (\R,0) rectangle (\R+\wall,\H);

    \draw (\wall,0) -- (\wall,\H);
    \draw (\R,0) -- (\R,\H);

    \draw[dashed] (\wall,0) -- (\R,0);

    \draw[->-] (2,\ya) -- (\R,\ya);
    \draw[->-] (\wall,\yb) -- (\R,\yb);

    \draw[->-]
        (\R,\ylow)
        .. controls (2.95,1.45) and (1.15,1.65) .. (1,2.3)
        .. controls (1.15,2.95) and (2.95,3.15) .. (\R,\yhigh);

    \draw[fill=black] (2,\ya) circle (0.05) node[above] {$\phi_a$};

    \node[above] at (3,\ya) {$a$};
    \node[above] at (2.65,\yb) {$b$};
    \node[above] at (2.15,\yhigh) {$b$};

    \node[right] at (\R,{0.5*(\yb+\ylow +1.25)}) {$m$};
    \node[right] at (\R,{0.5*(\yb+\ylow)}) {$s$};
    \node[right] at (\R,{0.5*(\ya+\yhigh)}) {$r$};

     \node[below] at (0,0) {$(m_L, \overline{m}_R)$};
    \node[above] at (0,\H) {$(n_L,\overline{n}_R)$};

    \node[below] at (\R+\wall/2,0) {$m$};
    \node[above] at (\R+\wall/2,\H) {$n$};
\end{tikzpicture}
\ee
In the last term of the equation, we recognize the action of the Strip algebra which maps $\bH_{mr}$ into $\bH_{sn}$. Using standard manipulations, the transformation rules can be computed from the action of the Tube algebra on $\phi_a$:
\bea
\begin{tikzpicture}[scale=0.75, baseline={(0,1.35)}, rotate=90,
    thick,
    ->-/.style={
        decoration={markings, mark=at position 0.55 with {\arrow{Latex}}},
        postaction={decorate}
    },
    dot/.style={
        circle, 
        draw=black, 
        fill=white, 
        inner sep=1.2pt
    }
]
    \filldraw[color=gray!10] (0,0) -- (4,0) -- (4,-0.75) -- (0,-0.75) -- cycle;
    \draw (0,0) -- (4,0);
    \draw[->-] (2,0) -- (2,2);
    \draw[fill=black] (2,2) circle (0.05) node[left] {$\phi_a$};
        \node[above] at (2,1) {$a$};
    \node[below] at (0,0) {$r$};
    \node[above] at (4,0) {$s$};
    \draw[->-] (1,0) arc (180:0: 1 and 3);
    \node[left] at (2,3) {$b$};
    \node at (1.5,-0.3) {$m$};
    \node at (2.5,-0.3) {$n$};
\end{tikzpicture} \quad =  \quad \sum_{c,x} \frac{d_c^{1/2}}{d_a^{1/2} d_b}
\begin{tikzpicture}
    [scale=0.75, rotate=90,
    thick,
    ->-/.style={
        decoration={markings, mark=at position 0.55 with {\arrow{Latex}}},
        postaction={decorate}
    },
    dot/.style={
        circle, 
        draw=black, 
        fill=white, 
        inner sep=1.2pt
    },baseline={(0,1.35)}
]
    \filldraw[color=gray!10] (0,0) -- (4,0) -- (4,-0.75) -- (0,-0.75) -- cycle;
    \draw (0,0) -- (4,0);
    \draw[->-] (2,0) -- (2,3);
    \draw[fill=black] (2,3) circle (0.05);
    \node at (2,3.35) {$\phi_a$};
        \node[above] at (2,1.5) {$c$};
    \node[below] at (0,0) {$r$};
    \node[above] at (4,0) {$s$};
    \draw[->-] (1,0) arc (180:0: 1 and 0.5);
    \draw[->-] (1.25,3) arc (180:-180: 0.75 and 0.75);
    \node[left] at (2,3.75) {$b$};
    \node at (2.75,0.65) {$b$};
    \node at (1.5,-0.3) {$m$};
    \node at (2.5,-0.3) {$n$};
    \node at (1.65,2.05) {\small$x$};
    \node at (1.75,0.9) {\small$x^\dagger$};
\end{tikzpicture}& \\
= \sum_{c,x, \phi_c} [\tau_{x,b}]_{\phi_a}^{\phi_c} \left[ \mu_{b a}^x \right]_{r n m} \left[ \mu_{x \overline{b}}^c \right]_{r s n} \, 
\begin{tikzpicture}[scale=0.75, baseline={(0,1.35)}, rotate=90,
    thick,
    ->-/.style={
        decoration={markings, mark=at position 0.55 with {\arrow{Latex}}},
        postaction={decorate}
    },
    dot/.style={
        circle, 
        draw=black, 
        fill=white, 
        inner sep=1.2pt
    }
]
    \filldraw[color=gray!10] (0,0) -- (4,0) -- (4,-0.75) -- (0,-0.75) -- cycle;
    \draw (0,0) -- (4,0);
    \draw[->-] (2,0) -- (2,2);
    \draw[fill=black] (2,2) circle (0.05) node[left] {$\phi_c$};
        \node[above] at (2,1) {$c$};
    \node[below] at (0,0) {$r$};
    \node[above] at (4,0) {$s$};
\end{tikzpicture}& \, .
\eea
Let us unpack the result: the first piece $[\tau_{x,b}]_{\phi_a}^{\phi_c}$ encodes the Tube algebra charge of $\phi_a$. The second piece, $\left[ \mu_{b a}^x \right]_{r n m} \left[ \mu_{x \overline{b}}^c \right]_{r s n}$ instead depends on  the realization of symmetry on the boundary: different defects/boundary conditions lead to inequivalent branching of the bulk charges. 

Intuitively, we should interpret $\mu$ as the symmetry charge carried by the topological junction. We will soon see that in the case of an invertible symmetry these are precisely the defect anomalies of Section \ref{sec:symdef}. 

We conclude that a Tube algebra representation can be translated into a Strip algebra one in the presence of a boundary. This can be expanded in irreps $K^v$ of the Strip algebra, leading to:
\be
f_\cM[\cL_\mu] = \sum_{v}  n_\mu^v \, K^v \, , \ \ \ n_\mu^v \in \bN\, . \label{eq: branch}
\ee
The branching \eqref{eq: branch} has a simple interpretation in the boundary SymTFT presentation \cite{Choi:2024tri,Bhardwaj:2024igy}: objects in $\cZ(\cS)$ are bulk line operators, and the module category $\cM$ corresponds to the choice of a second gapped boundary condition $B_\cM$ for $\cZ(\cS)$. The defects $K^v$ are the topological lines living on $B_\cM$. Pushing a bulk line $\cL_\mu$ on top of $B_\cM$ gives precisely equation \eqref{eq: branch}. We provide further details in Appendix \ref{app: symtft}.

This analysis leads to sufficient conditions for the overlap between different states $\phi_a |\Omega\rangle$ and $\phi_b |\Omega\rangle$ to be nonvanishing by symmetry. Let $\cL_\mu$ and $\cL_\nu$ be the Tube algebra irreps of $\phi_a$ and $\phi_b$ respectively. Then a sufficient condition for $\langle \Omega| \phi_a^\dagger(x,0) \phi_b(y,0)|\Omega\rangle$ to not be forced to vanish by symmetry is that both $\cL_\mu$ and $\cL_\nu$ contain the same Strip algebra representation in their branching. That is:
\be
\sum_v n_\mu^v n_\nu^v \neq 0 \, .
\ee
To understand the necessary conditions for these correlators to be non-vanishing, one must further decompose $K_v \otimes R_\Omega$ in irreps of the defect Strip algebra. Here let us just note that the tensor product above can be acted upon consistently by the Defect Strip algebra. Consider the following setup:
\be
\begin{tikzpicture}[baseline= {(0,1)}]
    \fill[gray!10] (0,0) -- (0,2) -- (0.5,2) -- (0.5,0) -- cycle;
    \draw (0,0) -- (0,2) node[above] {$B_{\calD}$};
    \fill[blue!10] (-1,0) arc (0:180: 0.5 and 1);
    \draw (-1,0) arc (0:180: 0.5 and 1);
    \node[above] at (-1.5,1) {$m$};
    \fill[blue!10] (-3,0) -- (-3,2) -- (-3.5,2) -- (-3.5,0) -- cycle;
    \draw (-3,0) -- (-3,2) node[above] {$n$};
\end{tikzpicture}
\ee
The bottom part of the diagram prepares a state in $\bH_{n,m}\otimes \bH_{\calD}^m$ and evolves it into a state in $\bH_{\calD}^n$. The bulk symmetry acts on the top by the Defect Strip algebra, while on the bottom via a product of Strip and dStrip representations:
\be \label{eq: tensorprodrep}
K^v \otimes R \in  \text{Rep}\left( \text{Strip}_{\cM}(\cS) \otimes \text{dStrip}_{\cM, \cM_{\calD}}(\cS) \right) \, ,
\ee
Graphically we represent the symmetry action as
\be
\begin{tikzpicture}[baseline={(0,1.5)}]
    \fill[gray!10] (0,0) -- (0,3) -- (0.5,3) -- (0.5,0) -- cycle;
    \draw (0,0) node[below] {$B_{\calD}'$} -- (0,3) node[above] {$B_{\calD}$};
    \fill[blue!10] (-1,0) arc (0:180: 0.5 and 1);
    \draw (-1,0) arc (0:180: 0.5 and 1);
    \node[above] at (-1.5,1) {$n$};
    \fill[blue!10] (-3,0) node[below] {$m$} -- (-3,3) -- (-3.5,3) -- (-3.5,0) -- cycle;
    \draw (-3,0) node[below] {$r$} -- (-3,3) node[above] {$s$};
    \draw[->-] (-3,2) -- (0,2); \node[above] at (-1.5,2) {$a$};
\end{tikzpicture}
\sim \sum_p \quad
\begin{tikzpicture}[baseline={(0,1.5)}]
    \fill[gray!10] (0,0) -- (0,3) -- (0.5,3) -- (0.5,0) -- cycle;
    \draw (0,0) node[below] {$B_{\calD}'$} -- (0,3) node[above] {$B_{\calD}$};
    \fill[blue!10] (-1,0) arc (0:180: 0.5 and 1);
    \draw (-1,0) node[below] {$n$} arc (0:180: 0.5 and 1) node[below] {$n$};
    \node[above] at (-1.5,1) {$p$};
    \fill[blue!10] (-3,0) -- (-3,3) -- (-3.5,3) -- (-3.5,0) -- cycle;
    \draw (-3,0) node[below] {$r$} -- (-3,3) node[above] {$s$};
    \draw[->-] (-3,0.5) -- (-1.95,0.5); \node[above] at (-2.425,0.5) {$a$};
    \draw[->-] (-1.05,0.5) -- (0,0.5); \node[above] at (-0.525,0.5) {$a$};
\end{tikzpicture}
\ee
This implies that the tensor product of representations \eqref{eq: tensorprodrep} can be decomposed in representations of $\text{dStrip}_{\cM, \cM_{\calD}}(\cS)$ only.
A full analysis of the selection rules for a generic symmetry is beyond the scope of this paper and will be pursued elsewhere.

\paragraph{Example} Let us provide one guiding example of this phenomenon, which first appears for $G= \bZ_2 \times \bZ_2$. There are two $G$-preserving boundary conditions, $\cM_{\text{triv}}$ and $\cM_{\text{SPT}}$, corresponding to the trivial and nontrivial $\bZ_2 \times \bZ_2$ SPT. As a boundary condition, the latter hosts a defect anomaly for $\bZ_2 \times \bZ_2$.

The bulk tube algebra representations are generated by (products of) the invertible lines:
\be
e_1, e_2, m_1, m_2 \, , \in \cZ(\bZ_2 \times \bZ_2)
\ee
$e_1$ and $e_2$ encode the charge under the two $\bZ_2$ symmetries, while $m_1$ and $m_2$ encode the symmetry operator twisting the field ($m_1$ describes a twist field for $U_1$ etc.). For both boundary conditions, the strip algebra has two representations $K_1$, $K_2$ of order two, which carry charge under the two $\bZ_2$ bulk symmetries:
\be
U_i \, K_j = (-1)^{\delta_{ij}}  K_j \, U_i \, .
\ee
However, the map between the representations of Tube and Strip algebra in the two cases are different:
\bea
&\cM_{\text{triv}} : &&e_1 \to K_1 \, , \quad e_2 \to K_2 \, , \quad m_1 \to \unit \, , \quad m_2 \to \unit \, , \\
&\cM_{\text{SPT}} : &&e_1 \to K_1 \, , \quad e_2 \to K_2 \, , \quad m_1 \to K_2 \, , \quad m_2 \to K_1 \, , 
\eea
the second line stemming from the $\bZ_2 \times \bZ_2$ defect anomaly. Thus, a local, charged particle, say $e_1$, is in the same strip algebra representation of a nonlocal, but neutral particle, $m_2$.

\paragraph{A physical picture: closed vs open Hilbert spaces} In massive QFTs it is sometimes more convenient to frame the discussion in terms of states rather than operators. Here we want to provide an alternative view on the relation between the Tube and Strip algebras irreps that uses this language. The Tube algebra naturally acts on the circle Hilbert spaces $\bH_{a}$ twisted by a simple line of the category $a$ of $\mathcal{C}$. Diagrammatically we may represent an element $\left[ U_c \right]_a^b$ of the Tube algebra as a topological network on the cylinder
\begin{equation}\label{eq:Tubeel}
\left[U_c\right]_a^b =
\begin{tikzpicture}[
    baseline={([yshift=-.5ex]current bounding box.center)},
    mid arrow/.style={
        postaction={
            decorate,
            decoration={
                markings,
                mark=at position 0.55 with {\arrow{Stealth}}
            }
        }
    },
    every node/.style={inner sep=1.5pt} 
]
    \def\R{1.2}
    \def\Ry{0.2}
    
    \draw[thick, dashed] (\R, 0.2) arc (0:180:{\R} and \Ry);
    
    \draw[thick] (-\R, 0.2) -- (-\R, 3.2);
    \draw[thick] (\R, 0.2) -- (\R, 3.2);
    
    \draw[thick] (-\R, 0.2) arc (180:360:{\R} and \Ry);
    
    \draw[thick] (0, 3.2) ellipse ({\R} and \Ry);

    \draw[thick, mid arrow] (0,0) -- (0,1) node[midway, above left =0.25pt] {$a$};
    \draw[thick, mid arrow] (0,1) -- (0,2) node[midway, right=1pt] {$d$};
    \draw[thick, mid arrow] (0,2) -- (0,3) node[midway, left=1pt] {$b$};

    \draw[thick, smooth] plot[variable=\t, domain=0:90, samples=30] ({\R*sin(\t)}, {1.2 + (\t/360) - \Ry*cos(\t)});
    \path[mid arrow] plot[variable=\t, domain=44:46, samples=2] ({\R*sin(\t)}, {1.2 + (\t/360) - \Ry*cos(\t)});
    
    \draw[thick, dashed, smooth] plot[variable=\t, domain=90:270, samples=40] ({\R*sin(\t)}, {1.2 + (\t/360) - \Ry*cos(\t)});
    
    \draw[thick, smooth] plot[variable=\t, domain=270:360, samples=30] ({\R*sin(\t)}, {1.2 + (\t/360) - \Ry*cos(\t)});
    \path[mid arrow] plot[variable=\t, domain=314:316, samples=2] ({\R*sin(\t)}, {1.2 + (\t/360) - \Ry*cos(\t)});

    \node[right=0pt] at ({\R}, 1.45) {$c$};

    \filldraw (0,1) circle (2pt) node[left=1pt] {$y$};
    \filldraw (0,2) circle (2pt) node[right=1pt] {$x$};

\end{tikzpicture} \qquad 
\, : \bH_{a}\rightarrow \bH_{b}\, .
\end{equation}
Now let us consider a bulk gapped theory. To make contact with the Strip algebra we work on a very large cylinder and insert a completeness relation in the vertical direction. In the limit of large radius the completeness relation is dominated by the vacuum states:
\be
\unit \sim \sum_{m \in \cM} |m\rrangle \llangle m| \, .
\ee
By using the NIM representation of $U_a$ on $\cM$ \eqref{eq: nimrep} on the left boundary we find that
\begin{equation}
     \bH_{a}= \bigoplus_{b,c} \widetilde{N}_{am}^n\bH_{mn}\, ,
\end{equation}
see figure \ref{fig: tubetostrip}.
In particular the untwisted Hilbert space decomposes as
\begin{equation}
     \bH_{\unit}= \bigoplus_{m}\bH_{mm}\, . 
\end{equation}
This construction is behind the intuition that kink states $\ket{K_{mn}}$ with $m\neq n$ must come from twisted sectors of the UV CFT, while untwisted sector states are associated with particles in $\bH_{mm}$. 
\begin{figure} \label{fig: tubetostrip}
    \centering
    \begin{equation*}
   \begin{tikzpicture}
       [
    baseline = {(0,1.5)},
    mid arrow/.style={
        postaction={
            decorate,
            decoration={
                markings,
                mark=at position 0.55 with {\arrow{Stealth}}
            }
        }
    },
    every node/.style={inner sep=1.5pt} 
]
    \def\R{1.2}
    \def\Ry{0.2}
    
    \draw[thick, dashed] (\R, 0.2) arc (0:180:{\R} and \Ry);
    
    \draw[thick] (-\R, 0.2) -- (-\R, 3.2);
    \draw[thick] (\R, 0.2) -- (\R, 3.2);
    
    \draw[thick] (-\R, 0.2) arc (180:360:{\R} and \Ry);
    
    \draw[thick] (0, 3.2) ellipse ({\R} and \Ry);

    \draw[thick, mid arrow] (0,0) -- (0,3) node[midway, above left =0.25pt] {$a$};
   \end{tikzpicture}
   \ = \quad \ds \sum_{m \in \cM} \
   \begin{tikzpicture}    [
    baseline = {(0,1.5)},
    mid arrow/.style={
        postaction={
            decorate,
            decoration={
                markings,
                mark=at position 0.55 with {\arrow{Stealth}}
            }
        }
    },
    every node/.style={inner sep=1.5pt} 
]
    \def\R{1.2}
    \def\Ry{0.2}

    \draw[thick, dashed] (\R, 0.2) arc (0:180:{\R} and \Ry);
    
    \draw[thick] (-\R, 0.2) -- (-\R, 3.2);
    \draw[thick] (\R, 0.2) -- (\R, 3.2);
    
    \draw[thick] (-\R, 0.2) arc (180:360:{\R} and \Ry);
    
    \draw[thick] (0, 3.2) ellipse ({\R} and \Ry);

    \draw[thick, mid arrow] (0,0) -- (0,3) node[midway, above left =0.25pt] {$a$};

 \begin{scope}[shift={(0,0.2)}]
  \filldraw[fill=blue!10] 
    ({\R*cos(220)}, {\Ry*sin(220)})        
    arc (220:250:{\R} and {\Ry})           
    -- ({\R*cos(250)}, {\Ry*sin(250) + 3}) 
    arc (250:220:{\R} and {\Ry}) -- cycle;
      \end{scope} 
     \node at (-\R +0.5, -0.3) {$|m\rrangle \llangle m |$};
          
   \end{tikzpicture} \ = \quad \ds \sum_{m, n} \widetilde{N}_{am}^n \
 \begin{tikzpicture}[baseline={(0,-0.2)},
    thick,
    dot/.style={
        circle,
        draw=black,
        fill=white,
        inner sep=1.2pt
    }
]
    \fill[blue!10] (1.5,-1.5) rectangle (1,1.5);
    \fill[blue!10] (-1.5,-1.5) rectangle (-1,1.5);

    \draw (-1,-1.5) -- (-1,1.5);
    \draw (1,-1.5) -- (1,1.5);

    \node[below] at (-1,-1.5) {$|m\rrangle$};
    \node[below] at (1,-1.5) {$\llangle n |$};
\end{tikzpicture}   
   \end{equation*}
    \caption{Going from the cylinder to the Strip Hilbert space by inserting a complete set of states. Left, the initial cylinder state in $\bH_a$, center, inserting a complete base of IR states in the large radius limit, right, decomposing into states on the Strip Hilbert space by fusing the topological line $U_a$.}
\end{figure}
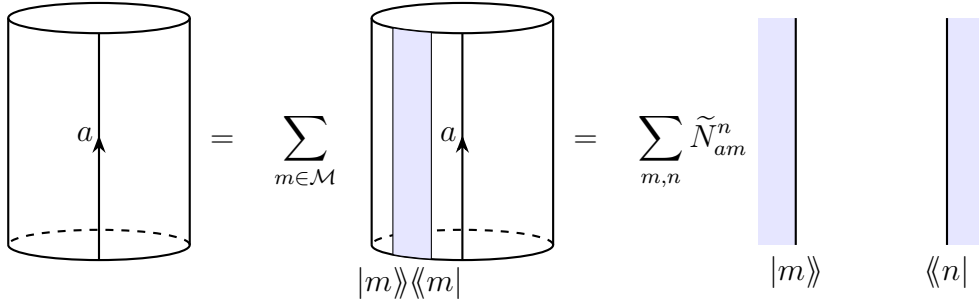
This also provides a concrete realization of the branching of Tube irreps into Strip algebra ones, let us make this precise. Let $\cL_\mu$ be a Tube irrep, this acts on the vector space
\be
V^\mu = \bigoplus_{a} V_a^\mu \, , \quad V^\mu_a \subset \bH_a \, ,
\ee
the spaces $V^\mu_a$ are most easily described in the SymTFT, see Appendix \ref{app: symtft}.
Inserting the completeness relation allows us to re-express:
\be
V^\mu_a = \bigoplus_{m,n} \widetilde{N}_{am}^n \, W^\mu_{mn} \, ,  \qquad W^\mu_{mn} \subset \bH_{mn}
\ee
which, given our decomposition, now carry a linear Strip algebra action, decomposable in irreps. Notice that the cutting operation introduces the choice of module category (forced by the TQFT describing the IR theory), which determines the decomposition.
The same construction can be applied to express Tube algebra elements in terms of Strip algebra ones: instead of cutting open the empty cylinder we cut directly the network \eqref{eq:Tubeel}. 
Equivalently, since a Tube algebra element is a map $\bH_{a}\rightarrow \bH_{b}$, we can exploit the decompositions of the twisted Hilbert spaces to decompose the map itself into component maps of the form $\bH_{mn}\rightarrow \bH_{rs}$, i.e. Strip algebra elements. 
Notice that, under this map, even the trivial representation of the Tube algebra can branch into a nontrivial i.e. not 1-dimensional representation of the Strip algebra. This is related to the breaking of (strong) symmetry by the boundary condition $\cM$.

\subsection{Defect Anomalies and Categorical Scattering} 
Let us reconnect the previous general discussion with the physics of defect anomalies. We will consider $\cS_L = G_L$ and $\cS_R = G_R$ to be groups and let us denote by $G \subseteq G_L \times G_R$ the group preserved by the interface $\calD$. 
$G$-preserving boundary conditions are in bijection with (1+1)d Symmetry Protected Topological (SPT) phases for $G$ \cite{Kapustin:2016jqm,Thorngren:2019iar,Inamura:2021szw}, which are classified by classes $\omega \in H^2(G,U(1))$: these are exactly the possible defect anomalies for a $G$-symmetric interface. We will also consider scattering on top of a single vacuum, for an anomaly-free symmetry $G$, so that $\cM = \text{Vec}$. Notice that we choose the asymptotic boundary condition to be the interface with the \emph{trivial} SPT. Changing this corresponds to stacking the bulk system with a decoupled $G$ SPT and shifts the correct notion of defect anomaly accordingly. This is correctly encoded in the representation of the Defect Strip algebra.
The results can be extended to cases in which the bulk symmetry is spontaneously broken, but we will not do so here. 
With these assumptions, the Defect Strip algebra takes the form:
\be
U_g^{\calD} \times U_h^{\calD} = \omega(g,h) U_{gh}^{\calD} \, ,
\ee
where $\omega(g,h)$ is the defect anomaly, see Figure \ref{fig:defect anomaly}. This immediately implies that states in $\bH_{\calD}$ must transform in a projective representation of $G$. In particular, the defect vacuum state is degenerate and the (topological) operators interpolating between the different vacua can be identified with the topological junctions between the bulk symmetry and the defect.

Let us now understand the quantum numbers carried by states obtained by acting with (twist) operators on the vacuum.
Representations of the Strip algebra in this case are just $G$ representations:
\be
\text{Rep}(\text{Strip})_{\cM_{\calD}}(G) = \text{Rep}(G) \, ,
\ee
while elements $\rho \in \text{Rep}(\text{Tube}(G))$ are described by a doublet $([h],\pi)$, where $[h]$ is a conjugacy class and $\pi \in \text{Rep}(C_g(G))$ is a representation of the centralizer of $h$. 
The defect anomaly implies: 
\be
U_g^{\calD} \times U_h^{\calD} = \chi(g,h) U_{g h g^{-1}}^{\calD} \times U_g^{\calD} \, , \qquad \chi(g,h) =  \frac{\omega(g,h)}{\omega(g h g^{-1}, g)} \, ,
\ee
see Figure \ref{fig:defect anomaly}. Indeed using the multiplication map $\mu_{g h}^{gh} = \omega(g,h)$ \eqref{eq:multiplication map} and the cocycle condition it is possible to prove that
\be
\mu_{g , h}^{gh} \mu_{gh , g^{-1}}^{g h g^{-1}} \simeq \chi(g,h) \, ,
\ee
where $\simeq$ indicates equivalence up to gauge transformations.
For fixed $h$, $\chi(g,h)$ provides a one-dimensional representation $\chi_h$ of the centralizer $C_h(G)$. Thus, the attached twist field actually carries a representation:
\be
\chi_h \otimes \pi  \, \in \,  \text{Rep}(C_h(G)) \, .
\ee
A representation $\pi$ of the centralizer $C_h(G)$ can be lifted to full-fledged representations $\text{Ind}_G(\pi)$ of $G$ by standard methods.\footnote{To lift the representation $\pi \in C_h(G)$, consider $|[h]|$ copies of the vector space $V$ on which $\pi$ acts $V^{\otimes |[h]|}$. Consider a vector $x \otimes v \in V^{\otimes |[h]|} $, so $x \in G/C_h(G)$ and let $k \in C_h(G)$ be the unique solution of $g x = y k $, then the full $G$ action is:
\be
\text{Ind}_G(g) (x \otimes v) = (y \otimes \pi(k) v) \, .
\ee
Notice that, if $g \in C_h(G)$, then $k = g$ and $y = x$, and we recover the starting representation of the centralizer.
}
Applying this procedure to $\chi \otimes \pi$ (since $\chi$ is always 1-dimensional, this is just the standard product) we describe the correct induced representation of the Strip algebra:
\be
\rho_h = \text{Ind}_G(\chi_h \otimes \pi) \, .
\ee
Notice in particular that, in the presence of a defect anomaly, this process accumulates a further $G$ charge $\chi_h$ on the defect, through its topological junction.

This has a crucial consequence on the selection rules: consider a state created by acting with a local field $\phi$ in the representation $\rho$ of $G$: $\phi |\Omega\rangle$. It's charge now can be matched by a state $\phi_h |\Omega\rangle $ created by a neutral particle, but such that:
\be
\text{Ind}_G(\chi_h) = \rho \, .
\ee
This is the fundamental principle allowing the scattering of non-local particles from an incoming state made up of purely local ones. The most general selection rule follows from considering 
states $\phi_h |\Omega\rangle$ , $\phi_{k}|\Omega\rangle$, with induced representations $\rho_h$, $\rho_k$ $\in$ Rep$(G)$. The projective representation $R_\Omega$ carried by the defect vacuum obeys the branching rules:
\be
\rho_h \otimes R_\Omega = \sum_R \, n_R^h \, R \, ,
\ee
so the correlation function is non-vanishing if:
\be
\sum_R n_R^h n_R^k \neq 0 \, .
\ee
In practice, in the concrete examples $G$ is an abelian group, and the tensor product of representations is trivial to compute.

We now provide two concrete examples of the branching rules and apply them to symmetry reflecting defects, and show that the defect anomaly is a necessary condition for these processes.

\paragraph{Examples} First we consider again $G = \bZ_2 \times \bZ_2$, which has a defect anomaly 
\be
\chi(a,b) = \exp\left( i \pi (a_1 b_2 + a_2 b_1) \right)\,.
\ee
The $\bZ_2 \times \bZ_2$ representations are $\mathbf{1}_{r s}$ with $r,s = \pm$ encoding the charge. The group is abelian, so there is no subtlety related to the centralizer and we have $\chi_{(1,0)} = \mathbf{1}_{+-}$, $\chi_{(0,1)} = \mathbf{1}_{- +}$ and $\chi_{(1,1)} = \mathbf{1}_{- -}$. A representation of the Tube algebra $(\rho, (b_1,b_2 ))$ branches into a Strip algebra representation:
\be
\rho \otimes \chi_{(b_1,b_2)} \, .
\ee
Thus, an uncharged ($\rho=1$) $(b_1,b_2)$ twist field has the same quantum numbers as a particle of charge $\chi_{(b_1,b_2)}$:
\be
\begin{tikzpicture}[scale=1.0]

\fill[gray!15] (0,-2.4) rectangle (1.0,2.5);

\draw[line width=1.1pt, black] (0,-2.4) -- (0,2.5);


    \draw[line width=1.1pt, black] (0,-2.4) -- (0,2.5);

    \draw[dashed, line width=0.9pt] (-2.2,-1.7) -- (0,0);

    \draw[red, line width=0.9pt, decorate,
      decoration={snake, amplitude=1.5pt, segment length=8pt}]
      (0,0) -- (-2.2,1.7);


    \node[below] at (-1.55,0.85) {$(b_1,b_2)$};
 
    \fill[black] (-2.2,-1.7) circle (2.2pt) node[below right]{$\mathbf{1}_{(b_2,b_1)}$};
    \fill[black] (-2.2,1.7) circle (2.2pt) node[above right]{$\mathbf{1}_{++}$};

    \fill[red] (0,0) circle (2.4pt) node[below right]{$\chi_{(b_1,b_2)}$};

    \draw[line width=1.0pt, ->] (-2.2,-1.7) -- (-2.2+0.3,-1.7+0.23);
    \draw[line width=1.0pt, ->] (-2.2,1.7) -- (-2.55,1.98);

\end{tikzpicture}
\ee
A second example is $G = D_8$, which also has a nontrivial SPT phase: $H^2(D_8,U(1))= \bZ_2$. The generators are 
\be
D_8 = \langle r: r^4=1, s: s^2 =1 , s r = r^3 s \rangle
\ee
and the representations are the one-dimensional $\mathbf{1}_{r s}$ with $r,s = \pm 1$ and the two dimensional $\mathbf{2}$. As there are several Tube algebra representations, for simplicity we fix the representation of the centralizer to $\pi = \mathbf{1}_{++}$.
The branching rules are:
\be
\begin{array}{c|c|c|c}
   {[ h ]} & C_g & \text{Ind}_G(\chi_h) , \, \cM_{\text{triv}}& \text{Ind}_G(\chi_h)  , \, \cM_{\text{SPT}} \\ \hline
    {[1]} & D_8  & \mathbf{1}_{++} & \mathbf{1}_{++} \\
    {[r^2]} & D_8  & \mathbf{1}_{++} & \mathbf{1}_{+-} \\
    {[r]} & \bZ_4 & \mathbf{1}_{++} \oplus \mathbf{1}_{+-} & \mathbf{1}_{++} \oplus \mathbf{1}_{+-} \\
    {[s]} & \bZ_2 \times \bZ_2 & \mathbf{1}_{++} \oplus \mathbf{1}_{--} & \mathbf{2} \\
    {[sr]} & \bZ_2 \times \bZ_2 & \mathbf{1}_{++} \oplus \mathbf{1}_{-+} & \mathbf{2}
\end{array}
\ee
Notice, that the two-dimensional representation can be matched purely by the defect anomaly, even if $\pi=\mathbf{1}_{++}$. This continues to hold true more in general: the $\mathbf{2}$ of $D_8$ carries a nontrivial charge under the center $r^2 = -1$, which can be matched by a twist operator when $\pi$ has trivial center charge:
\be
\begin{tikzpicture}[scale=1.0]

\fill[gray!15] (0,-2.4) rectangle (1.0,2.5);

\draw[line width=1.1pt, black] (0,-2.4) -- (0,2.5);


    \draw[line width=1.1pt, black] (0,-2.4) -- (0,2.5);

    \draw[dashed, line width=0.9pt] (-2.2,-1.7) -- (0,0);

     \draw[red, line width=0.9pt, decorate,
      decoration={snake, amplitude=1.5pt, segment length=8pt}]
      (0,0) -- (-2.2,1.7);
    

    \node[below] at (-1.3,0.85) {$[s]$};
 
    \fill[black] (-2.2,-1.7) circle (2.2pt) node[below right]{$\mathbf{2}$};
    \fill[black] (-2.2,1.7) circle (2.2pt) node[above right]{$\mathbf{1}_{++}$};

    \fill[red] (0,0) circle (2.4pt) node[below right]{$\ds \mathbf{2}$};

    \draw[line width=1.0pt, ->] (-2.2,-1.7) -- (-2.2+0.3,-1.7+0.23);
    \draw[line width=1.0pt, ->] (-2.2,1.7) -- (-2.55,1.98);

\end{tikzpicture} \, .
\ee
This shows, in practice, how the selection rules enforced by the realization of symmetry on a defect can lead to naively counterintuitive phenomena.

\subsection{Scattering on Symmetry Reflecting Interfaces} The constraints derived above are particularly strong in the presence of a symmetry-reflecting interface between theories $\cT_L$ and $\cT_R$, which preserves the $G_L \times G_R$ symmetry. 
Consider a setup in which a state $\phi_L |\Omega\rangle$ is excited from the defect vacuum, and $\phi_L$ carries a nontrivial $G_L$ representation $\rho_L$. We will also assume that both $G_L$ and $G_R$ do not have a self-defect anomaly in what follows, although this assumption is not crucial.
We consider a transmission event in which the $\phi_L$ particle becomes another local particle $\phi_R$ on the right. 

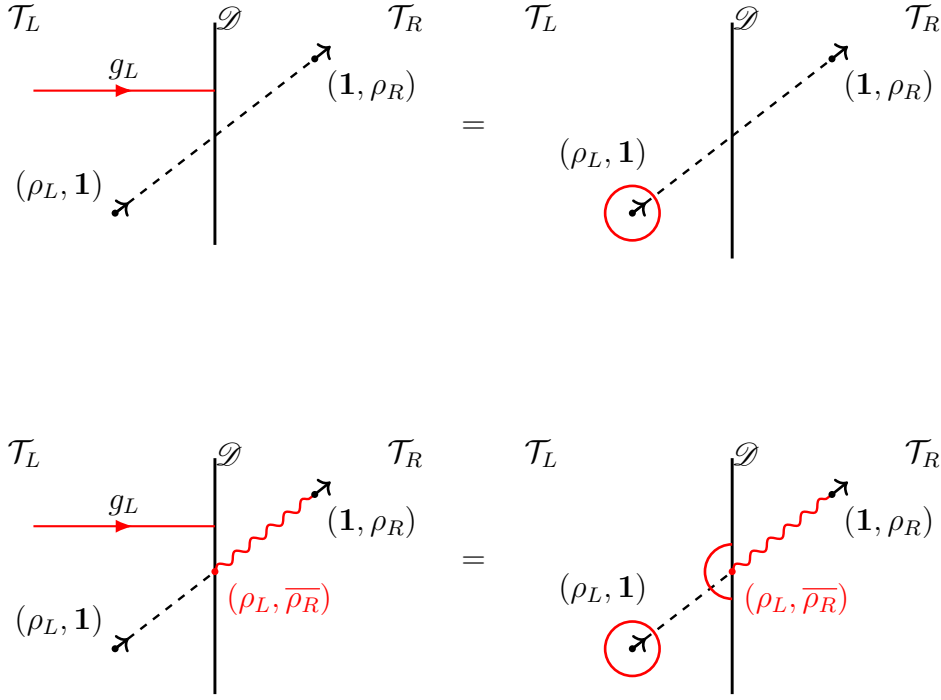
\begin{figure}
\centering
\begin{tikzpicture}[thick, scale=0.6,
    ->-/.style={
        decoration={markings, mark=at position 0.55 with {\arrow{Latex}}},
        postaction={decorate}
    },
    dot/.style={
        circle,
        draw=black,
        fill=white,
        inner sep=1.2pt
    }
]


\begin{scope}[xshift=-5.7cm, yshift=4.8cm]

    \node at (-4.2,2.6) {$\mathcal{T}_L$};
    \node at (4.2,2.6) {$\mathcal{T}_R$};
    \node at (0.3,2.6) {$\calD$};

    \draw[line width=1.1pt] (0,-2.4) -- (0,2.5);

    \draw[dashed, line width=0.9pt] (-2.2,-1.7) -- (0,0);
    \draw[dashed, line width=0.9pt] (0,0) -- (2.2,1.7);

    \fill[black] (-2.2,-1.7) circle (2.2pt)
        node[above left]{$(\rho_L,\mathbf{1})$};
    \fill[black] (2.2,1.7) circle (2.2pt)
        node[below right]{$(\mathbf{1},\rho_R)$};

    \draw[line width=1.0pt, ->] (-2.2,-1.7) -- (-1.9,-1.47);
    \draw[line width=1.0pt, ->] (2.2,1.7) -- (2.55,1.98);

    \draw[->-, red] (-4,1) -- (0,1);
    \node[above] at (-2,1) {$g_L$};

\end{scope}

\node at (0,5.0) {$=$};

\begin{scope}[xshift=5.7cm, yshift=4.8cm]

    \node at (-4.2,2.6) {$\mathcal{T}_L$};
    \node at (4.2,2.6) {$\mathcal{T}_R$};
    \node at (0.3,2.6) {$\calD$};

    \draw[line width=1.1pt] (0,-2.7) -- (0,2.5);

    \draw[dashed, line width=0.9pt] (-2.2,-1.7) -- (0,0);
    \draw[dashed, line width=0.9pt] (0,0) -- (2.2,1.7);

    \fill[black] (-2.2,-1.7) circle (2.2pt)
        node[above left, xshift=10pt, yshift=12pt]{$(\rho_L,\mathbf{1})$};
    \fill[black] (2.2,1.7) circle (2.2pt)
        node[below right]{$(\mathbf{1},\rho_R)$};

    \draw[line width=1.0pt, ->] (-2.2,-1.7) -- (-1.9,-1.47);
    \draw[line width=1.0pt, ->] (2.2,1.7) -- (2.55,1.98);

    \draw[red, line width=1pt] (-2.2,-1.7) circle (0.6);

\end{scope}


\begin{scope}[xshift=-5.7cm, yshift=-4.8cm]

    \node at (-4.2,2.6) {$\mathcal{T}_L$};
    \node at (4.2,2.6) {$\mathcal{T}_R$};
    \node at (0.3,2.6) {$\calD$};

    \draw[line width=1.1pt] (0,-2.7) -- (0,2.5);

    \draw[dashed, line width=0.9pt] (-2.2,-1.7) -- (0,0);
    
    \draw[red, line width=0.9pt, decorate,
      decoration={snake, amplitude=1.5pt, segment length=8pt}]
      (0,0) -- (2.2,1.7);
    

    \fill[black] (-2.2,-1.7) circle (2.2pt)
        node[above left]{$(\rho_L,\mathbf{1})$};
    \fill[black] (2.2,1.7) circle (2.2pt)
        node[below right]{$(\mathbf{1},\rho_R)$};
    \fill[red] (0,0) circle (2.2pt)
        node[below right]{$(\rho_L,\overline{\rho_R})$};

    \draw[line width=1.0pt, ->] (-2.2,-1.7) -- (-1.9,-1.47);
    \draw[line width=1.0pt, ->] (2.2,1.7) -- (2.55,1.98);

    \draw[->-, red] (-4,1) -- (0,1);
    \node[above] at (-2,1) {$g_L$};

\end{scope}

\node at (0,-4.6) {$=$};

\begin{scope}[xshift=5.7cm, yshift=-4.8cm]

    \node at (-4.2,2.6) {$\mathcal{T}_L$};
    \node at (4.2,2.6) {$\mathcal{T}_R$};
    \node at (0.3,2.6) {$\calD$};

    \draw[line width=1.1pt] (0,-2.7) -- (0,2.5);

    \draw[dashed, line width=0.9pt] (-2.2,-1.7) -- (0,0);
    
    \draw[red, line width=0.9pt, decorate,
      decoration={snake, amplitude=1.5pt, segment length=8pt}]
      (0,0) -- (2.2,1.7);
    

    \fill[black] (-2.2,-1.7) circle (2.2pt)
        node[above left, xshift=10pt, yshift=12pt]{$(\rho_L,\mathbf{1})$};
    \fill[black] (2.2,1.7) circle (2.2pt)
        node[below right]{$(\mathbf{1},\rho_R)$};
    \fill[red] (0,0) circle (2.2pt)
        node[below right]{$(\rho_L,\overline{\rho_R})$};

    \draw[line width=1.0pt, ->] (-2.2,-1.7) -- (-1.9,-1.47);
    \draw[line width=1.0pt, ->] (2.2,1.7) -- (2.55,1.98);

    \draw[red, line width=1pt] (-2.2,-1.7) circle (0.6);

    \draw[red, line width=1pt] (0,0.6)
        arc[start angle=90, end angle=270, radius=0.6cm];

\end{scope}

\end{tikzpicture}
\caption{Above: A topological line $g_L \subset G_L$ ending on a symmetry reflecting defect measures only the charge of the incoming radiation. Therefore, $G_L$ selection rules require $\rho_L=\mathbf{1}$ for this process to happen. Below: Because of the defect anomaly, a topological line $g_L \subset G_L$ ending on a symmetry reflecting defect can measure the charge of topological junctions of $g_R\subset G_R$ lines.}
\label{fig: selection rules}
\end{figure}

As the $G_L$ charges are measured only on the left of the interface, they cannot be matched by a state $\phi_R |\Omega\rangle$, see Figure \ref{fig: selection rules}.
Therefore, the overlap:
\be
\langle \Omega | \phi_R^\dagger(x_R,0) \, \phi_L(x_L,0) |\Omega \rangle = 0 \, 
\ee
is vanishing, unless $\rho_L=\rho_R = 1$. As these states would normally provide a basis for the transmitted radiation, we would conclude that symmetry-reflecting defects do not admit transmission of charged particles.
This conclusion is wrong: as we have discussed the outgoing state can also comprise particles created by twist fields $\phi_{a,R}$ on the right. This is not at odds with charge conservation, even though $\phi_{g_R,R}$ cannot carry $G_L$ charge: its topological junction can! A sufficient condition for the transmission to happen is that:\footnote{The necessary condition is that $\rho_L \otimes R_\Omega$ and $\text{Ind}_G(\chi_{g_R}) \otimes R_\Omega$ contain a common projective representation under the decomposition of the tensor product.}
\be
\text{Ind}_G(\chi_{g_R}) = \rho_L \, .
\ee
Thus, while energy is carried locally by the outgoing state, charge is not, and can be stored on the defect $\calD$. At the level of correlation functions we have:
\be
\langle \Omega | \phi^\dagger_L(x_L,0)  \, \phi_{g_R,R} (x_R,0) |\Omega\rangle \neq 0 \, .
\ee
Similar selection rules induced by interfaces have also been recently studied in \cite{Prembabu:2025qvi}. Versions of this mechanism are at the core of the results in \cite{vanBeest:2023dbu,Loladze:2025jsq,Ueda:2025ecm}, as we show in the next Sections.
\paragraph{Absorption and non-simple defects.} The transmission to a state created by a twist operator is one natural possibility compatible with the selection rules, but it is not the only one. Indeed, another option is that the scattering excites a charged mode $\xi$ localized on the defect $\calD$, possibly together with the charged particle $\phi_R$. The state:
\be
\xi \, \phi_R(x_R,0) |\Omega\rangle \, ,
\ee
has the correct quantum numbers to make the overlap $\langle \Omega| \phi^\dagger(x_L,0) \, \xi \, \phi_R(x_R,0) |\Omega\rangle$ non-vanishing. Let us make two comments about this situation: first, by tuning the local couplings on the defect $\calD$ it is often possible to make the excitations $\xi$ heavier than the energy scale at which we perform the scattering. In the example of the fermion-rotor system (Section \ref{sec: ferrot}) this is the mass $M$ of the rotor mode. In this way, these processes can be discarded. Second, the mechanism which allows the categorical scattering is also secretly of this type: the topological junction $x_{g_R}$ between $U_{g_R}$ and $\calD$ is precisely a defect-localized (topological) mode. Being entirely precise, the right state takes the form:
\be
x_{g_R} \phi_{g_R, R}(x_R,0) |\Omega\rangle \, ,
\ee
and $x_{g_R}$ carries charge under $G_L$ by the defect anomaly. To see that this is truly a local operator, recall that $U_{g_R}$ is an anomaly-free symmetry, so it can be realized by local unitaries. Far enough from the defect, the ground state is invariant under the $U_{g_R}$ action: the operator $U_{g_R} |\Omega\rangle$ reduces to a topological excitation localized in a neighbourhood of $\calD$, which is precisely $x_{g_R}$.

Finally, let us consider instead the case in which the starting defect is non-simple, i.e. it is given by a direct sum:
\be
\calD = \calD_1 \oplus \calD_2 
\ee
of indecomposable defects. In this case, the spectrum of defect operators includes defect-changing operators $x_{1 2}$ and $x_{2 1}$. If $\calD_1$ and $\calD_2$ are related by a topological manipulation, the defect-changing operators can be themselves topological and lead to localized, zero-energy defect modes which can absorb the bulk charge. 
As an example, let us consider $\bZ_2$ symmetry defect $\eta$ and 
\be
\calD = 1 \oplus \eta \, .
\ee
$\calD$ has two defect changing operators $x_{1 \eta}$ and $x_{\eta 1}$, which are exchanged by the action of the bulk $\bZ_2$ symmetry.
The operators:
\be
x^\pm = \frac{x_{1 \eta} \pm x_{\eta 1}}{\sqrt{2}} \, ,
\ee
carry $\bZ_2$ charge $\pm 1$ and are topological. Thus, a charged particle $\phi_L$ can be matched by an out-state of the form:
\be
x^-  \phi_{R}(x_R,0) |\Omega\rangle \, ,
\ee
where $\phi_R$ is a $\bZ_2$-odd operator localized on the right; notice that no topological line is needed. We will see an explicit example of this on the lattice in Section \ref{sec: lattice}.

\paragraph{Example: gauging interfaces} To conclude let us consider a paradigmatic example: a topological interface $\calD_{\abA}$ corresponding to the gauging of an abelian symmetry $\abA$. In this case, the interface system carries a symmetry $G = \abA \times \abA^\vee \simeq \abA^2$. Furthermore, the $\abA$ defects become trivial after passing the interface, which makes it symmetry-reflecting for both symmetries:
\be
\begin{tikzpicture}
    \fill[green!10] (0,0) -- (0,2) -- (2,2) -- (2,0) -- cycle;
    \draw (0,0) -- (0,2) node[above] {$\calD_{\abA}$};
    \node at (-1,1) {$\cT$};
    \node at (1,1) {$\cT/\abA$};
    \draw[->-,red] (-2,1.5) -- (-1,1.5) node[above] {$a$} -- (0,1.5);
    \draw[->-,red] (2,0.5) -- (1,0.5) node[below] {$a^\vee$} -- (0,0.5);
\end{tikzpicture} \, .
\ee
Furthermore, the interface carries a defect anomaly:
\be
\omega = 2 \pi i \int A^\vee \cup A \, , 
\ee
where $A^\vee$ and $A$ are gauge fields for $\abA^\vee$ and $\abA$, respectively. This stems from the minimal coupling $a \cup A^\vee$ for the discrete gauging:
\be
Z_{\cT/\abA}[A^\vee] = \sum_{a} Z_\cT[a] \exp\left(2 \pi i \int A^\vee \cup a \right) \, ,
\ee
subject to the Dirichlet boundary conditions at the interface. The simplest examples are KW-type interfaces, in which case $\abA=\bZ_2$, and we have the defect anomaly:
\be
\eta \times \eta^\vee = - \eta^\vee \times \eta \, .
\ee
A $\bZ_2$ charged particle $\sigma^-$ on the left can thus be transmuted into a $\bZ_2^\vee$-neutral disorder operator $\mu_{\eta^\vee}^+$, without violating charge conservation:
\be
\begin{tikzpicture}[scale=0.75]
   

    \fill[green!10] (0,-2.7) -- (0,2.5) -- (4.5,2.5) -- (4.5,-2.7) -- cycle ;
    \draw[line width=1.1pt] (0,-2.7) -- (0,2.5);

     \node at (-3.9,2.1) {$\cT$};
    \node at (3.9,2.1) {$\cT/\bZ_2$};
    \node[above] at (0,2.5) {$\calD$};

    \draw[dashed, line width=0.9pt] (-2.2,-1.7) -- (0,0);
    
    \draw[red, line width=0.9pt, decorate,
      decoration={snake, amplitude=1.5pt, segment length=8pt}]
      (0,0) -- (2.2,1.7);
    

    \fill[black] (-2.2,-1.7) circle (2.2pt)
        node[above left, xshift=10pt, yshift=12pt]{$\sigma^-$};
    \fill[black] (2.2,1.7) circle (2.2pt)
        node[below right]{$\mu_{\eta^\vee}^+$};
    \fill[red] (0,0) circle (2.2pt)
        node[below right]{$\mathbf{1}_{-+}$};

    \draw[line width=1.0pt, ->] (-2.2,-1.7) -- (-1.9,-1.47);
    \draw[line width=1.0pt, ->] (2.2,1.7) -- (2.55,1.98);

\end{tikzpicture}
\ee
It might seem that the requirement for a symmetry-reflecting defect is a strong one, and such defects are rare, or intimately topological. In Appendix \ref{app: gauging} we show that, using discrete gauging of the defect system, we can generate these examples from setups which are only symmetric, and, in Section \ref{sec: lattice} we show how to construct lattice examples which cannot be connected to topological interfaces by defect-localized deformations.

\section{Scattering of Free Massless Particles}\label{sec: massless scattering}

We now turn to the analysis of specific examples in which the general phenomenon described in the previous sections is concretely realized. The first class of examples we consider consists of models that have already been studied in the literature, and our goal here is to emphasize the role of defect anomalies in the corresponding scattering processes.

All these examples lie within the realm of two-dimensional massless free theories. These are, in fact, the only class of two-dimensional massless theories for which the $S$-matrix is properly defined. One motivation for studying these models is that they arise from an $s$-wave reduction of higher-dimensional scattering processes of particles off line defects, most notably the scattering of electrons off heavy monopoles. In particular, we will focus on two models: the 3450 model \cite{vanBeest:2023dbu} and the fermion--rotor system \cite{Polchinski:1984uw,Loladze:2024ayk,Loladze:2025jsq}.

\subsection{3450 Model}

The first model we want to analyze is the so--called 3450 model. It is a free theory of two left--moving and two right--moving Weyl fermions, which we dub $\psi_L^{i=1,2},\psi_R^{i=1,2}$, respectively. Having an equal number of left and right movers, the theory has vanishing gravitational anomaly hence can be placed on manifolds with boundaries. While there is a non-Abelian flavor symmetry, we just focus on a $U(1)\times U(1)'$ subgroup with the following charge assignment\footnote{The faithful symmetry is actually $\frac{U(1)\times U(1)'}{\bZ_5}$. In fact the rotation of angles $(\alpha,\beta)=\left(\frac{6\pi i}{5}, \frac{8\pi i}{5}\right)$, that is of order $5$, act trivially on all fermions.}:
\be
\begin{array}{c|cccc}
       & \psi_L^1  & \psi_L^2  & \psi_R^1  & \psi_R^2 \\   \hline 
  U(1) & 3 & 4 & 5 & 0  \\
  U(1)'& 4 & -3 & 0 & 5 
\end{array}
\ee
This $U(1)\times U(1)'$ is chosen to be anomaly-free, even though the charge assignment is chiral. The anomaly-free condition guarantees \cite{Smith:2019jnh, Smith:2020nuf} the existence of a symmetry preserving boundary condition, even though its construction is not trivial because of the chirality of charge assignment, that prevents the existence of symmetry preserving mass terms. Using the fact that $U(1)\times U(1)'$ embeds conformally in the full chiral algebra, one can use standard BCFT techniques to construct an explicit boundary state, see \cite{vanBeest:2023dbu}.

We want to first recall the puzzle concerning scattering off the $U(1)\times U(1)'$ symmetric boundary condition and its solution as explained in \cite{vanBeest:2023dbu} (for which we refer to more details). Then we provide a different viewpoint on the boundary condition that makes it clear that it has a defect anomaly, and then show that the proposal can be reinterpreted as a consequence of the general result of Section \ref{sec: defect sel rules}.

\paragraph{Categorical scattering.} The puzzle is extremely simple: if we scatter the state created by $\psi_L^1$ off the $U(1)\times U(1)'$, the outgoing state must have charges $(3,4)$ under the two symmetries, as they are preserved. However as the theory is free this outgoing state must be created by a free right-moving fermion, but there is no such fermion with charges $(3,4)$.

Using the explicit boundary state \cite{vanBeest:2023dbu} we can determine the operator that creates the out-state. It is convenient to use the bosonized picture in which
\begin{equation}
    \psi^i_L=e^{iX_i} \ , \ \ \ \ \psi ^j_R=e^{i\overline{X}_j} \ .
\end{equation}
We remind that a vertex operator $e^{is_iX_i +i\overline{s}_i\overline{X}_i}$ has conformal dimensions $h=\frac{1}{2}\sum _i s_i^2$, $\overline{h}=\frac{1}{2}\sum _i \overline{s}_i^2$. The out-going operator can be shown to be
\begin{equation}
    e^{i\left(\frac{3}{5}\overline{X}_1 +\frac{4}{5}\overline{X}_2\right)} \ .
\end{equation}
Notice that this is the only possible answer compatible with the symmetries. Clearly this is not a local operator, but lives at the end of a topological line generating a $\bZ_5$ symmetry. Nevertheless is a right-moving free fermion, as it has $(h,\overline{h})=(0,1/2)$. The $\bZ_5$ symmetry is the subgroup of $U(1)$, and therefore it is preserved by the boundary condition. The picture proposed in \cite{vanBeest:2023dbu} is that the outcome of the scattering is a state created by the twisted sector operator attached to the topological $\bZ_5$ line that terminates (topologically) on the boundary.  

\paragraph{The defect anomaly.}

To determine the defect anomaly of the boundary condition it is convenient to pass to the unfolded picture: we view the system as obtained from that of only 2 left-moving fermions $\psi^i_L(x)$. More precisely the boundary becomes an interface $\cI$ that separates $\psi^i_L(x)$ from $\widetilde{\psi}^i_L\equiv \psi_R^i(\widetilde{x})$ (where $\widetilde{x}$ is the coordinate $x$ reflected along the boundary). The interface is necessarily topological, as there are no right-moving degrees of freedom and hence no particles can be reflected. Notice that $\cI$ cannot be a simple invertible symmetry operator of the 2-fermions system. In fact, consider the $U(1)_r\times U(1)_r'$ symmetry of the 2-fermions system assigning charges $(3,4)$ and $(4,-3)$ respectively. By definition, after folding along $\cI$, we produce a boundary preserving $U(1)\times U(1)'$, but then the  $U(1)_r$ symmetry generator must be mapped to the generator of $U(1)_l$ that assigns charge $(5,0)$, and clearly this cannot happen if the interface is a simple invertible symmetry. In fact, we claim that 
\begin{equation}
    \cI=\cN
\end{equation}
where  $\cN$ is a non-invertible duality defect generating a $\text{TY}(\bZ_5)_{\chi}$ symmetry with bicharacter $\chi(a,b)=e^{\frac{6\pi i}{5} ab}$.\footnote{We remind that the bicharacter for a $\text{TY}(\bZ_N)_\chi$ fusion category determines how the duality defect $\cN$ acts on local operators. If $\chi(a,b) =e^{\frac{2\pi i k}{N}ab}$, with $k\in \bZ_N ^\times$, then the local operator $\cO_q$ of charge $q$ is mapped into an operator in the twisted sector $k^{-1}q$ of $\bZ_N$.} This exists because the 2-fermions model is self-dual under gauging the following non-anomalous $\bZ_5$ symmetry 
\begin{equation}
    \begin{array}{c|cc}
     & \psi^1_L & \psi^2_L \\ \hline
  \bZ_5   & 3 & 4 
\end{array}
\end{equation}
There is a simple intuition behind this: in the folded theory, the subgroup $\bZ_5\subset U(1)$ acts trivially on the right-movers, hence in the unfolded picture the corresponding $\bZ_5$ generator from the right must be allowed to terminate topologically on $\cI$. This is precisely a property characterizing duality defect.

To prove our claim it is useful to bosonize, $\psi_L^i=e^{iX_i}$. Denoting by $\eta$ the $\bZ_5$ generator, the twisted sector of $\eta^k$ contains the operators
\begin{equation}
    \cO_{n_1,n_2}^{(k)}=e^{i\left(\frac{3k+5n_1}{5}X_1+\frac{4k+5n_2}{5}X_2\right)}
\end{equation}
that have charge $3n_1+4n_2$ under $\bZ_5$. Therefore, the theory after gauging $\bZ_5$ contains all operators with $3n_1+4n_2=0 \, \text{mod}(5)$. To prove that the theory is self-dual it is enough to show that this spectrum contains 2 free fermions, namely such that
\begin{equation}
    h=\frac{1}{2}\left(\frac{3k+5n_1}{5}\right)^2+\frac{1}{2}\left(\frac{4k+5n_2}{5}\right)^2=\frac{1}{2} \ .
\end{equation}
In fact this is satisfied by
\begin{equation}
    \cO_{0,0}^{(1)}=e^{i\left(\frac{3}{5}X_1+\frac{4}{5}X_2\right)} \ ,\ \ \ \ \cO_{-1,-3}^{(3)}=e^{i\left(\frac{4}{5}X_1 -\frac{3}{5}X_2\right)}
\end{equation}
as well as their complex conjugates. By identifying $\psi_L^1 \equiv  \cO_{0,0}^{(1)}$ and $\psi _L^2\equiv \cO_{-1,-3}^{(3)}$ after gauging, the theory turns out to be self dual.\footnote{Notice that there is some degree of arbitrariness in the identification, because the duality can always be composed with some order-2 symmetry (e.g. here exchange of the fermions, or complex conjugation). The reason for our choice is merely to have that the boundary condition is obtained by folding around $\cN$, instead of $\cN$ composed with a further invertible symmetry.} This implies the existence of a topological defect $\cN$ such that
\begin{equation}
  \cN\times \eta =\cN \ , \ \ \   \cN \times \cN=\sum _{k=0}^4 \eta ^k \ .
\end{equation}
Moreover, because of the identification above $\psi_L^1$,  that has charge $3$ under $\bZ_5$, is mapped into $\cO^{(1)}_{0,0}$ by $\cN$, while $\psi ^2_L$ is mapped into $\cO_{-1,-3}^{(3)}$. This implies that the bicharacter is 
\begin{equation}
    \chi(a,b)=e^{\frac{6\pi i}{5}ab} \ .
\end{equation}
Finally we need to map $U(1)_r \times U(1)_r'$ across $\cN$. $U(1)_r$ assigns charges $(3,4)$ to $(\psi_L^1,\psi_L^2)$, hence it is mapped into a symmetry that assigns charges $(5,0)$. Similarly $U(1)_r'$ is mapped into a symmetry assigning $(0,5)$. This implies that folding around $\cI=\cN$ gives a boundary condition that preserves $U(1)\times U(1)'$, as we wanted.

What we earn from this perspective on the boundary is the defect anomaly. Indeed duality defects are symmetry reflecting with respect to the underlying Abelian symmetry, $\bZ_5$ in this case, and they have a defect anomaly for it corresponding to the F-symbol. After folding, the $\bZ_5$ that acts on the left is mapped into $\bZ_5'$ that only acts on right movers:
\begin{equation}
       \begin{array}{c|cccc}
     & \psi^1_L & \psi^2_L & \psi ^1_R & \psi ^2_R\\ \hline
  \bZ_5   & 3 & 4 & 0 & 0 \\
  \bZ_5' & 0 & 0 & 3 & 4 
\end{array}
\end{equation}
It is easy to check that $\bZ_5'$ is a subgroup of $U(1)\times U(1)'$, hence it is preserved by the boundary.
The defect anomaly of $\cI=\cN$ is then mapped to a boundary anomaly for $\bZ_5 \times \bZ_5'$, and the corresponding Ward identities as outlined in \ref{sec: defect sel rules} imply the scattering outcome illustrated above.
\subsection{Fermion--Rotor System}\label{sec: ferrot}

The next theory we study is the so-called fermion--rotor system. It consists of $N$ free right-moving Weyl fermions $\psi_i(t-x)$ coupled to a one-dimensional rotor degree of freedom $\alpha(t)\sim \alpha(t)+2\pi$, defining a localized impurity at $x=0$.\footnote{For some purposes, it can be useful to regularize this defect by opening a short width $\omega $, as done for instance in \cite{Loladze:2025jsq}. Here we will assume this regularization in the limit of $\omega \rightarrow 0$.}
The action of the model is
\be\label{eq:fermion-rotor}
S_{\theta}
=
\int d^2x \,\left(\sum_{i=1}^N \psi_i^\dagger\partial_+\,\psi_i \right)
+
\int dt \left(
i\alpha  \sum_i \psi_i^\dagger \psi_i
+
\frac{M}{2}\dot{\alpha}^2
+
\frac{i\theta}{2\pi}\dot{\alpha}
\right)\, ,
\ee
where $\partial_+ = \partial_x+\partial_t$. Therefore, the defect operator is
\be
\calD[\gamma] = \int D\alpha \exp{\left( \int_\gamma \left(
i\alpha  \sum_i \psi_i^\dagger \psi_i
+
\frac{M}{2}\dot{\alpha}^2
+
\frac{i\theta}{2\pi}\dot{\alpha}
\right)\right)}\,.
\ee
Notice that $\calD$ depends on an additional dimensionless parameter, namely a defect theta angle $\theta$ for the rotor. In contrast, $M$ is a dimensionful coupling that can be interpreted as the inertia of the rotor. Therefore it sets a scale and defines an (solvable) RG flow localized on the defect. At the infrared fixed point, understood as the limit $M\to\infty$, the defect is necessarily topological. Indeed, no radiation can be reflected, since the theory contains no left--moving particles, nor can it be absorbed, since in this limit the rotor degrees of freedom are frozen.\footnote{In the IR, it is also useful to fold the theory to obtain a boundary condition for $N$ Dirac fermions. This setup was first studied in \cite{Maldacena:1995pq}.}

As written, the action \eqref{eq:fermion-rotor} does not make the $\alpha$-periodicity manifest off-shell\footnote{It is, however, realized on-shell: $\sum_{i}\psi _i^\dagger \psi _i$ is the conserved current of a $U(1)$ symmetry, hence its integral is quantized to be an integer in any correlation function.}. An alternative way to present this theory that makes the $2\pi$ periodicity explicit, is in the bosonized picture, where we introduce $N$ right-moving compact scalars $X_R^i$. The bosonized action is
\be
S_{\theta}
= S_{2d}[X_R^i]
+
\int dt \left(
\frac{i}{2\pi}\alpha \sum_i dX_R^i
+
\frac{M}{2}\dot{\alpha}^2
+
\frac{i\theta}{2\pi}\dot{\alpha}
\right)\, .
\ee
Here $S_{2d}[X_R^i]$ should be understood schematically, since, as usual, writing a local action for purely chiral bosons involves a standard subtlety. Nevertheless, in the bosonized formulation the compatibility between the periodicity of $\alpha$ and the rotor coupling is manifest because of the off--shell conservation of the bosonic current $dX_R^i$.

\paragraph{Symmetries.}
Let us now discuss the symmetries of the model. The bulk theory has a manifest $U(N)$ symmetry acting on the $N$ 
Weyl fermions. The $SU(N)$ rotations are clearly preserved by the coupling to the rotor, since $\alpha$ couples only through the singlet operator $\sum_i \psi_i^\dagger \psi_i$. The remaining chiral $U(1)$ symmetry is affected by an ABJ anomaly \cite{Loladze:2025jsq} localized on the defect, which implies
\be\label{eq: anomaly fr}
\partial_+ J
=
\frac{N}{2\pi}\,\delta(x)\,\dot{\alpha}\, .
\ee
This relation can be understood by observing that the bulk-defect coupling acts as a coupling of the $U(1)$ current to a background gauge field, since it is of the form $A_\mu J^\mu$ with $A_{\mu} =\alpha(t)\delta(x)\delta_{\mu 1} $. The anomaly equation \eqref{eq: anomaly fr} then follows from the standard two-dimensional chiral anomaly.

Following \cite{Padayasi:2021sik}, we can identify $t = \frac{N}{2\pi}\dot{\alpha}$ with the so--called tilt operator associated with this $U(1)$ symmetry broken by the defect. Let us emphasize that, even though this seems a total derivative, it is not quite so as the field $\alpha$ is not a globally well-defined operator, and therefore a generic $U(1)$ topological operator is no longer topological in the presence of the rotor. In fact, this tilt operator is the one responsible for the appearance of the extra dimensionless parameter $\theta$ in the defect action \cite{Drukker:2022pxk}. 

Given the appearance of the factor $N$ in \eqref{eq: anomaly fr}, and the fact that $\oint \dot{\alpha}\in 2\pi \mathbb{Z}$, a $\bZ_N$ subgroup of $U(1)$ is preserved by the defect.\footnote{Actually, the situation is slightly more subtle, as the Weyl fermions are defined on a spin manifold. In this case one can show that the periodicity of $\alpha$ must be even:
\be
\frac{1}{2\pi}\oint \dot{\alpha} = 0 \mod 2 \, .
\ee
Thus, we learn that the defect coupling breaks the diagonal $U(1)_{\text{diag}}$ to $\bZ_{2N}$. Therefore, the full subgroup of $U(N)$ preserved by the defect is $\frac{SU(N) \times \bZ_{2N}}{\bZ_N} \subset U(N)$.
}. Moreover, we can also show that $\calD$ is symmetry reflecting under $\bZ_N$. To show this, we can compute the fusion of the $\bZ_N$ topological operator with the defect $\calD$. From the definition of the tilt operator we get that
\be
U_{\frac{2\pi k}{N}}[\gamma]\calD[\gamma] = \calD[\gamma]\exp{\left(i \frac{2\pi k}{N} \int_\gamma t \right)} = \calD[\gamma]\exp{\left(i k \int_\gamma \dot\alpha \right)} = \calD[\gamma]\,.
\ee
We see that the defect $\calD$ absorbs the $\bZ_N$ symmetry operator, thus being symmetry reflecting. For later purposes let us notice that effect of $\bZ_N$ transformation is to shift the defect $\theta$-angle by $2\pi k$, which is a trivial operation.

Finally, as noticed by Polchinski in \cite{Polchinski:1984uw}, the theory also enjoys a more subtle $U(1)_V$ symmetry which is not present in the bulk, acting on the fields as 
\be\label{eq: U(1)V action}
U(1)_V: \quad\psi^i \to e^{i \beta (\Theta(x)-\frac{1}{2})} \psi^i \, , \ \ \ \ \alpha \to \alpha -\beta \, .
\ee
where $\Theta(x)$ is the usual Heaviside function. This $U(1)_V$ symmetry acts with opposite charges on the fermions to the left and to the right of the defect.
Its origin is more transparent in the folded picture, in which the bulk theory is rewritten as a theory of $N$ free Dirac fermions. In this language, the symmetry is naturally identified with the $U(1)_V$ vector symmetry of the model.

For our purposes, let us notice that the topological operator implementing this symmetry can be written as
\be
V_{\beta}[\gamma] := U_{\beta}[\gamma_L]U_{-\beta}[\gamma_R]\exp{\left(i \beta j_{\text{shift}}(t_*)\right)}\quad,\; \partial \gamma_L = \partial \gamma_R = t_*
\ee
where $\gamma_{L(R)}$ is the part of the curve $\gamma$ lying of the left- (right-) hand side of the defect $\calD$. We see that the junction between the line $\gamma$ and the defect must be dressed by the shift current 
\be
j_{\text{shift}}(t_*) \equiv M\dot{\alpha}(t_*) + \frac{\theta}{2\pi}
\ee
in order to make it topological. Let us remark that, while $j_{\text{shift}}$ is conserved in a stand-alone rotor theory, the coupling with the fermions breaks its conservation, and only the combination above is conserved.

\paragraph{Defect Anomalies.} The presence of this $U(1)_V$ symmetry is precisely what underlies the scattering puzzle at the interface. Indeed, if we prepare an incoming state with positive $U(1)_V$ charge $q$, charge conservation requires the outgoing state to be either a left-moving particle with charge $q$ or a right-moving particle with charge $-q$. However, both of these states are absent from the untwisted Hilbert space. Recent studies \cite{Loladze:2024ayk, Loladze:2025jsq} have resolved this puzzle by arguing that the actual outgoing state is a right--moving particle created by an operator attached to a $\bZ_N$ topological line. As we will show, this resolution is consistent with the existence of a mixed defect anomaly between $U(1)_V$ and the symmetry associated with the $\bZ_N$ topological line. 

The appearance of the defect anomaly can be detected in several equivalent ways. For instance, we can show that the junction between the $U(1)_V$ topological line and the defect is charged under a $\bZ_N$ symmetry transformation. Indeed, as argued before, a $\bZ_N$ transformation has the effect of shifting $\theta \rightarrow \theta + 2\pi k$. Since the $U(1)_V$ junction is dressed with the defect current $j_{\text{shift}} = M\dot \alpha(t) + \frac{\theta}{2\pi}$, the $\bZ_N$ action has the effect:
\be
\bZ_N : \quad V_{\beta}[\widetilde{\gamma}] \calD[\gamma]\rightarrow \exp{\left(ik\beta \, (\gamma \cdot \widetilde{\gamma})\right)}\;V_{\beta}[\widetilde{\gamma}]\calD[\gamma]
\ee
where $(\gamma \cdot \widetilde{\gamma}):= \int_{\gamma}\text{PD}(\widetilde{\gamma})$ is the intersection number between the two curves. 

Another way to argue for the defect anomaly is to couple the theory with a $\bZ^V_N \subset U(1)_V$ background gauge field $C_V$ which we think of as a $U(1)$ connection subject to the constraint
\be
dC_V = 0 \quad,\quad \int C_V \in 0,\cdots , N-1\,.
\ee
Since the $U(1)_V$ symmetry shifts the rotor $\alpha \rightarrow \alpha + \beta$, the coupling to a background $C$ is implemented by $\dot\alpha\rightarrow\dot \alpha - \frac{2\pi}{N}C_V$ so that the action involves a term \footnote{There are also bulk couplings between $C_V$ and the fermion current $\sum_i \psi^\dagger_i \psi_i$. However, since the symmetry acts differently on the left- and right--hand side of the defect, such a coupling has different sign on the two sides.}
\be
S[C_V] \supset \int dt \frac{i\theta}{2\pi}\left(\dot \alpha -\frac{2\pi}{N}C_V\right)
\ee
arising from the defect $\theta$-term. Therefore, now fusing a $\bZ_N$ symmetry defect with $\calD$ produces a non-trivial anomaly term:
\be
U_{\frac{2\pi k}{N}}[\gamma]\calD[\gamma] = \exp{\left( \frac{2 \pi i k}{N} \int C_V \right)}\calD[\gamma]\,.
\ee

In accordance with our general arguments, the presence of the defect anomaly has important consequences on the scattering of charged particles off the interface. Indeed, if we scatter a right-moving fermion of charge $q$, the out--going state can be either created by a local operator of charge $\pm q$ (depending if it is a left/right mover respectively) or a right-moving particles state created by a twisted sector operator of charge $p<q$ and attached to an $U_{\frac{2\pi(q-p)}{N}}$ topological line. As discussed before, the latter possibility is the only one available in the fermion--rotor system.

\section{Scattering of Massive Particles} \label{sec:soliton}
We now turn to the analysis of two--dimensional gapped systems. This provides an ideal arena in which to search for new examples where scattering with exotic excitations can occur. Indeed, as is well known, in two--dimensional massive theories the scattering problem can sometimes be solved exactly, thanks to integrability. In the following, we present two examples in which the full scattering problem in the presence of defects can be solved exactly, and where exotic particle states are created when a standard particle state hits the defect.
\subsection{Ising Field Theory and Free Massive Majorana}
The first example we want to study is that of the Ising CFT in $1+1$ dimensions deformed by the energy operator $\epsilon$. The bulk theory is gapped and, depending on the sign of the coupling, the $\bZ_2$ symmetry is either spontaneously broken or preserved. In the fermionic picture, the theory is simply that of a free massive Majorana fermion, and the defect we want to study can be described by the action\footnote{Our conventions are $\eta=\diag(-1,1)$ and 
\begin{equation}
    \gamma^0=\sigma_2\, , \qquad \gamma^1=i\sigma_1\, , \qquad \overline{\Psi}= \Psi^{\dagger}\gamma^0\, ,\qquad \Psi^{T} = \left(
         \psi^R,
       \psi^L
  \right)\, , 
\end{equation}
with both $\psi^R$ and $\psi^L$ are real.}
\begin{equation}
      S=\int d^2 x \overline{\Psi}(i\gamma^{\mu}\partial_{\mu}- m\sign(x) - g \delta(x))\Psi\, ,
\end{equation}
where $m>0$. The defect (or interface) flips the sign of the fermion mass and supports a pinning field $\overline{\Psi}\Psi$ with a dimensionless coupling $g$. Similar defects, without the flip in the sign of the mass, have been studied in \cite{Delfino:1994nr, Delfino:1994nx}. 
\paragraph{Symmetries and Defect Anomalies.} Let us first comment on the symmetries and anomalies carried by the interface we are analyzing. In the fermionic frame, for vanishing mass $m=0$, the bulk theory has a $\bZ_2\times \bZ_2$ global symmetry, the fermion number and a chiral $\bZ_2$ symmetry, acting respectively as:
\be
(-1)^{F}:\Psi \mapsto -\Psi \quad,\;(-1)^{F_L}:\Psi\mapsto \sigma_3\Psi\,.
\ee
The two symmetries have a mixed anomaly $i\pi A \cup \text{Arf}(S)$, with $A$ the background for $(-1)^{F_L}$ and $S$ the spin structure. The mass flips sign under $(-1)^{F_L}$, implying that the partition function for mass $-m$ is related with the partition function for $m$ by a counterterm $\text{Arf}(S)$.

The interface is symmetric under $(-1)^F$, and crucially carries a defect anomaly for $(-1)^F$. To see this, fix the spin structure $S$ and the counterterms of the massive Majorana fermion so that the theory with $m>0$ has a trivially gapped bosonic vacuum. The phase with $m<0$ then differs by an additional topological Arf counterterm, so its vacuum realizes a non-trivial fermionic SPT phase \cite{Witten:2015aba}. Because the interface reverses the sign of the Majorana mass, it produces a non-trivial Arf SPT phase on one half-space. By our general definition, this precisely corresponds to a $(-1)^F$ defect anomaly.

This defect anomaly has an interesting implication in the Ising frame. Indeed, to get the bosonic theory we need to gauge $(-1)^F$. Following the general discussion of Appendix \ref{app: gauging}, we conclude that the non-trivial defect anomaly implies that the corresponding interface is now symmetry--reflecting under the $\bZ_2$ quantum symmetry and it carries the corresponding defect anomaly which makes the topological junctions charged under themselves. Therefore we can expect that the scattering of charged particles off the interface can transmit states created by twist operators. Moreover, since the fermionic presentation of the theory is quadratic, we are able to solve explicitly the scattering problem, as we are going to show.

\paragraph{Reflection and Transmission amplitudes.} Since the fermionic theory is manifestly free, it is convenient to work in this frame and then translate the result in the Ising picture if necessary. In the setup we chose the defect extends along time, piercing space at $x=0$, to describe the corresponding Hilbert space we Fourier transform in time and write
\begin{equation}
    \Psi(x,E)= \Psi_{-}(x,E)\Theta(-x)+\Psi_{+}(x,E)\Theta(x)\, .
\end{equation}
The equations of motion, together with the gluing condition forced by the delta function, are 
\begin{equation}
    \begin{split}
        &(i\sigma_3 \partial_x \pm m \sigma_2)\Psi_{\pm}(x,E)= E\Psi_{\pm}(x,E)\, , \\ & i\sigma_3 (\Psi_{+}(0,E)-\Psi_{-}(0,E))+ \frac{g}{2}\sigma_2(\Psi_{+}(0,E)+\Psi_{-}(0,E))=0\, ,
    \end{split}
\end{equation}
at $E=0$ we have the normalizable solution
\begin{equation}
    \Psi_{\pm}(x,0)= A^0_{\pm}\begin{pmatrix}
        1\\-1
    \end{pmatrix}e^{\mp m x}\, , 
\end{equation}
which describes a normalizable zero energy bound state tied to the defect and with width $m^{-1}$.  The presence of this zero mode tells us that the vacuum in the defect Hilbert space is doubly degenerate, signaling the defect anomaly for $(-1)^F$ discussed before.\footnote{Notice that the same structure is found in the Hilbert space of the massless theory twisted by $(-1)^{F_L}$. In this case, the defect anomaly can be interpreted as a mixed 't Hooft anomaly between $(-1)^F$ and $(-1)^{F_L}$, see \cite{Witten:2015aba}.} This zero mode also makes manifest the symmetry--reflecting nature of the interface in the bosonized picture. Indeed, since the fermionic field $\Psi$ is charged under $(-1)^F$, the zero mode is not gauge invariant in the Ising field theory: it must be attached to a topological $\bZ_2$ symmetry line. Since this state has zero energy, it describes the topological junction between $\calD$ and the symmetry line.

The remaining solutions are the usual bulk excited states and occur at $E=\sqrt{p^2+m^2}$. Fourier transforming back we have
\begin{equation}
    \Psi(x,t)= \Theta(x)\Psi_{+}(x,t)+\Theta(-x)\Psi_{-}(x,t)\, , 
\end{equation}
where
\begin{equation}
\begin{split}
    \Psi_{\pm}(x,t)&= A_{\pm}^0\begin{pmatrix}
        1\\-1
    \end{pmatrix}e^{\mp m x}\\ &+\int \frac{d\beta}{2\pi} u_{\pm}(\beta)A_{\pm}(\beta)e^{-im (t\cosh(\beta)-x\sinh(\beta))}+ v_{\pm}(\beta)A^{\dagger}_{\pm}(\beta)e^{im (t\cosh(\beta)-x\sinh(\beta))}\, ,
\end{split}
\end{equation}
and $\beta$ is the rapidity variable, defined as $E =  m\cosh(\beta)$, $p=m \sinh(\beta)$. The Majorana condition $\Psi^{\dagger}=\Psi$ gives $v_{\pm}(\beta)=u_{\pm}(\beta)^*$ while $A_0^{\pm}$ is self-adjoint. Fixing the normalization which gives canonical commutation relations for the creation and annihilation operators, the spinors $u_{\pm}$ and $v_{\pm}$ are:
\begin{equation}
    u_{\pm}(\beta)=\sqrt{\frac{m}{2}}\begin{pmatrix}
       e^{\beta/2} \\ \pm i e^{-\beta/2}
    \end{pmatrix} \, , \qquad v_{\pm}(\beta)=\sqrt{\frac{m}{2}}\begin{pmatrix}
       e^{\beta/2} \\ \mp i e^{-\beta/2}
    \end{pmatrix}\, .
\end{equation}
The operators $A_{\pm}$ are not independent, rather the gluing condition gives linear relations among the creation and annihilation operators at $x>0$ and $x<0$, allowing us to characterize uniquely the defect Hilbert space. We find 
\begin{equation}
    \begin{pmatrix}
       A^{\dagger}_{-}(\beta) \\ A^{\dagger}_{+}(-\beta)
    \end{pmatrix}= S(\beta)\begin{pmatrix}
       A^{\dagger}_{-}(-\beta) \\ A^{\dagger}_{+}(\beta)
    \end{pmatrix}\, ,\qquad A_{+}^0= \frac{2-g}{2+g} A_{-}^0
\end{equation}
where 
\begin{equation}
\begin{split}
    & S(\beta)= \begin{pmatrix}
        R(\beta) & T(\beta)\\ T(-\beta) & R(-\beta)
    \end{pmatrix} \, , \\&   R(\beta)=-\text{sech}(\beta )+\frac{i g }{\frac{g^2}{4}+1}\tanh (\beta )\, , \qquad T(\beta)= \frac{g^2-4}{g^2+4}\tanh(\beta)\, . 
\end{split}
\end{equation}
The reflection-transmission matrix $S$ is unitary, it becomes purely reflecting for low energy particles while it is purely transmissive at high energies. It displays a pole at $\beta=i\pi/2$, which matches with the explicit zero mode we found from the equations of motion.

On top of the vacua we have the Fock space of single particle states that scatter off the defect according to the reflection and transmission amplitude we derived above. For the special values $g=\pm 2$ the interface becomes purely reflective, i.e. a product of boundary conditions. When this happens the zero mode survives only on one side of the defect. In usual $S$-matrix theory one relates the residue at the poles with the coupling between the fields that create the asymptotic state and the zero mode. In our case we have 
\begin{equation}
    S\left(\frac{i\pi}{2}+x\right)= \frac{i}{x}\begin{pmatrix}
         \frac{(g+2)^2}{g^2+4} & -i\frac{g^2-4}{g^2+4}\\  i\frac{g^2-4}{g^2+4}&  \frac{(g-2)^2}{g^2+4}
    \end{pmatrix} = \frac{i}{x} V V^{\dagger}
\end{equation}
where we have written the rank $1$ residue matrix in terms of the vector
\begin{equation}
    V= \frac{1}{\sqrt{g^2+4}}\begin{pmatrix}
        g+2\\ i(g-2)
    \end{pmatrix}\, , 
\end{equation}
which is interpreted as the on-shell vertex between the particles and the defect bound state. Note that the ratio of the components of $V$ is related to the ratio of $A^0_{\pm}$ from the gluing conditions. In particular for $g=\pm 2$ only the particles on one side couple to the zero mode. 

\paragraph{Correlation functions and Twist operators.} Interesting observables in presence of the defect are correlation functions. There is a simple trick we can use for those that leverages the fact that the bulk theory is completely solved \cite{Delfino:1994nr}. We rotate the defect by $90$ degrees $t\rightarrow i x$, $x\rightarrow -i t$, so that from being a defect it becomes an operator acting on the Hilbert space of the bulk theory. In the rapidity plane this rotation corresponds to $\beta \mapsto \frac{i\pi}{2}-\beta$. Therefore, denoting $\calD$ the defect operator, correlation functions are
\begin{equation}
\bra{0}T\left[\Phi_1(x_1,t_1)..\Phi_m(x_m,t_m)\calD\Phi_{m+1}(x_{m+1},t_{m+1})..\Phi_{m+n}(x_{m+n},t_{m+n})\right]\ket{0}
\end{equation}
where $T$ is time ordering. Notice that the former pole at $\beta=i\pi/2$ now occurs at $\beta=0$, which corresponds to the threshold value $E=m$ for the energy. The advantage of this formulation is that correlation functions can be expressed in terms of the form factors of the bulk operators and of the defect operator with respect to the asymptotic states of the bulk theory. The former are well known (see e.g. \cite{Mussardo:2010mgq}), while the latter can be derived from the matrix $S$ we have computed. The operators defined at $x<0$ before the rotation create ingoing states, while those on the other side create outgoing states,
\begin{equation}
\begin{split}
     &A_{-}^{\dagger}(i\pi/2-\beta)= A_{in}(\beta)\,, \qquad   A_{-}^{\dagger}(-(i\pi/2-\beta))= A_{in}^{\dagger}(-\beta)\\ &A_{+}^{\dagger}(i\pi/2-\beta)= A_{out}(\beta)\,, \qquad   A_{+}^{\dagger}(-(i\pi/2-\beta))= A_{out}^{\dagger}(-\beta)\, .
\end{split}
\end{equation}
In the defect picture the reflection-transmission matrix gives linear relations among the creation operators on the two spatial sides of the defect, here such relations have to be interpreted as the result of commuting with the defect operator $\calD$, in particular
\begin{equation}
\begin{split}
     &\calD A_{in}(\beta)= \widehat{R}(\beta)\calD A_{in}^{\dagger}(-\beta) + \widehat{T}(\beta) A_{out}(\beta)\calD\\ 
    &A_{out}^{\dagger}(-\beta)\calD = \widehat{R}(\beta)^*A_{out}(\beta) \calD -\widehat{T}(\beta)\calD A^{\dagger}_{in}(-\beta)\, .
\end{split}
\end{equation}
where we defined $\widehat{R}(\beta)=R(i\pi/2-\beta)$ and $\widehat{T}(\beta)=T(i\pi/2-\beta)$. These equations, together with the bulk scattering matrix, which in this case is just a sign, can be used to solve for all the form factors of the defect operator $\calD$, for instance it is easy to see that
\begin{equation}
    \bra{0}\calD \ket{\beta_1,\beta_2}= 2 \pi \langle \calD\rangle \widehat{R}(-\beta_1)^*\delta(\beta_1+\beta_2)\, , \qquad \bra{\beta_1,\beta_2}\calD \ket{0}= 2 \pi \langle \calD\rangle  \widehat{R}(\beta_1)\delta(\beta_1+\beta_2)\, .
\end{equation}
Similarly we can obtain equations for the overlaps with single particle states
\begin{equation}
    \begin{split}
        &0=\widehat{R}(\beta)\bra{0}\calD\ket{-\beta}+ \widehat{T}(\beta)\bra{\beta}\calD\ket{0}\\ & 0=\widehat{R}(\beta)^*\bra{\beta}\calD\ket{0}-\widehat{T}(\beta)\bra{0}\calD\ket{-\beta}\, .
    \end{split}
\end{equation}
Something interesting happens here. If we try to solve those for general values of $\beta$ one easily sees that unitarity of $S$ implies that the only solutions are trivial
\begin{equation}
    \bra{0}\calD\ket{-\beta}=\bra{\beta}\calD\ket{0}=0\, .
\end{equation}
However the situation changes if we zoom in $\beta\sim 0$. Dividing the two equations by $R(\beta)$ and its complex conjugate, in the $\beta\rightarrow 0$ limit we get
\begin{equation}
    \begin{split}
        0=\bra{0}\calD A_{in}^{\dagger}(0)\ket{0}-i\frac{g-2}{g+2}\bra{0}A_{out}(0)\calD\ket{0}\, ,
    \end{split}
\end{equation}
which fixes the ratio of the two matrix elements. We see that this result links the effect of scattering a standing fermion off the defect at $t=0$ to the Majorana zero mode we previously found. Up to a normalization we can write the solutions
\begin{equation}
    \bra{0}\calD \ket{\beta}= C i(g-2)\delta(\beta)\,, \qquad \bra{\beta}\calD\ket{0}= C (g+2)\delta(\beta)\, .
\end{equation}
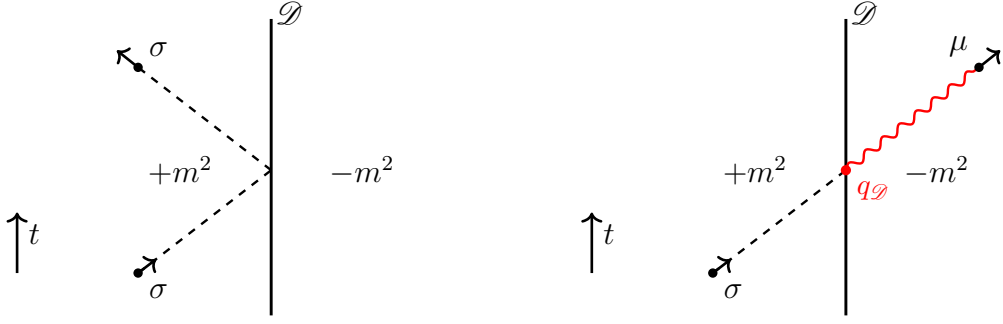
\begin{figure}[t]
\centering
\begin{tikzpicture}[scale=0.8]

\begin{scope}[xshift=0cm]

    \node at (-1.5,0.) {$+m^2$};
    \node at (1.5,0.) {$-m^2$};
    \node at (0.3,2.6) {$\calD$};

    \draw[line width=1.1pt, black] (0,-2.4) -- (0,2.5);

    \draw[dashed, line width=0.9pt] (-2.2,-1.7) -- (0,0);
    \draw[dashed, line width=0.9pt] (0,0) -- (-2.2,1.7);

    \fill[black] (-2.2,-1.7) circle (2.2pt) node[below right]{$\sigma$};
    \fill[black] (-2.2,1.7) circle (2.2pt) node[above right]{$\sigma$};

    \draw[line width=1.0pt, ->] (-2.2,-1.7) -- (-1.9,-1.47);
    \draw[line width=1.0pt, ->] (-2.2,1.7) -- (-2.55,1.98);

    \draw[line width=1.0pt, ->] (-4.2,-1.7) -- (-4.2,-0.7) node[below right]{$t$};

\end{scope}

\begin{scope}[xshift=9.5cm]

      \node at (-1.5,0.) {$+m^2$};
    \node at (1.5,0.) {$-m^2$};
    \node at (0.3,2.6) {$\calD$};

    \draw[line width=1.1pt, black] (0,-2.4) -- (0,2.5);

    \draw[dashed, line width=0.9pt] (-2.2,-1.7) -- (0,0);
   
    \draw[red, line width=0.9pt, decorate,
      decoration={snake, amplitude=1.5pt, segment length=8pt}]
      (0,0) -- (2.2,1.7);

    \fill[black] (-2.2,-1.7) circle (2.2pt) node[below right]{$\sigma$};
   
    \fill[black] (2.2,1.7) circle (2.2pt) node[above left]{$\mu$};

    \fill[red] (0,0) circle (2.4pt) node[below right]{\small$q_{\calD}$};

    \draw[line width=1.0pt, ->] (-2.2,-1.7) -- (-1.9,-1.47);
    
    \draw[line width=1.0pt, ->] (2.2,1.7) -- (2.55,1.98);

    \draw[line width=1.0pt, ->] (-4.2,-1.7) -- (-4.2,-0.7) node[below right]{$t$};

\end{scope}
\end{tikzpicture}
\caption{The two possible channels of an in--going right--moving single--particle state created by $\sigma$ scattering off the interface $\calD$ described in this Section. When transmitted, the out--going single--particle state must be created by a twist operator $\mu$ and the global charge is stored on the defect.}
\label{fig: ising scattering}
\end{figure}

\paragraph{Scattering in the bosonized picture.} Let us argue that in the bosonic Ising frame, the scattering process described above realizes an example of massive categorical scattering. Recall that, in the twisted sector of $(-1)^{F}$ (i.e. the Ramond sector on the cylinder) there are two operators $\Phi_{\pm}$ distinguished by their charge under $(-1)^{F}$. Upon bosonization, depending on the phase we want to describe, one of them is interpreted as $\sigma$ and the other as $\mu$. The form factors\footnote{defined as the overlaps 
\begin{equation}
    f_{n}^{\Phi}(\beta_1, ..,\beta_n)= \bra{0}\Phi(0)\ket{\beta_1,..,\beta_n}\, .
\end{equation}} of those operators are well known \cite{Mussardo:2010mgq}, the fermion parity requires
\begin{equation}
    f^{\Phi_{+}}_{2n+1}=0\, , \qquad f^{\Phi_{-}}_{2n}=0\, ,
\end{equation}
and the explicit solution is
\begin{equation}
   \begin{split}
       &f_{n}^{\Phi_{\pm}}(\beta_1,..,\beta_{n})=i^n \langle\Phi_+\rangle \prod_{1\le i<j\le n}\tanh\left(\frac{\beta_i-\beta_j}{2}\right) \, , \qquad n\ge 2 \\ & f_0^{\Phi_+}= f_1^{\Phi_-}=\langle\Phi_+\rangle\, ,
   \end{split}
\end{equation}
where $n$ even refers to $\Phi_{+}$ and $n$ odd to $\Phi_{-}$. The fact that the vev of $\Phi_{+}$ also controls the overlap of $\Phi_{-}$ with the single particle state is non-trivial, and derives from matching with the CFT behaviour in the UV. Notice that the defect has non-trivial $1$-particle overlap, even at zero momentum, implying that the correlation functions involving the operators $\Phi_{\pm}$ in the twisted sector of $(-1)^F$ are non-vanishing. In particular, the leading contribution in the IR to the mixed two point function is
\begin{equation}
    \langle \Phi_{-}(x,-t) \calD \Phi_{+}(0,t)\rangle = C \frac{i (g-2)}{2\pi} |\langle \Phi_{+}\rangle |^2 + O( e^{-m |x|})\, .
\end{equation}
After bosonization, this corresponds to a non-zero correlation function $\langle \sigma \calD \mu \rangle$. Rotating back to have the defect extended along time, this two point function is mapped to a scattering process of the single particle state created by $\sigma$ to the twisted sector state created by $\mu$. We can interpret this process as a single particle excitation created on top of the disordered vacuum of the Ising field theory that, after scattering off the interface, is partially transmitted and becomes a kink-like excitation of the ordered phase, see Figure \ref{fig: ising scattering}.

\subsection{Integrable Models with a Boundary}

Another class of theories which can provide examples are integrable massive QFTs with a boundary. Let us briefly review the framework. The bulk theory is gapped and has many vacua, which form a module category of the fusion category symmetry $\mathcal{C}$ preserved along the RG flow. To each pair of vacua $a,b$ we can associate an Hilbert space on the line $\bH_{ab}$ in the usual way: massive excitations in Hilbert spaces with $a\neq b$ are kinks, while for $a=b$ we have particles/breathers. To each vacuum we can associate a state $\ket{a}$ in the circle Hilbert space of the theory. However since the theory is massive the properties of these states depend on the quantity $L m$, with $L$ the length of the spatial circle and $m$ the smallest mass scale of the theory. For $Lm \rightarrow \infty$ the states $\ket{a}$ are the ground states of the Hamiltonian and are all degenerate, as we decrease $L m$ going in the UV they split up and their energy asymptotes to a value determined by the conformal dimension of the CFT operator that creates them. 

One caveat has to be taken into account. In the large volume limit, while states corresponding to a primary operator in the UV CFT that, in the IR, flow to a ground state furnish a particular basis of the ground state space, they are not guaranteed to provide the correct clustering vacua, as there may be a basis change relating them. For simplicity we are going to focus on the case in which the vacua are described by the regular module category, which physically corresponds to the symmetry $\mathcal{C}$ being fully broken. 

To describe the scattering we denote the excited states by $\ket{K_{ab}(\beta)}$ with $a,b$ labelling the vacua, and $\beta$ the rapidity. The scattering $S$ matrix is defined by ($\beta_1>\beta_2$)
\begin{equation}
    \ket{K_{dc}(\beta_1)K_{cb}(\beta_2)}= \sum_{a}S^{ab}_{cd}(\beta_1-\beta_2) \ket{K_{da}(\beta_2)K_{ab}(\beta_1)}\,.
\end{equation}
\begin{figure}[t]
\centering
\begin{tikzpicture}[x=1cm,y=1cm]

\node at (0,0) {$S^{ab}_{dc}(\beta_1-\beta_2)\,=$};

\begin{scope}[shift={(4.6,0)}]

\fill[red!65!black!30] (0,0) circle (0.35);

\draw[red, line width=0.9pt, decorate,
  decoration={snake, amplitude=1.2pt, segment length=7pt}]
  (0.25,0.25) -- (1.45,1.45);

\draw[red, line width=0.9pt, decorate,
  decoration={snake, amplitude=1.2pt, segment length=7pt}]
  (-0.25,0.25) -- (-1.45,1.45);

\draw[red, line width=0.9pt, decorate,
  decoration={snake, amplitude=1.2pt, segment length=7pt}]
  (0.25,-0.25) -- (1.45,-1.45);

\draw[red, line width=0.9pt, decorate,
  decoration={snake, amplitude=1.2pt, segment length=7pt}]
  (-0.25,-0.25) -- (-1.45,-1.45);

\node at (0,0.78) {$a$};
\node at (0.78,0) {$b$};
\node at (0,-0.78) {$c$};
\node at (-0.78,0) {$d$};

\node at (1.78,1.72) {$\beta_1$};
\node at (-1.78,1.72) {$\beta_2$};
\node at (1.78,-1.72) {$\beta_2$};
\node at (-1.78,-1.72) {$\beta_1$};

\end{scope}

\node at (10.5,0) {$R^{a}_{bc}(\beta)\,=$};

\begin{scope}[shift={(13.2,0)}]

\fill[red!65!black!30] (0,0) circle (0.35);

\fill[gray!15] (0,-1.35) rectangle (0.55,1.35);
\draw[black, line width=1pt] (0,-1.35) -- (0,1.35);
\node[above, black] at (0.275,1.35) {$B$};

\draw[red, line width=0.9pt, decorate,
  decoration={snake, amplitude=1.2pt, segment length=7pt}]
  (-0.25,0.25) -- (-1.45,1.45);

\draw[red, line width=0.9pt, decorate,
  decoration={snake, amplitude=1.2pt, segment length=7pt}]
  (-0.25,-0.25) -- (-1.45,-1.45);

\node at (-0.72,0) {$a$};
\node at (-0.35,0.78) {$c$};
\node at (-0.35,-0.78) {$b$};

\node at (-1.72,1.65) {$\beta$};
\node at (-1.72,-1.65) {$\beta$};

\end{scope}

\end{tikzpicture}
\caption{Pictorial representation of the S-matrix (left) and Reflection matrix (right) in the presence of kink states interpolating between different vacua.}
\label{fig:scattering}
\end{figure}
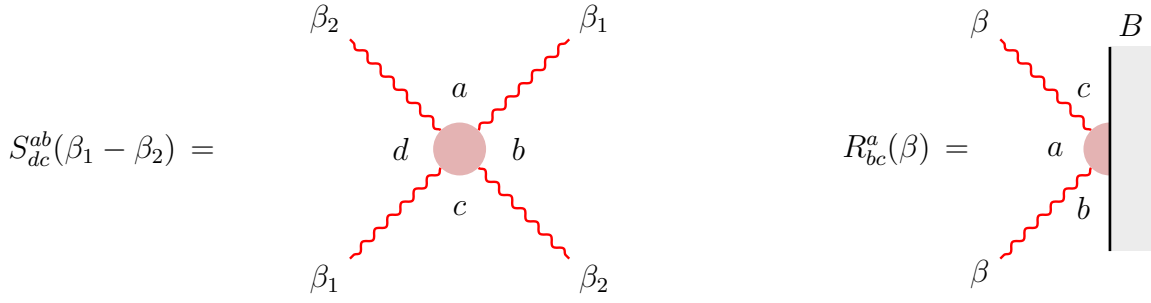
For integrable scatterings, the corresponding S-matrix is constrained by the following three axioms:
\begin{itemize}
    \item Unitarity
    \begin{equation}
        \sum_{a}S^{ab}_{dc}(\beta)S^{eb}_{da}(-\beta)= \delta_{d}^e\, , 
    \end{equation}
    \item Yang-Baxter equation
    \begin{equation}
        \sum_{g}S^{gd}_{fe}(\beta_{12})S_{gd}^{bc}(\beta_{13})S_{fg}^{ab}(\beta_{23})=\sum_{g}S^{gc}_{ed}(\beta_{23})S^{ag}_{fe}(\beta_{13})S^{bc}_{ag}(\beta_{12})\, , 
    \end{equation}
    \item Modified crossing symmetry \cite{Copetti:2024rqj, Copetti:2024dcz}
    \begin{equation}
        S^{ab}_{dc}(\beta)= \sqrt{\frac{d_a d_c}{d_b d_d}}S^{bc}_{ad}(i\pi-\beta)\, ,
    \end{equation}
    with $d_a$ the quantum dimension of the line $a$ of $\mathcal{C}$. 
\end{itemize}
Therefore, any solution of this system of equations defines a non-perturbative and generically interacting scattering matrix.

Now let's consider the addition of a boundary $B$ which is weakly symmetric under the symmetry $\mathcal{C}$ \cite{Choi:2023xjw}. Scattering in integrable QFTs with boundaries has been formalized in \cite{Ghoshal:1993tm}, see also \cite{Shimamori:2025ntq}. In presence of a boundary, since the spatial momentum is no longer conserved, the sign of the rapidity matters: states with $\beta>0$ represent particles moving towards the boundary, while for $\beta<0$ particles move away from it. In this sense in-states have positive rapidity while out-states have $\beta<0$. The scattering setup we want to consider has a boundary at a fixed location in space, say $x=0$, and extending along time. At $x=-\infty$ we must select an asymptotic vacuum corresponding to a fixed boundary condition $a$. The allowed Hilbert spaces are therefore denoted by $\bH_{a B}$. We assume that the boundary condition is compatible with all vacua (namely, as a boundary state, it has non-zero overlaps with all the states $\ket{a}$).

The central quantity in the scattering processes is the reflection matrix defined as a unitary between in and out states,
\begin{equation}
    \ket{K_{ab}(\beta)}= \sum_{c}R^{a}_{bc}(\beta)\ket{K_{ac}(-\beta)}\, , 
\end{equation}
with $\beta>0$. We can think of the label $a$ as specifying the Hilbert space in which the scattering takes place (i.e. the one on which $R^a$ acts) while $b,c$ specify the states that are acted upon. Assuming that bulk integrability is preserved, the reflection amplitude satisfies
\begin{itemize}
    \item Unitarity
    \begin{equation}
        \sum_{d}R^{b}_{cd}(\beta)R^{b}_{da}(-\beta)= \delta_{ac}
    \end{equation}
    \item Boundary Yang-Baxter
    \begin{equation}
    \begin{split}
         \sum_{fg}S^{ga}_{cb}(\beta_{1}-\beta_2) R^{g}_{af}(\beta_1)&S^{df}_{cg}(\beta_1+\beta_2)R^{d}_{fe}(\beta_2)= \\&\sum_{fg}R^{b}_{ag}(\beta_2)S^{fg}_{cb}(\beta_1+\beta_2)R^{f}_{ge}(\beta_1)S^{de}_{cf}(\beta_1-\beta_2)  \, .
    \end{split}
    \end{equation}
    \item Modified crossing\cite{Shimamori:2025ntq}
    \begin{equation}
        R^{b}_{ca}\left(\frac{i\pi}{2}-\beta\right) =\sum_{d}\sqrt{\frac{d_d}{d_b}}S^{ba}_{cd}(2\beta)R^{d}_{ca}\left(\frac{i\pi}{2}+\beta\right)\, .
    \end{equation}
\end{itemize}
In presence of a fixed weakly symmetric boundary the symmetry $\cC$ is realized slightly differently than in infinite space. Indeed we no longer have access to the full strip algebra that acts on the infinite line, rather we can consider only configurations in which the vacuum can change but the boundary is fixed, this is the Defect-Strip algebra discussed in \ref{sec:half-strip}. In pictures 
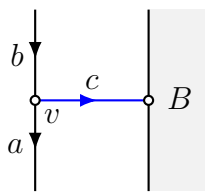
\begin{center}
\begin{tikzpicture}[
    thick,
    ->-/.style={
        decoration={markings, mark=at position 0.55 with {\arrow{Latex}}},
        postaction={decorate}
    },
    dot/.style={
        circle, 
        draw=black, 
        fill=white, 
        inner sep=1.2pt
    }
]
    \fill[gray!10] (1.5, -1.2) rectangle (2.3, 1.2);

    \draw[->-] (0, 1.2) -- (0, 0) node[midway, left] {$b$};
    \draw[->-] (0, 0) -- (0, -1.2) node[midway, left] {$a$};

    \draw[-] (1.5, -1.2) -- (1.5, 0) ;
    \draw[-] (1.5, 0) -- (1.5, 1.2) ;

    \draw[blue, ->-] (0, 0) -- (1.5, 0) node[midway, above, text=black] {$c$};

    \node[dot] at (0,0) {};
    \node[dot] at (1.5,0) {};

    \node[below right, inner sep=3pt] at (0,0) {$v$};
    \node[below left, inner sep=3pt] at (2.2,0.3) {$B$};

\end{tikzpicture}
\end{center}
where the left boundary is at infinity. The bulk symmetry lines (together with a choice of morphism) provide maps $\bH_{aB}\rightarrow \bH_{bB}$ and the Ward Identities for $\mathcal{C}$ impose that the reflection matrix commutes with those maps. Our objective in this section is to find examples in which a breather, which we assume is created by a local operator, scatters into a kink (which is necessarily created by a twisted sector operator).

\paragraph{Deformed tricritical Ising with a boundary.} 
The simplest example is that of the Tricritical Ising model deformed by $\sigma'$ which preserves only the Fibonacci line $W$ of the UV CFT. In the Table below we recall the primary field content of the Tricritical Ising CFT and their charges under $W$
\begin{equation}
    \begin{array}{|r|c|c|c|c|c|c|}
    \hline
    & 1 & \varepsilon & \varepsilon' & \varepsilon'' & \sigma & \sigma' \\\hline
    W : & \zeta & -\zeta^{-1} & -\zeta^{-1} & \zeta & -\zeta^{-1} & \zeta \\\hline
    h : &0 &\frac{1}{10} & \frac{3}{5}& \frac{3}{2}& \frac{3}{80} &\frac{7}{16} \\\hline
    \overline{h} : &0 &\frac{1}{10} &\frac{3}{5} &\frac{3}{2} & \frac{3}{80}& \frac{7}{16}\\\hline
\end{array}
\end{equation}
See Appendix \ref{app: tricrit} for more details on the UV CFT and the Fibonacci symmetry. The sign of the coupling is immaterial since $\sigma'$ is charged under an invertible $\bZ_{2}$ symmetry and the IR is gapped with $2$ vacua $\ket{\unit}$ and $\ket{W}$. The Fibonacci line acts on the vacua according to the fusion rules of the category, 
\begin{equation}
    W\ket{\unit}= \ket{W}\, , \qquad W\ket{W}= \ket{\unit}+ \ket{W}\, . 
\end{equation}
Diagonalizing this action we find the eigenvalues $\zeta$ and $-\zeta^{-1}$, with $\zeta= \frac{1+\sqrt{5}}{2}$ the Golden ratio. Comparing with the eigenvalues of $W$ on the operators of the UV CFT (see e.g. \cite{Chang:2018iay}) it is natural to relate the ground states with the states created by the identity operator and $\sigma$, which is the lightest of the UV primaries (see also \cite{Klassen:1992qy}). The spectrum of massive excitations contains a breather in $\bH_{WW}$ a kink in $\bH_{\unit W}$ and an anitkink in $\bH_{W\unit}$, while the Hilbert space $\bH_{\unit \unit}$ does not contain single particle states (the lightest excitation being the $2$-particle state kink-antikink). In the circle Hilbert space, besides the two vacua, only the breather appears as a single particle state and the corresponding UV operator is $\varepsilon$, the second lightest primary  \cite{Klassen:1992qy}. 

The theory is integrable and the $2\rightarrow 2$ scattering matrix is known explicitly\footnote{The function $R(\beta)$ is given by 
\begin{equation}
    R(\beta)=-\frac{f_{-2/5}(9\beta/5)f_{3/5}(9\beta/5) F_{-1/9}(\beta)F_{2/9}(\beta)}{\sinh\left(\frac{9}{5}(\beta+i\pi)\right)}
\end{equation}
where
\begin{equation}
    f_{\alpha}(x)= \frac{\sinh\left(\frac{x+i\alpha \pi}{2}\right)}{\sinh\left(\frac{x-i\alpha \pi}{2}\right)}\, , \qquad F_{\alpha}(x)= -f_{\alpha}(x)f_{\alpha}(i\pi-x)\, .
\end{equation}
}\cite{Zamolodchikov:1990xc,Smirnov:1991uw,Colomo:1991gw, Colomo:1992nr, Copetti:2024rqj}
\begin{equation}
    S^{ab}_{dc}(\beta)= R(\beta)\left[\sqrt{\frac{d_a d_c}{d_d d_b}} \sinh\left(\frac{9}{5}\beta\right)\delta_{ac}+\sinh\left(\frac{9}{5}(i\pi-\beta)\right)\delta_{bd}\right]\, ,
\end{equation}
where $a,b,c,d$ can be either $\unit$ or $W$ with the constraint that adjacent vacua cannot both be $\unit$.
Now let's consider the addition of a fixed boundary. In this case we have two simple boundary conditions $B_{\unit}$ and $B_{W}$, but only the latter is weakly symmetric under $W$. The scattering setup we want to consider is with the boundary at a fixed location in space, say $x=0$, and extending along time. At $x=-\infty$ we need to choose a vacuum, therefore, with a fixed boundary condition, we have two Hilbert spaces we can consider $\bH_{\unit B_W}$ and $\bH_{W B_W}$. Another useful perspective is to view $B_{W}$ as a boundary state $\ket{B_W}$ in the Hilbert space of the bulk theory. Since there are superselection sectors in infinite volume, $\ket{B_W}$ can have components in all of them, overlapping in particular with both the vacua $\ket{\unit}$ and $\ket{W}$. At the level of single particle excitations the Hilbert space $\bH_{\unit B_{W}}$ contains the kink while $\bH_{W, B_W}$ contains both the antikink and the breather. The symmetry algebra is very simple in this example as there are only $3$ non-trivial maps
\begin{equation*}
    M_1=\begin{tikzpicture}[baseline,
    thick,
    ->-/.style={
        decoration={markings, mark=at position 0.55 with {\arrow{Latex}}},
        postaction={decorate}
    },
    dot/.style={
        circle, 
        draw=black, 
        fill=white, 
        inner sep=1.2pt
    }
]
    \fill[gray!10] (1.5, -1.2) rectangle (2.3, 1.2);

    \draw[-] (0, 1.2) -- (0, 0) node[midway, left] {$W$};
    \draw[-] (0, 0) -- (0, -1.2) node[midway, left] {$W$};

    \draw[-] (1.5, -1.2) -- (1.5, 0) ;
    \draw[-] (1.5, 0) -- (1.5, 1.2) ;

    \draw[blue, -] (0, 0) -- (1.5, 0) node[midway, above, text=black] {$W$};

    \node[dot] at (0,0) {};
    \node[dot] at (1.5,0) {};

    \node[below left, inner sep=3pt] at (2.4,0.3) {$B_W$};

\end{tikzpicture}\,  : \bH_{WB_W}\rightarrow \bH_{WB_W}
\end{equation*}
\begin{equation*}
    M_2=\begin{tikzpicture}[baseline,
    thick,
    ->-/.style={
        decoration={markings, mark=at position 0.55 with {\arrow{Latex}}},
        postaction={decorate}
    },
    dot/.style={
        circle, 
        draw=black, 
        fill=white, 
        inner sep=1.2pt
    }
]
    \fill[gray!10] (1.5, -1.2) rectangle (2.3, 1.2);

    \draw[-] (0, 1.2) -- (0, 0) node[midway, left] {$W$};
    \draw[dashed] (0, 0) -- (0, -1.2) ;

    \draw[-] (1.5, -1.2) -- (1.5, 0) ;
    \draw[-] (1.5, 0) -- (1.5, 1.2) ;

    \draw[blue, -] (0, 0) -- (1.5, 0) node[midway, above, text=black] {$W$};

    \node[dot] at (0,0) {};
    \node[dot] at (1.5,0) {};

    \node[below left, inner sep=3pt] at (2.4,0.3) {$B_W$};

\end{tikzpicture}\,  : \bH_{\unit B_W}\rightarrow \bH_{WB_W}
\end{equation*}
\begin{equation}\label{eq: maps tric}
    M_3=\begin{tikzpicture}[baseline,
    thick,
    ->-/.style={
        decoration={markings, mark=at position 0.55 with {\arrow{Latex}}},
        postaction={decorate}
    },
    dot/.style={
        circle, 
        draw=black, 
        fill=white, 
        inner sep=1.2pt
    }
]
    \fill[gray!10] (1.5, -1.2) rectangle (2.3, 1.2);

    \draw[dashed] (0, 1.2) -- (0, 0);
    \draw[-] (0, 0) -- (0, -1.2) node[midway, left] {$W$};

    \draw[-] (1.5, -1.2) -- (1.5, 0) ;
    \draw[-] (1.5, 0) -- (1.5, 1.2) ;

    \draw[blue, -] (0, 0) -- (1.5, 0) node[midway, above, text=black] {$W$};

    \node[dot] at (0,0) {};
    \node[dot] at (1.5,0) {};

    \node[below left, inner sep=3pt] at (2.4,0.3) {$B_W$};

\end{tikzpicture}\,  : \bH_{W B_W}\rightarrow \bH_{\unit B_W}
\end{equation}
Their full algebra is easy to compute using the bulk $F$-symbols. We find
\begin{equation}\label{eq:algstr}
    \begin{split}
        &M_1 \circ M_1= \zeta \unit + \zeta^{-1}M_1\, , \quad M_2\circ M_3 = \unit -\zeta^{-1}M_1 \\ & M_1\circ M_2= -M_2\, , \quad M_3\circ M_1 =-M_3 \, , \quad M_3\circ M_2= \zeta \unit
    \end{split}
\end{equation}
where $\unit$ is the identity line operator (in the appropriate Hilbert space). Restricting to the space of single particle states it is easy to describe explicitly the representation. Since $\bH_{W, B_W}$ contains both the breather and the antikink, we take $M_1$ to be a $2\times 2$ matrix, while $M_2$ and $M_3$ can be thought of as $2\times 1$ and $1\times 2$ matrices. From the algebra one can see that $M_1$ has eigenvalues $-1$ and $\zeta$. Let us denote the eigenspaces by $E_{-1}$ and $E_{\zeta}$. From $M_1 \circ M_2 =-M_2$ we see that the image of $M_2$ coincides with the $E_{-1}$, while $M_3 \circ M_1 =-M_3$ tells us that the kernel of $M_3$ is $E_\zeta$. The map $M_2\circ M_3$ is an unnormalized projector onto the eigenspace $E_{-1}$. One can easily get the explicit matrices choosing a basis for the one-particle subspaces of $\bH_{W, B_W}$ and $\bH_{\unit B_W}$.

We are now in position to study the integrability equation for the reflection coefficients. In this case we can organize them into $2$ matrices, depending in which Hilbert space the scattering process takes place. In particular $R^{\unit}_{WW}$ is a single function describing the reflection off the boundary of the kink, while $R^{W}_{ab}$, with $a,b=\unit,W$, is a $2\times 2$ matrix describing the breather and the antikink interacting with $B_W$. In other words we can think of the reflection amplitudes as maps $R^{\unit}:\bH_{\unit B_W}\rightarrow \bH_{\unit B_W}$ and $R^{W}:\bH_{W B_W}\rightarrow \bH_{W B_W}$, then the selection rules of the Fibonacci symmetry are 
\begin{equation}
    M_1 \circ R^{W} = R^{W}\circ M_1\, , \qquad M_2\circ R^{\unit}= R^{W}\circ M_2\, , \qquad M_3 \circ R^{W}= R^{\unit}\circ M_3\, .
\end{equation}
In order to translate those to explicit constraints on the components of the reflection matrix we need to choose a basis, or, better, we need to compute the matrices $M_{1,2,3}$ in the kinks and breather basis. To do so, let us define the single particle states via the path integrals
\begin{equation*}
\ket{K_{WW}}=\frac{1}{\zeta}\begin{tikzpicture}[baseline={(1,-0.5)},
    thick,
    ->-/.style={
        decoration={markings, mark=at position 0.55 with {\arrow{Latex}}},
        postaction={decorate}
    },
    dot/.style={
        circle, 
        draw=black, 
        fill=white, 
        inner sep=1.2pt
    }
]
    \fill[gray!10] (1.5, -1.2) rectangle (2.3, 0.5);

    \draw[-] (0, 0.5) -- (0, 0);
    \draw[-] (0, 0) -- (0, -1.2) node[midway, left] {$W$};

    \draw[-] (1.5, -1.2) -- (1.5, 0) ;
    \draw[-] (1.5, 0) -- (1.5, 0.5) ;

    \draw[-] (0, -1.2) -- (1.5, -1.2);

    \draw[-] (1.5/2,-1.2) -- (1.5/2, -1.9) node[midway,left] {$W$};
     \node[dot] at (1.5/2,-1.2) {};
    \node[below left, inner sep=3pt] at (2.4,0.3) {$B_W$};

\end{tikzpicture}\, , \quad \ket{K_{W\unit}}=\frac{1}{\zeta^{1/2}}\begin{tikzpicture}[baseline={(1,-0.5)},
    thick,
    ->-/.style={
        decoration={markings, mark=at position 0.55 with {\arrow{Latex}}},
        postaction={decorate}
    },
    dot/.style={
        circle, 
        draw=black, 
        fill=white, 
        inner sep=1.2pt
    }
]
    \fill[gray!10] (1.5, -1.2) rectangle (2.3, 0.5);

    \draw[-] (0, 0.5) -- (0, 0);
    \draw[-] (0, 0) -- (0, -1.2) node[midway, left] {$W$};

    \draw[-] (1.5, -1.2) -- (1.5, 0) ;
    \draw[-] (1.5, 0) -- (1.5, 0.5) ;

    \draw[-] (0, -1.2) -- (1.5/2, -1.2);
    \draw[dashed] (1.5/2, -1.2) -- (1.5, -1.2);

    \draw[-] (1.5/2,-1.2) -- (1.5/2, -1.9) node[midway,left] {$W$};
     \node[dot] at (1.5/2,-1.2) {};
    \node[below left, inner sep=3pt] at (2.4,0.3) {$B_W$};

\end{tikzpicture}
\end{equation*}
\begin{equation*}
\ket{K_{\unit W}}=\frac{1}{\zeta^{1/2}}\begin{tikzpicture}[baseline={(1,-0.5)},
    thick,
    ->-/.style={
        decoration={markings, mark=at position 0.55 with {\arrow{Latex}}},
        postaction={decorate}
    },
    dot/.style={
        circle, 
        draw=black, 
        fill=white, 
        inner sep=1.2pt
    }
]
    \fill[gray!10] (1.5, -1.2) rectangle (2.3, 0.5);

    \draw[dashed] (0, 0.5) -- (0, 0);
    \draw[dashed] (0, 0) -- (0, -1.2) ;

    \draw[-] (1.5, -1.2) -- (1.5, 0) ;
    \draw[-] (1.5, 0) -- (1.5, 0.5) ;

    \draw[dashed] (0, -1.2) -- (1.5/2, -1.2);
    \draw[-] (1.5/2, -1.2) -- (1.5, -1.2);

    \draw[-] (1.5/2,-1.2) -- (1.5/2, -1.9) node[midway,left] {$W$};
     \node[dot] at (1.5/2,-1.2) {};
    \node[below left, inner sep=3pt] at (2.4,0.3) {$B_W$};

\end{tikzpicture}
\end{equation*}
where the prefactors are needed to achieve unit normalized states. The $W$ line at the bottom is artificial (it should be thought of as extending in the SymTFT bulk \cite{Copetti:2024dcz}) and is needed to single out the strip algebra representation in which the triplet of single particle states transforms. With these definitions we can explicitly compute the maps $M_1, M_2, M_3$ in the single particle states basis. We find, for $M_1$,
\begin{equation}
    M_1 \ket{K_{WW}}= \zeta^{1/2} \ket{K_{W\unit}} +\zeta^{-1}\ket{K_{WW}}\, ,\quad M_1 \ket{K_{W\unit}}= \zeta^{1/2}\ket{K_{WW}}
\end{equation}
one can check that, consistently, the corresponding $2\times 2$ matrix $M_1$ has eigenvalues $-1$ and $\zeta$. Similarly we have, 
\begin{equation}
\begin{split}
     & M_3\ket{K_{WW}}= - \zeta^{-1/2}\ket{K_{\unit W}}\, ,\quad  M_3\ket{K_{W\unit}}=  \ket{K_{\unit W}}\, , \\ & \quad M_{2}\ket{K_{\unit W}}= \ket{K_{W\unit}}-\zeta^{-1/2}\ket{K_{WW}}
\end{split}
\end{equation}
one can easily check that these explicit maps satisfy the algebra \eqref{eq:algstr}. We can now explicitly solve the Ward Identities, we work in the single-particle basis so that the physical meaning of the reflection coefficients is transparent. Out of the initial $5$ independent reflection amplitudes only two are independent; we choose those to be $R^{W}_{\unit W}$ and $R^{W}_{\unit \unit}$ representing the breather either scattering into the antikink or just bouncing off the boundary. The full reflection amplitude is parametrized as
\begin{equation}
\begin{split}
     & R^{W}= R^{W}_{\unit \unit}\begin{pmatrix}
       1 & 0\\ 0 & 1    \end{pmatrix} + R^{W}_{\unit W}\begin{pmatrix}
        \zeta^{-3/2} & 1\\ 1 & 0
    \end{pmatrix}\, , \\ &
    R^{\unit}_{WW}= R^{W}_{\unit \unit} -\zeta^{-1/2} R^{W}_{\unit W}\, .
\end{split}
\end{equation}
We now plug this form in the boundary Yang-Baxter equations. One can see that only $4$ are linearly independent. To solve them, we expand for $\beta_2\rightarrow \beta_1$ and pick the leading order term. Defining the ratio
\begin{equation}
    r(\beta)= \frac{R^{W}_{\unit W}(\beta)}{R^{W}_{\unit \unit}(\beta)}
\end{equation}
the limit $\beta_2\rightarrow \beta_1$ produces a first order differential equation, the same for all boundary YB equations 
\begin{equation}
\begin{split}
    -\frac{1}{18} \left(5+\sqrt{5}\right) \sinh \left(\frac{18 \beta}{5}\right) r'(\beta)&+\frac{2 r(\beta)^2 \sin \left(\frac{1}{5} (\pi -18 i
   \beta)\right)}{5^{3/4}}\\ &+\frac{1}{5} \left(5+\sqrt{5}\right) r(\beta) \cosh \left(\frac{18 \beta}{5}\right)=0\, , 
\end{split}   
\end{equation}
with solution
\begin{equation}
    r(\beta)= \frac{\sinh \left(\frac{18 \beta}{5}\right)}{k+\zeta^{-1}5^{-1/4}\sinh \left(\frac{3}{10} (i \pi -12 \beta)\right)}
\end{equation}
where $k$ is an arbitrary constant. The next constraint to consider is unitarity, which imposes
\begin{equation}
    \begin{split}
        &\left(1+  r(\beta)r(-\beta)\right) R^{W}_{\unit \unit}(\beta)R^{W}_{\unit \unit}(-\beta)=1\, ,\\
        &\left(1+ \zeta^{-3/2}( r(\beta)+r(-\beta))+2\zeta^{-1} r(\beta)r(-\beta) \right) R^{W}_{\unit \unit}(\beta)R^{W}_{\unit \unit}(-\beta)=1\, ,\\
        &  \left(r(\beta) + r(-\beta) + \zeta^{-3/2} r(\beta)r(-\beta) \right)=0\, .
    \end{split}
\end{equation}
The third equation is identically satisfied by our solution $r(\beta)$, while the other two coincide when evaluated on the solution and fix
\begin{equation}
    R^{W}_{\unit \unit}(\beta)R^{W}_{\unit \unit}(-\beta)= \frac{1}{1+r(\beta)r(-\beta)}\, .
\end{equation}
The final constraints come from the modified crossing equations which fix $R^{W}_{\unit \unit}$ up to CDD ambiguities. We find that the solution for $r(\beta)$ is compatible with crossing, unitarity, and integrability for all values of the constant $k$, which is therefore a free parameter of the scattering solution. The non-trivial solution for $r(\beta)$ shows that there is indeed a non-vanishing reflection coefficient $R^W_{\unit W}$ for the (anti)kink scattering off the boundary and becoming a breather.
\paragraph{An interpretation of the parameter $k$.} As already noticed, we actually found a $1$-parameter family of solutions for $r$, and none of the integrability equations fix it. Also notice that, various values of $k$ parametrize different positions for the poles for $r(\beta)$, giving rise to different resonances. A possible explanation of this parameter is related to the existence of a relevant boundary coupling compatible with both the Fibonacci symmetry and integrability that is generated along the RG flow. The parameter $k$ would then be related to an appropriate dimensionless ratio of the bulk and boundary couplings.

To test this hypothesis, we may look at the boundary UV operators that are both relevant and neutral under the Fibonacci symmetry. To this aim, the first step is to classify the UV boundary conditions that are weakly symmetric with respect to $W$. In the tricritical Ising model there are $3$ of those, corresponding to the bulk primaries $\varepsilon, \varepsilon'$ and $\sigma$ \cite{Choi:2023xjw}. Since we cannot rule out a composite boundary in the UV we shall consider as candidate operators all those that correspond to states in the strip Hilbert spaces $\bH_{AB}$ with $A,B$ any of the three weakly symmetric boundary conditions (which we denote with the same symbol as the primary they correspond to). The annulus partition functions are
\begin{equation}
    Z_{AB}(q)= \sum_{C}N_{AB}^C \chi_C(q)
\end{equation}
with $N_{AB}^C$ the bulk fusion coefficients. The line $W$ has a natural action on the Hilbert spaces $\bH_{AB}$ and fuses according to $W\times W=\zeta\unit+ \zeta^{-1}W$ (it is analogous to $M_1$ of the defect strip algebra), hence it can only act with eigenvalues $-1, \zeta$ on the primary states of $\bH_{AB}$. To find the correct assignments of eigenvalues we proceed as follows. We first let $W$ act on $\bH_{AB}$ and parametrize the resulting partition function as
\begin{equation}
    Z^W_{AB}(q)= \sum_{C}N_{AB}^C X_{AB}^C\chi_C(q)\, , 
\end{equation}
then using an $S$ transformation we pass to the closed channel
\begin{equation}
     Z^W_{AB}(\widetilde{q})= \,_W\bra{A} \widetilde{q}^{L_0+\overline{L}_0-c/12}\ket{B}_W\, ,
\end{equation}
where $\ket{A}_W, \ket{B}_W$ are the components of the Cardy states in the twisted sector of $W$. These admit a decomposition in twisted Ishibashi states (i.e. Ishibashi states built out of primary states in the twisted sector of $W$)
\begin{equation}
    \ket{A}_W= \sum_{D}M_{A}^D \ket{D}\rangle\, ,
\end{equation}
so that the bootstrap equations read
\begin{equation}
     \sum_{C}N_{AB}^C X_{AB}^C S_{CD}= M_A^D M_{B}^D\, .
\end{equation}
We find only one consistent solution, the corresponding spectrum is reported in Tab.\ref{tab:eigenvalues}.
\begin{table}[]
\centering
\renewcommand{\arraystretch}{1.8}
\resizebox{\textwidth}{!}{%
\begin{tabular}{|l|c|c|c|c|c|c|c|c|c|c|c|c|c|c|}
\hline
\textbf{$(A, B)$} & $(\varepsilon, \varepsilon)$ & $(\varepsilon, \varepsilon)$ & $(\varepsilon', \varepsilon')$ & $(\varepsilon', \varepsilon')$ & $(\sigma, \sigma)$ & $(\sigma, \sigma)$ & $(\sigma, \sigma)$ & $(\sigma, \sigma)$ & $(\varepsilon, \varepsilon')$ & $(\varepsilon, \varepsilon')$ & $(\varepsilon, \sigma)$ & $(\varepsilon, \sigma)$ & $(\varepsilon', \sigma)$ & $(\varepsilon', \sigma)$ \\
\hline
$h$ & $0$ & $3/5$ & $0$ & $3/5$ & $0$ & $1/10$ & $3/5$ & $3/2$ & $1/10$ & $3/2$ & $7/16$ & $3/80$ & $7/16$ & $3/80$ \\
\hline
$X$ & $\zeta$ & $-1$ & $\zeta$ & $-1$ & $\zeta$ & $-1$ & $-1$ & $\zeta$ & $-1$ & $\zeta$ & $\zeta$ & $-1$ & $\zeta$ & $-1$ \\
\hline
\end{tabular}%
}
\vspace{0.3cm}
\caption{Eigenvalue assignment for open channel operators in the tricritical Ising CFT.}
\label{tab:eigenvalues}
\end{table}
To determine whether an operator is neutral under $W$ we compare its eigenvalue with that of the identity, which is $\zeta$ is our choice of normalizations. We see that there are two neutral relevant operators, both of them are boundary changing operators living in $\bH_{\varepsilon \sigma}$ and $\bH_{\varepsilon' \sigma}$ and have weight $h=7/16$. It is therefore possible that an appropriate combination of these deformations preserves boundary integrability and provides a parametrization of the family of solutions we have found.

\section{Lattice Impurities}\label{sec: lattice}
We now turn our attention to a different set of theories, namely quantum lattice systems defined on a chain. Indeed, these are a natural arena to study the phenomenon of categorical scattering beyond QFT, as recently noticed in \cite{Ueda:2025ecm}. The aim of this Section is first to show that (one of) the defects presented in \cite{Ueda:2025ecm} is actually symmetry reflecting and enjoys a defect anomaly, thus fitting in our general discussion, and then we construct generalizations of this model where physics is expected to be richer.

We consider two quantum many--body systems $\cT_L$ and $\cT_R$ with Hamiltonians
\be 
H_L = \sum_j h_{L,j} \, , \qquad H_R = \sum_{j} h_{R,j} \, ,
\ee
where $h_{L,j}$ and $h_{R,j}$ are quasilocal terms.
In the presence of an impurity at site $j=m$ the total Hamiltonian is modified to
\be
H_{\calD} = \sum_{j}^{m-1} h_{L,j} + \sum_{k=m+1} h_{R,k} + H_I \equiv H_L(j<m) + H_R(j>m) + H_{I} \, ,
\ee
where $H_I$ is a quasilocal Hamiltonian localized at the impurity location $j=m$. The Hilbert space at the impurity can in principle be different from the one in either spin chain, notably by the addition of further degrees of freedom.

Our construction of symmetric (and symmetry reflecting) defects is very transparent in a lattice setting. Let us assume that the two systems $\cT_{L,R}$ are symmetric with respect to two on--site symmetries generated by unitary operators $U_{L,R}$, respectively. As the Hamiltonian is made up of quasi-local pieces, the original symmetries $U_L$, $U_R$ still commute with it far from the impurity:
\be
U_L h_{L, j} U^{-1}_L = h_{L,j} \, , \quad  U_R h_{R, j} U^{-1}_R = h_{R,j}\,.
\ee
However, at the impurity, the definition of the defect $U = U_L(j<m) \otimes U_R(j>m)$ must be improved by a local unitary $u_{\calD}$ localized at $j=m$:
\be
\widetilde{U} = U_L(j<m) \otimes u_{\calD} \otimes U_R(j>m) \, ,
\ee
to ensure commutativity with the full Hamiltonian.
Interestingly, the resulting symmetry operators might now fail to commute at the location of the interface: this signals the defect anomaly. 
The study of symmetry defects on quantum chains, especially in the presence of boundaries, has been subject of several works \cite{Cheng:2022sgb,Seifnashri:2023dpa,Seiberg:2023cdc,Bhardwaj:2024kvy,Pace:2024oys,Inamura:2024jke,Franco-Rubio:2025qss,Argurio:2026txf}.
In the case of a symmetry-reflecting defect, we simply end the symmetry defect at the impurity site by choosing an adequate decoration.

The purpose of this section is to show in practice how this leads to a vast landscape of impurities carrying defect anomalies, which can lead to transmission of non-local particles from incident charged wavepackets according to our arguments. Furthermore, we will see that in the XYZ model we can construct in this manner nontrivial interfaces which are not connected to topological (transmissive) ones by deformations localized on the defect. 

\subsection{Order/Disorder Interface in Ising} A paradigmatic example is provided by the Ising model, whose interfaces have been classified in \cite{Oshikawa:1996dj}. We will focus on an interface between the ordered and disordered descriptions of the Ising model \cite{Ueda:2025ecm}. This is also the lattice regularization of the model presented in Section \ref{sec:soliton}. Recall that the Ising Hamiltonian:
\be\label{eq: Ising hamiltonian}
H_{\text{Ising}} = -\sum_j  X_j X_{j+1} - h \sum_j  Z_j \, , 
\ee
flows to a $\bZ_2$ ordered/disordered phase for $h > 1$ and $h < 1$, respectively. In \eqref{eq: Ising hamiltonian}, $X$ and $Z$ denote respectively the X and Z Pauli matrices acting on the two-dimensional Hilbert space localized at site $j$.
Following \cite{Ueda:2025ecm}, we consider the following interface at site $j=m$:
\bea
&H &&= H_L + H_R + H_I \, , \\
&H_L &&= -\sum_{j}^{m-2} X_j X_{j+1} - h_L \sum_{j}^{m-1} Z_j \, , \\
&H_R &&= - \sum_{j=m+1} X_j - h_R \sum_{j=m+1} Z_j Z_{j+1} \, , \\
&H_I &&= - X_{m-1} X_m  - \gamma Z_{m} Z_{m+1} \, .
\eea
For $h_L = h_R = \gamma$ the interface is topological, while, for $h_L=h_R = h$ and $\gamma \neq h$ it describes the $\epsilon$ deformation of the KW interface between ordered and disordered Ising theories we have previously analyzed. The mass of the single particle state is:
\be
M = 2 \sqrt{h^2 -1} \, ,
\ee
so choosing different $h_L$ and $h_R$ can be understood as different magnitudes of \emph{bulk} $\epsilon$ perturbations in the Ising CFT.

Let us look at the symmetry of this system. The left and right Hamiltonians are invariant under the $\bZ_2$ symmetries
\be
U_L = \prod_j Z_j \, , \quad U_R = \prod_j X_j \, ,
\ee
respectively. The interface is symmetry reflecting under $\bZ_2 \times \bZ_2$, since the truncated operators
\be
\widetilde{U}_L = \prod_j^m Z_j \, , \quad \widetilde{U}_R = \prod_{j=m} X_j \, , 
\ee
both commute with the defect Hamiltonian:
\be
[\widetilde{U}_L, H] = [\widetilde{U}_R, H] = 0 \, ,
\ee
for all values of $h_L, h_R, \gamma$. However the two symmetries have a defect anomaly, as they fail to commute on site $j=m$ due to the presence of the defect:
\be
\widetilde{U}_L \times \widetilde{U}_R = - \widetilde{U}_R \times \widetilde{U}_L \, .
\ee
From this it is clear that this type of impurity allows for transmission between local charged particles ($\sigma_j = X_j$) and twist defects ($\mu_j = \prod_{k=m}^{j} X_k$). This was explicitly shown in \cite{Ueda:2025ecm}.

\subsection{Interfaces in the XYZ Model}
A much richer playground for this type of defects can be found in the XYZ model, defined by the Hamiltonian
\be
H_{XYZ} = - \sum_{j} J_X  X_{j} X_{j+1} + J_Y Y_j Y_{j+1} + J_Z Z_j Z_{j+1} \, .
\ee
In this case we look for defects, i.e. localized modifications of $H_{XYZ}$. In particular, we would like to construct symmetry-reflecting impurities for the following two symmetries:
\be
\widetilde{U}_L = \prod_j^{m-1}(-)^j Z_j \, X_m \, , \quad \widetilde{U}_R = Y_m \prod_{j=m+1} (-)^j Z_j \, , \label{eq: ulur}
\ee
generating a $\bZ_2^L\times \bZ_2^R$ symmetry. Again, an impurity commuting with $\widetilde{U}_{L,R}$ will have the same $\bZ_2 \times \bZ_2$ defect anomaly following from the lack of commutativity of these two operators at site $j=m$. There are several other choices for these symmetries, which can be mapped to the above by a change of Pauli basis. Notice that:
\be
\widetilde{U}_L \widetilde{U}_R = i \prod_j (-)^j Z_j \, , 
\ee
is an order 4 operator, as implied by the defect anomaly.

To find a suitable defect Hamiltonian we first remove the kinetic terms which cross the site $m$. Apart from the obvious boundary terms $Z_{m-1}$ and $Z_{m+1}$ there are several interesting quadratic terms\footnote{We only report here the ones which are bi-linear in fermions after the Jordan-Wigner transformation.}
\be
Y_{m-1} Y_m \, , \ X_m X_{m+1} \, , \ X_{m-1} Y_m \, , \ X_m Y_{m+1} 
\ee
and cubic terms:
\be
X_{m-1} Z_m X_{m+1} \, , \ X_{m-1} Z_m Y_{m+1} \, , \ Y_{m-1} Z_m X_{m+1} \, , \ Y_{m-1} Z_m Y_{m+1} \, ,
\ee
which we are free to add to the defect Hamiltonian.
To prune this further, let us impose time reversal and parity symmetry on the defect, too. In the bulk, time reversal $\cT$ acts by complex conjugation 
\be
T\;:\quad X_j \to X_j \, , \ Y_j \to - Y_j \, , Z_j \to Z_j\quad, j \not = m\,,
\ee
but on the defect, in order to commute with both $\widetilde{U}_L$ and $\widetilde{U}_R$ we must conjugate by $X_m$, thus 
\be
T\; : \quad X_m \to X_m \, , \ Y_m \to Y_m \, , \ Z_m \to - Z_m\,.
\ee
Similarly, parity exchanges sites $j \leftrightarrow 2m-j$ while acting by $X_m \leftrightarrow Y_m$ and $Z_m \to - Z_m$ on the defect's site. In this way, there are no anomalies with the two $\bZ_2$ symmetries. With this, the defect Hamiltonian takes the form:
\be
H_{I} = \gamma \left(X_{m-1} Y_m + X_{m} X_{m+1} \right) + \delta \left( Y_{m-1} Z_m X_{m+1} - X_{m-1} Z_m Y_{m+1} \right) \, .
\ee
This defines a two-parameter family of defects $\calD(\gamma,\delta)$. Notice that, for $\gamma=0$, we can diagonalize $Z_m$, leading to a direct sum of two defects:
\be \label{eq: decomposition}
\calD(0,\delta) = \calD_{\uparrow}(\delta) \oplus \calD_{\downarrow}(-\delta) \, .
\ee
We now make use of the Jordan-Wigner transformation to make the physics more transparent.
On a single site we have a complex fermion $c_j = a_j + i b_j$ and the Jordan Wigner map reads 
\be
a_j = \begin{cases} -X_j \, \prod_{k<j} (-)^k Z_k \quad &j \, \text{even} \\  
    Y_j \, \prod_{k<j} (-)^k Z_k \quad &j \, \text{odd}  
\end{cases} \, , \qquad  b_j = \begin{cases} Y_j \, \prod_{k<j} (-)^k Z_k \quad &j \, \text{even} \\  
    X_j \, \prod_{k<j} (-)^k Z_k \quad &j \, \text{odd}  
\end{cases} \, , \quad i a_j b_j = Z_j \, .
\ee
In the fermionic variables the defect Hamiltonian is quadratic and reads (for simplicity, we take $m$ to be odd):
\be
H_{I} = H_\gamma + H_\delta = - i \gamma \left(b_{m-1} a_m + a_m a_{m+1} \right) - i \delta \left( a_{m-1} a_{m+1} + b_{m-1} b_{m+1} \right) \, 
\ee
Notice that $b_m$ does not appear in the Hamiltonian. This reflects the fact that the defect ground state is doubly degenerate and spanned by:
\be
|\Omega \rangle \, , \quad b_m |\Omega \rangle \, .
\ee
In the fermionic picture this is the Majorana zero mode imposed by the Arf defect anomaly.
Now let us understand these defects in two different known regimes.

\paragraph{XX model.} The first model we analyze is the $XX$ model, obtained by setting $J_X = J_Y = J \, , \ J_Z = 0$ and which flows to a critical $c=1$ boson at $R=\sqrt{2}$ (i.e. a free Dirac fermion).\footnote{In our conventions, the self-dual radius is $R=1$.}
The theory has a KW duality symmetry \cite{Thorngren:2021yso} (see also \cite{Pace:2024oys} for a derivation of this symmetry on the lattice) corresponding to the gauging of the $\bZ_2^m$ shift symmetry. However, the KW duality defect is not time reversal invariant on the lattice ($\calD_{KW}^\dagger \neq \calD_{KW}$), so we cannot find it as a limit of $H_{\calD}$. However, the corresponding continuum topological defect will appear in the long distance limit of our analysis.

The model is dual to free fermions $c_j = a_j + i b_j$, with the Hamiltonian:
\be
H_{XX}^F = - i J \sum_j a_j a_{j+1} + b_j b_{j+1} \, .
\ee
We analyze first the limit in which $H_{\calD}$ is a topological defect. This is $(\gamma, \delta) = (0,J)$, which corresponds to a defect $\mathscr{T}$ implementing a unit lattice translation, and can be understood by applying the lattice translation on sites $j \geq m$ only. Around this point, we can use the expansion of the lattice field in low-energy modes, which reads:
\be \label{eq: contfermion}
c_j = \begin{cases}
    \psi_L(x) + (-)^j \psi_R(x) \, , \ \ \ &j<m \\
    \psi_L(x) + (-)^{j+1} \psi_R(x) \, , \ \ \ &j>m \\
    a_m + i b_m \, , \ \ \ &j = m \, ,
\end{cases} \, , \quad x = a j \, ,
\ee
where $\psi_L = a_L + i b_L, \psi_R = a_R + i b_R$ are the continuum Weyl fermions, the shift comes from the lattice translation and we have isolated the fermion zero modes $a_m, b_m$. In these variables, the lattice translation acts as chiral fermion parity $(-)^{F_R}$ on the right Weyl fermion \cite{Seiberg:2023cdc}. This is an example of an emanant symmetry \cite{Cheng:2022sgb}. Since the Weyl fermion has two Majorana components $a_R$ and $b_R$, there are indeed two separate Majorana zero modes, corresponding to the edge modes of a Kitaev phase for the two separate chiral fermion parity symmetries. 
If we bosonize the system (in the continuum limit), this leads to a non-simple topological defect $CU^m_{\pi/2} \oplus C U^m_{\pi/2}$ (as we remarked, the zero mode $i a_m b_m = Z_m$ is decoupled). The direct sum decomposition remains true as long as we keep $\gamma=0$.

\begin{figure}
    \centering
     \begin{tikzpicture}
         \draw[blue] (-3,0) -- (3,0) ;
         \draw[very thick] (-3,-0.1) -- (-3,0.1);
                  \draw[very thick] (3,-0.1) -- (3,0.1);
         \draw[fill=black] (0,0) node[above] {$C U_{\pi/2} \oplus U_{\pi/2} C$} circle (0.05);
         \draw[<->] (-1,1) -- (1,1); \node[above] at (0,1) {$ \delta\int_{\calD} \epsilon$};
         \begin{scope}[shift={(0,-3)}]
         \draw[blue] (-3,0) -- (3,0) ;
           \draw[very thick] (-3,-0.1) -- (-3,0.1);
                  \draw[very thick] (3,-0.1) -- (3,0.1);
         \draw[fill=black] (0,0) node[below] {$\cN$} circle (0.05);
         \draw[<->] (-1,-1) -- (1,-1); \node[below] at (0,-1) {$ \delta\int_{\calD} \epsilon$};    
         \end{scope}
         \draw[->] (0,-0.5) -- (0,-2.5); \node[left] at (0,-1.5) {$ \gamma\int_{\calD} a_m \, \psi $};
     \end{tikzpicture}
    \caption{Schematic structure of the symmetry reflecting defects at the $XX$ point: starting from the topological defect $\calD(0,J) = C U_{\pi/2} \oplus U_{\pi/2} C$ we have a conformal manifold generated by the mass deformation $\epsilon$. Deforming by $\gamma$ leads to the $\epsilon$ deformation of the duality defect, instead. The conformal manifolds have the topology of a segment, due to the breaking of a phantom symmetry \cite{Antinucci:2025uvj}, the end-points being reflective walls.}
    \label{fig: flowlat}
\end{figure}
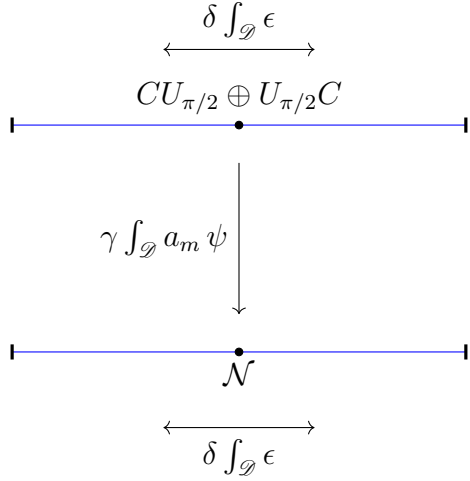

It is simple to see using \eqref{eq: contfermion} that the $\delta$ deformation becomes an exactly marginal mass deformation, while the $\gamma$ deformation lifts the $a_m$ zero mode via:
\be
H_\gamma = - i \gamma a_m \left( a_L - b_L - a_R - b_R \right) \, .
\ee
This is a relevant deformation, driving the defect to a new fixed point. Related deformations coupling Majorana zero modes to bulk fermions have been studied in \cite{Runkel:2007wd,Kormos:2009sk,Smith:2020rru,Gaiotto:2020fdr,Popov:2025cha}. Using the methods of \cite{Smith:2020rru}, one can see that the IR defect corresponds to chiral fermion parity for a single Weyl-Majorana mode only, which under bosonization is mapped to the duality defect of the $c=1$ compact boson.
This defect is stable, as there are no relevant bulk operators (including $\bZ_2$-twisted fields) which preserve both the left and right $\bZ_2$ action, but also has an exactly marginal mass deformation. The schematic flow is described in Figure \ref{fig: flowlat}.
Notice however that the topological defect is not stable under bulk deformation: if we add a $ZZ$ term in the bulk (reducing the radius of the IR scalar) the energy operator $\epsilon=V_{0,1}$ becomes relevant and drives the system to the stable (pinning) configuration: the two hard walls.

\paragraph{Heisenberg chain} Another interesting limit is the Heisenberg chain $J_X = J_Y = J_Z = J$, in which the system at long distances is described by the SU(2)$_1$ CFT.  The $SU(2)_1$ CFT (i.e. the free boson at $R=1$) is not self-dual under gauging the $\bZ_2^m$ symmetry, and as such there is no perfectly transmissive impurity to deform available.\footnote{The $SU(2)_1$ has another $\bZ_2$ symmetry, acting as the center of the chiral $SU(2)_L$ symmetry under whose gauging the theory is self-dual. Our lattice symmetry instead is part of the vector symmetry, but not central.}

However, we can consider as a starting point the direct product of two $\bZ_2$ symmetric boundary conditions:
\be
\calD_0 = | f \rrangle \llangle f | \, ,
\ee
in free boson notation, $|f \rrangle$ is a Neumann boundary condition.
This defect has no defect anomaly. To remedy this, we stack a decoupled QM describing a spin 1/2 particle $z$:
\be
S_{\text{QM}} = i \int z^\dagger_i  \dot{z}^i \, , \quad  z^\dagger z = 2 S \, .
\ee
which carries a $\bZ_2 \times \bZ_2 \subset SO(3)$ anomaly \cite{Komargodski:2025jbu}
\be
\omega(A) = i \pi \int  \omega_2(A) = i \pi \int A_1 \cup A_2 \, ,
\ee
and couple the two $\bZ_2$ symmetries to the bulk $U_L$ and $U_R$, respectively
\be
\begin{tikzpicture}
    \fill[gray!15] (0,-1) rectangle (1.0,2);
    \draw (0,-1) -- (0,2) node[above] {$|f\rrangle$};
    \draw (1,-1) -- (1,2) node[above] {$\llangle f |$};
    \draw[thick] (0.5,-1) node[below] {$S=1/2 \ \text{QM}$} -- (0.5,2);
    \draw[red] (-2,1) node[left,black] {$U_L$} -- (0,1); \draw[red!50] (0,1) -- (0.5,1);
     \draw[red] (3,0) node[right,black] {$U_R$} -- (1,0); \draw[red!50] (1,0) -- (0.5,0);
     \node at (-2,2) {$SU(2)_1$}; \node at (3,2) {$SU(2)_1$};
\end{tikzpicture}
\ee
The $SU(2)$ generators are $S^a = \frac{1}{2} z^\dagger \sigma^a z$. This recovers exactly the expression \eqref{eq: ulur} for the defect symmetry. Recall that, at the critical point $SU(2)_1$, a bulk spin excitation $S^a_j$ ($a=X,Y,Z)$ decomposes in the $\Delta = 1$ current $J^a$ and the $\Delta=1/2$ $SU(2)_1$ spin field $n^a$ via:
\be
S^a_j \sim J^a(x) + (-)^j n^a(x) \, , \quad x = a j \, ,
\ee
up to some non-universal coefficients. Plugging this expression in the defect interaction leads to:\footnote{Notice also that the other exactly marginal current deformations $J_\perp^a$ are forbidden by parity symmetry. }
\be
H_{I}^{SU(2)_1} \sim \widetilde{\delta} \left( n^X_L S^Y + S^X n^X_R \right) + \widetilde{\gamma} S^Z \left( n^X_L n^Y_R - n^Y_L n^X_R \right) \, . 
\ee
The second interaction is marginal and needs to be studied in conformal perturbation theory. In the free boson language $n^X = \cos(\phi)$, $n_Y=\sin(\phi)$, and the interaction is $S^Z \sin(\phi_L - \phi_R) \equiv S^Z \cO$.
The first term in the $\cO \times \cO$ OPE is the kinetic term $\partial \phi_L \partial \phi_R = \frac{1}{2}\left( \partial \phi_+^2 - \partial \phi_-^2 \right)$. Thus, the perturbation changes the radius of the boson at the defect, and, with the correct sign, the perturbation can be made relevant by the BKT mechanism. The overall situation differs from the one at the $XX$ point, where the second deformation is instead exactly marginal. However, also in this case, $\calD(0,\gamma)$ is the direct sum of two separate defects.
These are simple to discuss,  by minimizing (maximizing) the sine, they impose
\be
\phi_L - \phi_R = \pm \pi/2 \, ,
\ee
at the defect. Thus, a charged vertex operator on the left $e^{i \phi_L}$ becomes a charged and local vertex operator $e^{i \phi_R}$ on the right. 
To recover the $\bZ_2 \times \bZ_2$ charge conservation this must be dressed by a topological defect operator, which is precisely the zero-mode $x = Z_m$ which exists as a consequence of \eqref{eq: decomposition}. Indeed, the presence of the defect zero mode can be immediately verified by time evolution with the defect Hamiltonian.

The first interaction instead is strongly relevant $(\Delta = 1/2)$ and drives the defect to a new interacting fixed point. To solve it we take a mean field approach and diagonalize the spin variable in a fixed scalar background. The Hamiltonian $A \cdot S$, with $A= (A_X, A_Y))$ planar, has eigenvalues $\pm \sqrt{A_X^2 + A_Y^2}$ and eigenvector along the $(A_X,A_Y)$ direction. At $R=1$ $n^X \sim \cos(\phi)$, so $(A_X,A_Y) = (\pm,\pm)$ and the IR defect is a direct sum of 4 Dirichlet bc:
\be
\calD = |D,0\rrangle \llangle D, 0| \oplus  |D,\pi\rrangle \llangle D, 0| \oplus  |D,0\rrangle \llangle D, \pi| \oplus  |D,\pi\rrangle \llangle D, \pi| \, .
\ee
The $\bZ_2^L \times \bZ_2^R$ symmetry exchanges the four by flipping the coordinates of the eigenvector.


\paragraph{Acknowledgment} We thank Clay Cordova, Gabriel Cuomo, Davide Gaiotto, Marco Meineri, Brandon Rayhaun, Yifan Wang for discussions. The work of A.A. is supported by the UKRI Frontier Research Grant, underwriting the ERC Advanced Grant ``Generalized Symmetries in Quantum Field Theory and Quantum Gravity". C.C. is supported by STFC grant ST/X000761/1. The research of G.G. is funded through an ARC advanced project, and further supported by IISN- Belgium (convention 4.4503.15). C.C. and G.G. thank the Isaac Newton Institute for Mathematical Sciences, Cambridge, for support and hospitality during the program ``Quantum field theory with boundaries, impurities, and defects". A.A. acknowledges the hospitality of Perimeter Institute for Theoretical Physics, where part of this work was completed.

\appendix

\section{Tube to Strip in the SymTFT}\label{app: symtft} In this appendix, we explain the map: $\text{Rep}(\text{Tube})(\cS) \to \text{Rep}(\text{Strip})_{\calM}(\cS)$ from the boundary SymTFT \cite{Bhardwaj:2024igy,Choi:2024tri,Copetti:2024onh}. See \cite{Freed:2022qnc,Gaiotto:2020iye,Apruzzi:2021nmk} for details regarding the SymTFT construction.
To describe a system with a boundary, we supply the standard SymTFT with three topological boundary conditions: \begin{itemize}
    \item $B_{\text{phys}}$, the dynamical boundary condition encoding the input QFT.
    \item $B_\cS$, the first topological boundary condition encoding the symmetry of the theory.
    \item $B_{\cM}$ the second topological boundary condition describing the action of symmetry on the boundary.
\end{itemize}
The three boundaries are connected in the following manner to generate the SymTFT description of a boundary condition for the theory $\cT$:
\be
\begin{tikzpicture}[baseline={(0,0)}]
\filldraw[fill=blue!10] (0,0) -- (0,2) -- (2,2) -- (2,0) -- cycle;
 \node[above right] at (0,0) {$B_{\cS}$};
\filldraw[fill=gray!10, opacity=0.75] (2,0) -- (2,2) -- (5,0) -- (5,-2) -- cycle;
     \node[below right] at (2,1.5) {$B_{\cM}$};
\node[below] at (2,0) {$m$};
\draw[opacity=0.25] (0,0) -- (3,-2); \draw[opacity=0.25] (0,2) -- (3,0);  \draw[opacity=0.25] (2,2) -- (5,0); \draw (2,0) -- (5,-2);
\begin{scope}[shift={(3,-2)}]
    \filldraw[fill=white, opacity=0.2] (0,0) -- (0,2) -- (2,2) -- (2,0) -- cycle;
    \node[above right] at (0,0) {$B_{\text{phys}}$};
    \end{scope}
\end{tikzpicture}
\quad = \quad 
\begin{tikzpicture}[baseline= {(0,2)}]
\fill[gray!15] (0,0) rectangle (1.0,4.0);
\draw (0,0) -- (0,4);
\node[below] at (0,0) {$m$};
\node at (-1,2) {$\cT$};
\end{tikzpicture}
\ee
A generalized charge \cite{Bhardwaj:2023ayw} corresponds to a representation of the Tube algebra \cite{Lin:2022dhv}, and is described via a bulk line $\cL_\mu$ ending on $B_{\cS}$:
\be
    \begin{tikzpicture}[baseline={(0,0)}]
\filldraw[fill=blue!10] (0,0) -- (0,2) -- (2,2) -- (2,0) -- cycle;
 \node[above right] at (0,0) {$B_{\cS}$};
\filldraw[fill=gray!10, opacity=0.75] (2,0) -- (2,2) -- (5,0) -- (5,-2) -- cycle;
     \node[below right] at (2,1.5) {$B_{\cM}$};
\node[below] at (2,0) {$m$}; \node[above] at (2,2) {$n$}; \node[above] at (1.5,1) {$a$};

\draw[opacity=0.25] (0,0) -- (3,-2); \draw[opacity=0.25] (0,2) -- (3,0);  \draw[opacity=0.25] (2,2) -- (5,0); \draw (2,0) -- (5,-2);
\begin{scope}[shift={(3,-2)}]
    \filldraw[fill=white, opacity=0.2] (0,0) -- (0,2) -- (2,2) -- (2,0) -- cycle;
    \node[above right] at (0,0) {$B_{\text{phys}}$};
    \end{scope}
    \draw[very thick, ->-]  (4,-1) -- (1,1); \draw[blue,thick,->-] (1,1) -- (2,1) node[black,above left] {\small$i$};
    \draw[fill=black] (1,1) node[left] {\small$x$} circle (0.05);     \draw[fill=black] (4,-1) circle (0.05);
    \node[above] at (3.75,-0.75) {$\cL_\mu$};
\end{tikzpicture}
\quad = \quad 
\begin{tikzpicture}[baseline= {(0,2)}]
\fill[gray!15] (0,0) rectangle (1.0,4.0);
\draw (0,0) -- (0,4);
\node[below] at (0,0) {$m$};
\node[above] at (0,4) {$n$};
\draw[blue, ->-] (-2,2) -- (0,2) node[black, above left] {$i$} ;
\node[above] at (-1,2) {$a$};
\draw[fill=black] (-2,2) node[above] {$\phi_{a,x}^\mu$} circle (0.05);
\end{tikzpicture}
\ee
The tube algebra acts linearly on the vector space:
\be
V_\mu = \bigoplus_{a} \text{Hom}(\mu,a) \, .
\ee
In the presence of the topological boundary $B_\cM$, the bulk symmetry instead acts, via the Strip algebra, on the space:
\be
W_{\mu} = \bigoplus_{m,n} W_\mu^{m,n} \, , \qquad W_\mu^{m,n} = \bigoplus_{a} \text{Hom}(\mu,a) \otimes \text{Hom}(a\otimes m,n) \, .
\ee
On the other hand, irreps of the Strip algebra are encoded in the topological lines $K^v$ confined on $B_\cM$ \cite{Bhardwaj:2024igy,Cordova:2024iti,Copetti:2024dcz}:
\be
    \begin{tikzpicture}[baseline={(0,0)}]
\filldraw[fill=blue!10] (0,0) -- (0,2) -- (2,2) -- (2,0) -- cycle;
 \node[above right] at (0,0) {$B_{\cS}$};
\filldraw[fill=gray!10, opacity=0.75] (2,0) -- (2,2) -- (5,0) -- (5,-2) -- cycle;
     \node[below right] at (2,1.5) {$B_{\cM}$};
\node[below] at (2,0) {$m$}; \node[above] at (2,2) {$n$};
\draw[opacity=0.25] (0,0) -- (3,-2); \draw[opacity=0.25] (0,2) -- (3,0);  \draw[opacity=0.25] (2,2) -- (5,0); \draw (2,0) -- (5,-2);
\begin{scope}[shift={(3,-2)}]
    \filldraw[fill=white, opacity=0.2] (0,0) -- (0,2) -- (2,2) -- (2,0) -- cycle;
    \node[above right] at (0,0) {$B_{\text{phys}}$};
    \end{scope}
    \draw[->-, very thick]  (5,-1) node[below left] {$K^v$} -- (2,1); 
    
\end{tikzpicture}
\quad = \quad 
\begin{tikzpicture}[baseline= {(0,2)}]
\fill[gray!15] (0,0) rectangle (1.0,4.0);
\draw (0,0) -- (0,4);
\draw[->-] (1,2) -- (0,2);
\node[above] at (0.5,2) {$K^v$};
\draw[fill=black] (0,2) circle (0.05);
\node[below] at (0,0) {$m$};
\node[above] at (0,4) {$n$};
\end{tikzpicture}
\ee
which act on the vector space:
\be
K^v = \bigoplus_{m,n} K^v_{m,n} \, , \qquad K^v_{m,n} = \text{Hom}(v\otimes m, n) \, .
\ee
Given a topological boundary condition $B$, bulk lines $\cL_\mu$ have a bulk-to-boundary expansion:
\be
\begin{tikzpicture}[baseline= {(0,0)}]
\filldraw[fill=gray!10, opacity=0.75] (2,0) -- (2,2) -- (5,0) -- (5,-2) -- cycle;
\draw[very thick,->-] (5,-1.6666666)  -- (1,1) node[above] {$\cL_\mu$};
\node at (2.25,1.5) {$B$};
\end{tikzpicture} = \sum_{v,x} \qquad
\begin{tikzpicture}[baseline= {(0,0)}]
    \filldraw[fill=gray!10, opacity=0.75] (2,0) -- (2,2) -- (5,0) -- (5,-2) -- cycle;
    \node at (2.25,1.5) {$B$};
 \draw[->-]  (4,-0.3333333) -- (3,0.3333333);

    \draw[very thick, ->-]
        (5,-1.66666666)
       --
        (4,-0.33333333);

    \draw[very thick, ->-]
        (3,0.33333333)
      --
        (1,1)  node[above] {$\cL_\mu$};
        \draw[fill=black]   (4,-0.33333333) node[above] {$z$} circle (0.05);
           \draw[fill=black]  (3,0.33333333) node[above] {$z^\dagger$} circle (0.05);
           \node[above] at (3.55,0.1) {$K^v$};
\end{tikzpicture}
\ee
leading to the OPE of lines
\be
t_\cM[\cL_\mu] = \sum_v n_{\mu}^v K^v \, , \quad n_\mu^v = \sum_x 1 \, .
\ee
We can now apply this to $B=B_\cM$ to convert a Tube algebra representation $\cL_\mu$ into a (reducible) Strip algebra representation on $B_\cM$: given a vector
\be
|w_a^{x,i}\rangle \in W^\mu_{m,n} \, ,
\ee
the bulk to-boundary expansion of $\cL_\mu$ allows to express it as a linear combination of vectors in $K^v_{m,n}$. This shows the claimed map from the perspective of the SymTFT.

\paragraph{Example: Fibonacci} The simplest example is the Fibonacci category Fib. It's center is $\cZ(\text{Fib}) = \text{Fib}^2$. There is only one topological boundary condition, corresponding to the condensation of the diagonal anyon $(\unit,\unit) + (W,W)$. Bulk topological lines $(\unit,\unit), \, (\unit,W) , \, (W,\unit) , \, (W,W)$ have the following expansion:
\bea
&(\unit,\unit) \to \unit \, , \\
&(\unit,W) \to W \, , \qquad (W,\unit) \to W \, , \\
&(W,W) \to \unit \oplus W \, .
\eea
The Tube algebra thus has three 1-dimensional representations $(\unit,\unit), \, (\unit,W) , \, (W,\unit)$ and one 2-dimensional representation $(W,W)$. 
The Strip algebra instead has two irreps, $\unit$, which is two dimensional (both $K^\unit_{\unit\unit}$ and $K^\unit_{WW}$ are 1-dimensional), and $W$, which is three dimensional ($K^W_{\unit W}, \, K^W_{W\unit} , \, K^W_{WW}$ are 1-dimensional).

Let us look at the branching: $(\unit,\unit)$ carries a 1d Tube representation, but it acts on both $\bH_{11}$ and $\bH_{WW}$. Thus it branches in a 2d Strip algebra representation. Similarly, $(\unit,W)$ has a 1d Tube representation, and branches into the 3d Strip representation by acting on $\bH_{\unit W}, \, \bH_{W\unit} , \, \bH_{WW}$. Finally, $(W,W)$ carries the 2d Tube representation, and the total space $W_W$ on which the Strip algebra acts has dimension:
\be
\text{dim}(W_W) = 1 \times 2 + 1 \times 3 = 5 \, ,
\ee
which is compatible only with the branching $W_W = K_\unit \oplus K_W$.

\section{Symmetry-Reflecting Defect via Discrete Gauging}\label{app: gauging}
In \cite{Antinucci:2024izg} we have shown that symmetry reflecting surface defects can be constructed by coupling anomalous (1+1)d matter to the bulk gauge field magnetically.

We describe a similar construction which applies to (1+1)d interfaces. We focus on the case in which the bulk theory has the same anomaly-free $\bZ_n$ symmetry both on the left and on the right. We do this starting from a $\bZ_n$ symmetric defect, with a nontrivial defect anomaly. After folding, we gauge the full $G = \bZ_n \times \bZ_n$ symmetry and produce a new $\bZ_n \times \bZ_n$ symmetry reflecting defect $\widetilde{\calD}$ with the nontrivial defect anomaly.

\paragraph{Gauging and folding} We consider the folded system. A defect can be described as a symmetric interface between the dynamical theory $\cT = \cT_L \times \overline{\cT}_R$ and a symmetric gapped phase $\cF$ (a TQFT):
\be
\begin{tikzpicture}
 \draw[color=white, fill = white!90!cyan] (0,0) -- (0,2) -- (1,2) -- (1,0) -- cycle;
     \draw[blue, very thick] (0,0) node[below] {$B_\cF$} -- (0,2);
     \node at (-1,1) {$\cT$};
     \node at (0.5,1) {$\cF$};
\end{tikzpicture}
\ee
$\cF$ captures the symmetry breaking pattern on the interface. Defect anomalies for $B_\cF$ arise when $\cF$ contains a nontrivial SPT phase for the relevant symmetry. We leverage this to understand the symmetry realization on $B_{\cF/\cA}$ after gauging a symmetry $\cA$:
\be
\begin{tikzpicture}[baseline={(0,1)}]
 \draw[color=white, fill = white!90!cyan] (0,0) -- (0,2) -- (1,2) -- (1,0) -- cycle;
     \draw[blue, very thick] (0,0) node[below] {$B_\cF$} -- (0,2);
     \node at (-1,1) {$\cT$};
     \node at (0.5,1) {$\cF$};
\end{tikzpicture}
\quad \longrightarrow \quad 
\begin{tikzpicture}[baseline={(0,1)}]
 \draw[color=white, fill = white!90!green] (0,0) -- (0,2) -- (1,2) -- (1,0) -- cycle;
     \draw[green!80!black, very thick] (0,0) node[below] {$ B_{\cF/\cA}$} -- (0,2);
     \node at (-1,1) {$\cT/\cA$};
     \node at (0.5,1) {$\cF/\cA$};
\end{tikzpicture}
\ee

\paragraph{Internal symmetries} Consider $G= \bZ_2 \times \bZ_2$ and a symmetric defect $\calD$, with nontrivial defect anomaly. We fold the system and gauge $G \times G$. 
Before gauging, the defect was described by a TQFT which preserves the diagonal subgroup $G_{\text{diag}}$ via a nontrivial SPT $\alpha$, while breaking the quotient:
\be
\cF = \text{SPT}_\alpha (G_{\text{diag}}) \otimes \text{SSB}(G \times G/ G_{\text{diag}}) \, .
\ee
The partition function for the SSB phase is $Z_{SSB} = \delta (A_L = A_R)$, thus we get:
\bea
Z_{\cF/G\times G} [A_L,A_R,B_L,B_R] = \sum_{a, b} \exp\left(i \pi \int a b + a (A_L + A_R) + b (B_L + B_R) \right)& \\
= \exp \left( i \pi \int (A_L + A_R)(B_L + B_R) \right) = \exp\left( i \pi \int A_L B_R + A_R B_L \right) = \text{SPT}_{\widehat{\alpha}}& \, .
\eea
Thus the defect after gauging preserves $\widehat{G} \times \widehat{G}$ (is symmetry reflecting) with the defect anomaly $\widehat{\alpha}$. The computation can be generalized to $G=\bZ_n \times \bZ_n$ in the obvious way.

\paragraph{Fermion parity} Another key case is that of a fermionic system, in which the fermion parity symmetry is preserved by the defect with a nontrivial defect anomaly given by the Arf-phase:
\be
\alpha = \text{Arf} \, .
\ee
The computation is similar. $\cF$ breaks $(-)^{F_L} \times (-)^{F_R}$ to the diagonal subgroup, with an Arf SPT. 
The bosonization of this system is
\be
Z_{\text{bos}}[B] = \text{Arf}(B\cdot \rho) \text{Arf}(\rho) \sum_{b} Z_f[b] e^{i \pi \int b B} \, .
\ee
Using that \cite{Karch:2019lnn}:
\be
\sum_b \text{Arf}(b \cdot \rho) e^{i \pi \int b \cup B} = \text{Arf}(B\cdot \rho) \text{Arf}(\rho) \, ,
\ee
and 
\be
\text{Arf}((A+B)\cdot \rho) =\text{Arf}(A \cdot \rho) \text{Arf}(B \cdot \rho) e^{i \pi \int A B} \, . 
\ee
We find:
\bea
Z_{\cF_{bos}}[A_L,A_R] &= \sum_a \text{Arf}(a \cdot \rho) e^{i \pi \int a (A_L + A_R)} \text{Arf}(A_L\cdot \rho)  \text{Arf}(A_R\cdot \rho) \\
&= e^{i \pi \int A_L A_R} = \text{SPT}_{\alpha} \, .
\eea
Thus bosonizing a fermionic defect with an Arf defect anomaly gives a symmetry reflecting $\bZ_2 \times \bZ_2$ defect with defect anomaly $\alpha = i \pi \int A_L A_R$.

Both examples are known from previous studies of non-invertible symmetries:
\begin{itemize}
    \item The first case is realized by the duality defect $\cN$ of Rep$(D_8)$. This can be obtained by (twisted) gauging $\bZ_2 \times \bZ_2$ in a $\bZ_2^3$ symmetric theory, with type III anomaly $i \pi \int A_1 A_2 A_3 $. The $A_1$ defect has defect anomaly $\alpha = i \pi \int A_2 A_3$ and becomes $\cN$.
    \item In the second case, the chiral fermion parity $(-)^{F_\chi}$ of the Majorana fermion has a mixed anomaly with $(-)^F$. This leads to a defect anomaly for $(-)^{F_\chi}$. Upon bosonization the system becomes the Ising CFT, and $(-)^{F_\chi}$ becomes the KW duality defect $\cN$.
\end{itemize}
These techniques can be applied to many other cases, giving rise to a plethora of symmetry reflecting defects with the wanted defect anomaly.

\section{Details on the Tricritical Ising Flow and the  Defect Strip Algebra}\label{app: tricrit}
In the main text we have analyzed the integrable boundary flow obtained by deforming the tricritical Ising CFT with the relevant deformation $\sigma'$. In this appendix we provide some extra details on this analysis.

As well known, among the full set of the topological lines present in the tricritical Ising CFT, a distinguished subset is the one generating the so--called Fibonacci subcategory, also dubbed Fib. The Fibonacci fusion category is a rank-2 category with objects $\{\unit, W\}$, fusion rules
\be
W \times W = \unit + W\,,
\ee
and one only non-trivial F-symbol
\be
F^W_{WWW}=\begin{pmatrix}
    \zeta^{-1} & \zeta^{-1} \\1&-\zeta^{-1} 
\end{pmatrix}\,,
\ee
where $\zeta = \frac{1+\sqrt{5}}{2}$ is the golden ratio.

To compute various observables, it is usually convenient to employ a diagrammatic calculus in which the object $W$ is represented by an unlabeled line, while $\unit$ is represented either by the absence of a line or, depending on the context, by a dashed line. In this notation, some of the identities we use are the following:
\be
\begin{tikzpicture}[baseline=-0.5ex, line cap=round, line width=1.2pt]

\draw[black] (0.7,0) circle (0.45);
\node at (1.6,0) {$=$};
\node at (2.1,0) {$\zeta$};

\draw[black] (4.2,0) -- (4.85,0);
\draw[black] (5.3,0) circle (0.45);
\draw[black] (5.75,0) -- (6.4,0);

\node at (7.2,0) {$=$};
\node at (8.0,0) {$\zeta$};

\draw[black] (8.9,0) -- (10.1,0);

\begin{scope}[yshift=-1.8cm]
\draw[black] (0,0) -- (1.6,0);
\draw[black] (0.45,0) -- (0.45,0.9);
\draw[black] (1.15,0) -- (1.15,0.9);

\node at (2.1,0.35) {$=$};

\draw[black] (4.2,0) -- (5.8,0);
\draw[black] (4.6,0.9) arc (180:360:0.4);

\node at (7.2,0.35) {$-$};
\node at (8.0,0.35) {$\zeta^{-1}$};

\draw[black] (8.9,0) -- (10.5,0);
\draw[black] (9.7,0) -- (9.7,0.45);
\draw[black] (9.3,0.9) arc (180:360:0.4);
\end{scope}

\begin{scope}[yshift=-3.2cm]
\draw[black] (0,0) -- (0.45,0);
\draw[black] (0.45,0) -- (0.45,0.9);
\draw[black] (1.15,0.9) -- (1.15,0);
\draw[black] (1.15,0) -- (1.6,0);

\node at (2.1,0.35) {$=$};
\node at (2.9,0.35) {$\zeta^{-1}$};

\draw[black] (4.2,0) -- (5.8,0);
\draw[black] (4.6,0.9) arc (180:360:0.4);

\node at (7.2,0.35) {$+$};
\node at (8.0,0.35) {$\zeta^{-1}$};

\draw[black] (8.9,0) -- (10.5,0);
\draw[black] (9.7,0) -- (9.7,0.45);
\draw[black] (9.3,0.9) arc (180:360:0.4);
\end{scope}

\end{tikzpicture}
\ee

Let's consider how the primary fields organize into representations of the Tube algebra of Fibonacci category. First we recall that the local primaries (i.e. in the untwisted Hilbert space), with their charge under $W$ and its holomorphic weights, are 
\begin{equation}
    \begin{array}{|r|c|c|c|c|c|c|}
    \hline
    & 1 & \varepsilon & \varepsilon' & \varepsilon'' & \sigma & \sigma' \\\hline
    W : & \zeta & -\zeta^{-1} & -\zeta^{-1} & \zeta & -\zeta^{-1} & \zeta \\\hline
    h : &0 &\frac{1}{10} & \frac{3}{5}& \frac{3}{2}& \frac{3}{80} &\frac{7}{16} \\\hline
    \overline{h} : &0 &\frac{1}{10} &\frac{3}{5} &\frac{3}{2} & \frac{3}{80}& \frac{7}{16}\\\hline
\end{array}
\end{equation}
while the content of the $W$ twisted sector is
\begin{equation}
   \begin{array}{|c|c|c|c|}
\hline
\text{\textbf{Defect Operator }} & h & \bar{h} & \text{\textbf{Spin}} \\
\hline
\overline{\psi} & 0 & \frac{3}{5}  & +\frac{2}{5} \\
\hline
\psi & \frac{3}{5} & 0 & -\frac{2}{5} \\
\hline
\tilde{\varepsilon}' & \frac{3}{5} & \frac{3}{5} & 0 \\
\hline
\tilde{\varepsilon} & \frac{1}{10} & \frac{1}{10} & 0 \\
\hline
\Phi & \frac{1}{10} & \frac{3}{2} & -\frac{2}{5} \\
\hline
\overline{\Phi} & \frac{3}{2} & \frac{1}{10} &  +\frac{2}{5} \\
\hline
\tilde{\sigma} & \frac{3}{80} & \frac{3}{80} & 0 \\
\hline
\Lambda & \frac{3}{80} & \frac{7}{16} &-\frac{2}{5} \\
\hline
\overline{\Lambda} & \frac{7}{16} & \frac{3}{80}  & +\frac{2}{5} \\
\hline
\end{array}
\end{equation}
\paragraph{Fibonacci's tube algebra.}Since Fib is an MTC, its Drinfeld center is simply the product $\cZ(\text{Fib})= \text{Fib}  \times \overline{\text{Fib}}$, therefore there are $4$ simple anyons 
\be
\cZ(\text{Fib})=\{(\unit, \unit)\;,\; (\unit, W)\;,\; (W, \unit) \;,\;(W,W)\}\,,
\ee
meaning that the Tube algebra has only $4$ irreducible representations. Of these, the three representations $(\unit, \unit), (\unit, W), (W, \unit)$ are $1$-dimensional, while $(W,W)$ is $2$-dimensional. 
 
Now let's map this to the field content of the Tricritical Ising model. The lagrangian algebra that corresponds to the Fib-symmetric boundary condition (which in this case is also the unique one) is $\cL = (\unit, \unit) \oplus (W,W)$, and contains the only two spinless anyons of the Drinfeld center (the spin of $W$ is $2/5$ in Fib and $-2/5$ in $\overline{\text{Fib}}$). This means that the primaries of the untwisted sector transform in either the trivial $(\unit, \unit)$ irrep, or the non-trivial $2$d irrep $(W,W)$. Looking at the eigenvalues of the $W$ line on the untwisted sector primaries we see that the identity, $\varepsilon''$ and $\sigma'$ all transform in the trivial irrep, while $\varepsilon, \varepsilon'$ and $\sigma$ transform in the $2$d irrep. This is consistent with the content of the twisted Hilbert space: for all three untwisted primaries $\varepsilon, \varepsilon'$ and $\sigma$ there is a corresponding twisted sector primary with the same conformal dimensions which completes the doublet. The remaining operators in the $W$-twisted sector transform either in the $(\unit, W)$ or the $(W, \unit)$ $1$-dimensional irreps depending on the sign of their spin. More formally the Tube algebra is generated by $5$ non-trivial elements, the simplest is the $W$ line acting on the untwisted Hilbert space $\bH_{\unit}$

\begin{equation}
     U_5 = \begin{tikzpicture}[
        baseline={([yshift=-.5ex]current bounding box.center)},
        mid arrow/.style={
            postaction={
                decorate,
                decoration={
                    markings,
                    mark=at position 0.5 with {\arrow{Stealth}}
                }
            }
        }
    ]
        
        \draw[thick] (0,1) -- (1,1);
        
        \draw[thick] (0,1) -- (-1,1);
        
        \draw[thick] (0.95, 0.8) -- (1.05, 1.2);
        \draw[thick] (1.05, 0.8) -- (1.15, 1.2);
        
        \draw[thick] (-1.05, 0.8) -- (-0.95, 1.2);
        \draw[thick] (-1.15, 0.8) -- (-1.05, 1.2);
    \end{tikzpicture}\, : \bH_{\unit}\rightarrow \bH_{\unit}\, , 
\end{equation}
we then have maps acting within the twisted Hilbert space $\bH_{W}$, 
\begin{equation}
 U_1 = \begin{tikzpicture}[
        baseline={([yshift=-.5ex]current bounding box.center)},
        mid arrow/.style={
            postaction={
                decorate,
                decoration={
                    markings,
                    mark=at position 0.5 with {\arrow{Stealth}}
                }
            }
        }
    ]
        \draw[thick] (0,0) -- (0,3);
        
        \draw[thick] (0,1) -- (1,1);
        
        \draw[thick] (0,2) -- (-1,2);
        
        \draw[thick] (0.95, 0.8) -- (1.05, 1.2);
        \draw[thick] (1.05, 0.8) -- (1.15, 1.2);
        
        \draw[thick] (-1.05, 1.8) -- (-0.95, 2.2);
        \draw[thick] (-1.15, 1.8) -- (-1.05, 2.2);
    \end{tikzpicture}\, : \bH_{W}\rightarrow \bH_{W}\, , \qquad  U_2 = \begin{tikzpicture}[
        baseline={([yshift=-.5ex]current bounding box.center)},
        mid arrow/.style={
            postaction={
                decorate,
                decoration={
                    markings,
                    mark=at position 0.5 with {\arrow{Stealth}}
                }
            }
        }
    ]
        \draw[thick] (0,0) -- (0,1);
        \draw[dashed] (0,1) -- (0,2);
        \draw[thick] (0,2) -- (0,3);
        
        \draw[thick] (0,1) -- (1,1);
        
        \draw[thick] (0,2) -- (-1,2);
        
        \draw[thick] (0.95, 0.8) -- (1.05, 1.2);
        \draw[thick] (1.05, 0.8) -- (1.15, 1.2);
        
        \draw[thick] (-1.05, 1.8) -- (-0.95, 2.2);
        \draw[thick] (-1.15, 1.8) -- (-1.05, 2.2);
    \end{tikzpicture}\, : \bH_{W}\rightarrow \bH_{W} \, , 
\end{equation}
and maps between the two sectors
\begin{equation}
    U_3 = \begin{tikzpicture}[
        baseline={([yshift=-.5ex]current bounding box.center)},
        mid arrow/.style={
            postaction={
                decorate,
                decoration={
                    markings,
                    mark=at position 0.5 with {\arrow{Stealth}}
                }
            }
        }
    ]
        \draw[thick] (0,1) -- (0,2);
        
        \draw[thick] (0,1) -- (1,1);
        
        \draw[thick] (0,1) -- (-1,1);
        
        \draw[thick] (0.95, 0.8) -- (1.05, 1.2);
        \draw[thick] (1.05, 0.8) -- (1.15, 1.2);
        
        \draw[thick] (-1.05, 0.8) -- (-0.95, 1.2);
        \draw[thick] (-1.15, 0.8) -- (-1.05, 1.2);
    \end{tikzpicture}\, : \bH_{\unit}\rightarrow \bH_{W}\, , \qquad  U_4 = \begin{tikzpicture}[
        baseline={([yshift=-.5ex]current bounding box.center)},
        mid arrow/.style={
            postaction={
                decorate,
                decoration={
                    markings,
                    mark=at position 0.5 with {\arrow{Stealth}}
                }
            }
        }
    ]
        \draw[thick] (0,0) -- (0,1);
        
        \draw[thick] (0,1) -- (1,1);
        
        \draw[thick] (0,1) -- (-1,1);
        
        \draw[thick] (0.95, 0.8) -- (1.05, 1.2);
        \draw[thick] (1.05, 0.8) -- (1.15, 1.2);
        
        \draw[thick] (-1.05, 0.8) -- (-0.95, 1.2);
        \draw[thick] (-1.15, 0.8) -- (-1.05, 1.2);
    \end{tikzpicture}\, : \bH_{W}\rightarrow \bH_{\unit}\, .
\end{equation}
Their algebra easily follows from the $F$-symbols. Depending on which representation we want to describe only some of these elements will be non-trivial. Clearly in the trivial irrep only $U_5$ acts, while in the two non-trivial $1d$ irreps only $U_1$ and $U_2$ can act. Instead in the $2d$ irrep all $5$ elements are non-trivial. 

\paragraph{The Strip algebra.} The Fibonacci Strip algebra is easy to describe. It is generated by $9$ non-trivial elements (on top of those we have the identity in each of the $4$ strip Hilbert spaces), these can be loosely divided in two classes. The first acts on strip Hilbert spaces with equal boundary conditions
\begin{equation}
   N_1 =\begin{tikzpicture}[baseline,
    thick,
    ->-/.style={
        decoration={markings, mark=at position 0.55 with {\arrow{Latex}}},
        postaction={decorate}
    },
    dot/.style={
        circle, 
        draw=black, 
        fill=white, 
        inner sep=1.2pt
    }
]
    
    \draw[dashed] (0, 0) -- (0, 1.2);
    \draw[-] (0, 0) -- (0, -1.2);

    \draw[-] (1.5, -1.2) -- (1.5, 0);
    \draw[dashed] (1.5, 0) -- (1.5, 1.2);

    \draw[black, -] (0, 0) -- (1.5, 0);

    \node[dot] at (0,0) {};
    \node[dot] at (1.5,0) {};

    \node[below left, inner sep=3pt] at (2.4,0.3) {};

\end{tikzpicture}\, , N_2 = \begin{tikzpicture}[baseline,
    thick,
    ->-/.style={
        decoration={markings, mark=at position 0.55 with {\arrow{Latex}}},
        postaction={decorate}
    },
    dot/.style={
        circle, 
        draw=black, 
        fill=white, 
        inner sep=1.2pt
    }
]
   
    \draw[-] (0, 1.2) -- (0, 0);
    \draw[dashed] (0, 0) -- (0, -1.2);

    \draw[dashed] (1.5, -1.2) -- (1.5, 0);
    \draw[-] (1.5, 0) -- (1.5, 1.2);

    \draw[black, -] (0, 0) -- (1.5, 0);

    \node[dot] at (0,0) {};
    \node[dot] at (1.5,0) {};

    \node[below left, inner sep=3pt] at (2.4,0.3) {};

\end{tikzpicture}\, , N_3 = \begin{tikzpicture}[baseline,
    thick,
    ->-/.style={
        decoration={markings, mark=at position 0.55 with {\arrow{Latex}}},
        postaction={decorate}
    },
    dot/.style={
        circle, 
        draw=black, 
        fill=white, 
        inner sep=1.2pt
    }
]

    \draw[-] (0, 1.2) -- (0, 0) node[midway, left]{};
    \draw[-] (0, 0) -- (0, -1.2) node[midway, left]{};

    \draw[-] (1.5, -1.2) -- (1.5, 0) ;
    \draw[-] (1.5, 0) -- (1.5, 1.2) ;

    \draw[black, -] (0, 0) -- (1.5, 0) node[midway, above, text=black]{} ;

    \node[dot] at (0,0) {};
    \node[dot] at (1.5,0) {};

    \node[below left, inner sep=3pt] at (2.4,0.3){};

\end{tikzpicture}\, 
\end{equation}
while the second maps spaces with different boundary conditions 
\bea
   &  N_4 =\begin{tikzpicture}[baseline,
    thick,
    ->-/.style={
        decoration={markings, mark=at position 0.55 with {\arrow{Latex}}},
        postaction={decorate}
    },
    dot/.style={
        circle, 
        draw=black, 
        fill=white, 
        inner sep=1.2pt
    }
]
    
    \draw[-] (0, 0) -- (0, 1.2);
    \draw[dashed] (0, 0) -- (0, -1.2);

    \draw[-] (1.5, -1.2) -- (1.5, 0);
    \draw[dashed] (1.5, 0) -- (1.5, 1.2);

    \draw[black, -] (0, 0) -- (1.5, 0);

    \node[dot] at (0,0) {};
    \node[dot] at (1.5,0) {};

    \node[below left, inner sep=3pt] at (2.4,0.3) {};

\end{tikzpicture}\, , N_5 = \begin{tikzpicture}[baseline,
    thick,
    ->-/.style={
        decoration={markings, mark=at position 0.55 with {\arrow{Latex}}},
        postaction={decorate}
    },
    dot/.style={
        circle, 
        draw=black, 
        fill=white, 
        inner sep=1.2pt
    }
]
   
    \draw[dashed] (0, 1.2) -- (0, 0);
    \draw[-] (0, 0) -- (0, -1.2);

    \draw[dashed] (1.5, -1.2) -- (1.5, 0);
    \draw[-] (1.5, 0) -- (1.5, 1.2);

    \draw[black, -] (0, 0) -- (1.5, 0);

    \node[dot] at (0,0) {};
    \node[dot] at (1.5,0) {};

    \node[below left, inner sep=3pt] at (2.4,0.3) {};

\end{tikzpicture}\, , N_6 = \begin{tikzpicture}[baseline,
    thick,
    ->-/.style={
        decoration={markings, mark=at position 0.55 with {\arrow{Latex}}},
        postaction={decorate}
    },
    dot/.style={
        circle, 
        draw=black, 
        fill=white, 
        inner sep=1.2pt
    }
]

    \draw[-] (0, 1.2) -- (0, 0) node[midway, left]{};
    \draw[-] (0, 0) -- (0, -1.2) node[midway, left]{};

    \draw[-] (1.5, -1.2) -- (1.5, 0) ;
    \draw[dashed] (1.5, 0) -- (1.5, 1.2) ;

    \draw[black, -] (0, 0) -- (1.5, 0) node[midway, above, text=black]{} ;

    \node[dot] at (0,0) {};
    \node[dot] at (1.5,0) {};

    \node[below left, inner sep=3pt] at (2.4,0.3){};

\end{tikzpicture}\\
& N_7 =\begin{tikzpicture}[baseline,
    thick,
    ->-/.style={
        decoration={markings, mark=at position 0.55 with {\arrow{Latex}}},
        postaction={decorate}
    },
    dot/.style={
        circle, 
        draw=black, 
        fill=white, 
        inner sep=1.2pt
    }
]
    
    \draw[-] (0, 0) -- (0, 1.2);
    \draw[dashed] (0, 0) -- (0, -1.2);

    \draw[-] (1.5, -1.2) -- (1.5, 0);
    \draw[-] (1.5, 0) -- (1.5, 1.2);

    \draw[black, -] (0, 0) -- (1.5, 0);

    \node[dot] at (0,0) {};
    \node[dot] at (1.5,0) {};

    \node[below left, inner sep=3pt] at (2.4,0.3) {};

\end{tikzpicture}\, , N_8 = \begin{tikzpicture}[baseline,
    thick,
    ->-/.style={
        decoration={markings, mark=at position 0.55 with {\arrow{Latex}}},
        postaction={decorate}
    },
    dot/.style={
        circle, 
        draw=black, 
        fill=white, 
        inner sep=1.2pt
    }
]
   
    \draw[-] (0, 1.2) -- (0, 0);
    \draw[-] (0, 0) -- (0, -1.2);

    \draw[dashed] (1.5, -1.2) -- (1.5, 0);
    \draw[-] (1.5, 0) -- (1.5, 1.2);

    \draw[black, -] (0, 0) -- (1.5, 0);

    \node[dot] at (0,0) {};
    \node[dot] at (1.5,0) {};

    \node[below left, inner sep=3pt] at (2.4,0.3) {};

\end{tikzpicture}\, , N_9 = \begin{tikzpicture}[baseline,
    thick,
    ->-/.style={
        decoration={markings, mark=at position 0.55 with {\arrow{Latex}}},
        postaction={decorate}
    },
    dot/.style={
        circle, 
        draw=black, 
        fill=white, 
        inner sep=1.2pt
    }
]

    \draw[dashed] (0, 1.2) -- (0, 0) node[midway, left]{};
    \draw[-] (0, 0) -- (0, -1.2) node[midway, left]{};

    \draw[-] (1.5, -1.2) -- (1.5, 0) ;
    \draw[-] (1.5, 0) -- (1.5, 1.2) ;

    \draw[black, -] (0, 0) -- (1.5, 0) node[midway, above, text=black]{} ;

    \node[dot] at (0,0) {};
    \node[dot] at (1.5,0) {};

    \node[below left, inner sep=3pt] at (2.4,0.3){};

\end{tikzpicture}
\eea
The algebra again follows from the $F$-symbols. There are two irreducible representations of dimension $2$ and $3$, we can describe them explicitly defining the states on which they act. For the $2$-dimensional we set
\begin{equation*}
\ket{v_{W}}=\frac{1}{\zeta^{1/2}}\,\,\begin{tikzpicture}[baseline={(1,-0.5)},
    thick,
    ->-/.style={
        decoration={markings, mark=at position 0.55 with {\arrow{Latex}}},
        postaction={decorate}
    },
    dot/.style={
        circle, 
        draw=black, 
        fill=white, 
        inner sep=1.2pt
    }
]

    \draw[-] (0, 0.5) -- (0, 0);
    \draw[-] (0, 0) -- (0, -1.2);

    \draw[-] (1.5, -1.2) -- (1.5, 0) ;
    \draw[-] (1.5, 0) -- (1.5, 0.5) ;

    \draw[-] (0, -1.2) -- (1.5, -1.2);

    \draw[dashed] (1.5/2,-1.2) -- (1.5/2, -1.9);

\end{tikzpicture}\, , \quad \ket{v_{\unit}}=\,\,\begin{tikzpicture}[baseline={(1,-0.5)},
    thick,
    ->-/.style={
        decoration={markings, mark=at position 0.55 with {\arrow{Latex}}},
        postaction={decorate}
    },
    dot/.style={
        circle, 
        draw=black, 
        fill=white, 
        inner sep=1.2pt
    }
]

    \draw[dashed] (0, 0.5) -- (0, 0);
    \draw[dashed] (0, 0) -- (0, -1.2);
    \draw[dashed] (1.5, -1.2) -- (1.5, 0) ;
    \draw[dashed] (1.5, 0) -- (1.5, 0.5) ;
    \draw[dashed] (0, -1.2) -- (1.5/2, -1.2);
    \draw[dashed] (1.5/2, -1.2) -- (1.5, -1.2);
    \draw[dashed] (1.5/2,-1.2) -- (1.5/2, -1.9);

\end{tikzpicture}
\end{equation*}
 while for the $3d$ one we have
\begin{equation*}
\ket{v_{WW}}=\frac{1}{\zeta}\,\,\begin{tikzpicture}[baseline={(1,-0.5)},
    thick,
    ->-/.style={
        decoration={markings, mark=at position 0.55 with {\arrow{Latex}}},
        postaction={decorate}
    },
    dot/.style={
        circle, 
        draw=black, 
        fill=white, 
        inner sep=1.2pt
    }
]
    \draw[-] (0, 0.5) -- (0, 0);
    \draw[-] (0, 0) -- (0, -1.2);

    \draw[-] (1.5, -1.2) -- (1.5, 0) ;
    \draw[-] (1.5, 0) -- (1.5, 0.5) ;

    \draw[-] (0, -1.2) -- (1.5, -1.2);

    \draw[-] (1.5/2,-1.2) -- (1.5/2, -1.9) ;
\end{tikzpicture}\, , \quad \ket{v_{W\unit}}=\frac{1}{\zeta^{1/2}}\,\,\begin{tikzpicture}[baseline={(1,-0.5)},
    thick,
    ->-/.style={
        decoration={markings, mark=at position 0.55 with {\arrow{Latex}}},
        postaction={decorate}
    },
    dot/.style={
        circle, 
        draw=black, 
        fill=white, 
        inner sep=1.2pt
    }
]
    \draw[-] (0, 0.5) -- (0, 0);
    \draw[-] (0, 0) -- (0, -1.2);
    \draw[dashed] (1.5, -1.2) -- (1.5, 0) ;
    \draw[dashed] (1.5, 0) -- (1.5, 0.5) ;
    \draw[-] (0, -1.2) -- (1.5/2, -1.2);
    \draw[dashed] (1.5/2, -1.2) -- (1.5, -1.2);

    \draw[-] (1.5/2,-1.2) -- (1.5/2, -1.9);
   
\end{tikzpicture}
\end{equation*}
\begin{equation*}
\ket{v_{\unit W}}=\frac{1}{\zeta^{1/2}}\,\,\begin{tikzpicture}[baseline={(1,-0.5)},
    thick,
    ->-/.style={
        decoration={markings, mark=at position 0.55 with {\arrow{Latex}}},
        postaction={decorate}
    },
    dot/.style={
        circle, 
        draw=black, 
        fill=white, 
        inner sep=1.2pt
    }
]
    \draw[dashed] (0, 0.5) -- (0, 0);
    \draw[dashed] (0, 0) -- (0, -1.2) ;
    \draw[-] (1.5, -1.2) -- (1.5, 0) ;
    \draw[-] (1.5, 0) -- (1.5, 0.5) ;
    \draw[dashed] (0, -1.2) -- (1.5/2, -1.2);
    \draw[-] (1.5/2, -1.2) -- (1.5, -1.2);

    \draw[-] (1.5/2,-1.2) -- (1.5/2, -1.9);

\end{tikzpicture}\, .
\end{equation*}
It is now easy to obtain the explicit representations. For example the in the $2d$ irrep only the $3$ maps $N_1, N_2$ and $N_3$ are non-trivial and we get
\begin{equation}
N_1 \ket{v_W}= \zeta^{1/2}\ket{v_{\unit}}\, , \qquad N_2 \ket{v_{\unit}}= \zeta^{1/2}\ket{v_{W}}\, , \qquad N_{3} \ket{v_W}= \zeta  \ket{v_W}\, .
\end{equation}
One can similarly obtain the explicit maps of the $3d$ irrep.

\paragraph{Tubes and Strips.} In this example we can describe more explicitly the relation between Tube and Strip irreps. First we consider the trivial Tube irrep, which acts on $\bH_{\unit}= \bH_{\unit \unit}\oplus \bH_{WW}$. At the level of the Strip only $N_1, N_2$ and $N_3$ act within $\bH_{\unit}$, so the corresponding irrep is the $2d$ one. The element $U_5$ of the Tube is the only one acting non-trivially and its explicit realization in terms of the Strip algebra is (we fix the coefficients imposing the Tube algebra fusion rules)
\begin{equation}
    U_5=\frac{1}{\zeta^{1/2}}\left(N_1+N_2\right) + \frac{1}{\zeta}N_3\, .
\end{equation}
It acts by 
\begin{equation}
    U_5\ket{v_{\unit}}=\ket{v_{W}}\, , \qquad U_5\ket{v_{W}}=\ket{v_{\unit}}+\ket{v_{W}}\, ,
\end{equation}
which is what one expects for the vacua of the theory. Let's now consider the other Tube irreps. The non-trivial $1d$ ones act within $\bH_{W}=\bH_{W\unit}\oplus\bH_{\unit W}\oplus \bH_{WW}$ which also carries the $3d$ irrep of the strip algebra. Finally the $2d$ Tube irrep acts on
\begin{equation}
    \bH_{\unit}\oplus \bH_{W}=\bH_{\unit \unit}\oplus \bH_{WW}\oplus \bH_{W\unit}\oplus\bH_{\unit W}\oplus \bH_{WW}\, , 
\end{equation}
which carries both the $2d$ and $3d$ Strip algebra irreps.

\small{
\bibliographystyle{ytphys}
\baselineskip=0.75\baselineskip
\bibliography{bibdef}

}

\end{document}